\documentclass{revtex4-2}

\usepackage[utf8]{inputenc}
\usepackage[english]{babel}
\usepackage[T1]{fontenc}

\usepackage[hidelinks]{hyperref}

\usepackage{graphicx}
\usepackage{subcaption}
\usepackage{parskip}

\usepackage{float}
\usepackage{booktabs}
\usepackage{makecell}
\usepackage{multirow}

\usepackage[figurename=Fig.]{caption}
\usepackage[tablename=Tab.]{caption}
\usepackage[font=small]{caption}

\usepackage{amsmath}
\usepackage{amssymb}
\usepackage{amsfonts}
\usepackage{mathtools}

\usepackage{upgreek}
\usepackage{xcolor}
\usepackage[normalem]{ulem}

\begin{document}

\newcommand{\bh}[1]{{\textcolor{blue}{{#1}}}}
\newcommand{\jk}[1]{{\textcolor{green}{{#1}}}}


\title{Accretion Structures around Kerr Black Holes in a Swirling Background}

\date{\today}

\author{Kristian Gjorgjieski}
 \email{kristian.gjorgjieski@uol.de}
\affiliation{Department of Physics, Carl von Ossietzky University of Oldenburg, 26111 Oldenburg, Germany}

\author{Rog\'{e}rio Capobianco}
 \email{rogerio.capobianco@gmail.com}
\affiliation{Instituto de F\'{ı}sica de S\~{a}o Carlos, Universidade de S\~{a}o Paulo, S\~{a}o Carlos, S\~{a}o Paulo 13560-970, Brazil}

\begin{abstract}
In this paper, we investigate thick accretion structures around a Kerr black hole in a swirling background, that is, a rotating black hole immersed in a rotating background. This is a novel solution characterized by the black hole mass in addition to two distinct rotational parameters, the Kerr parameter $a$, which identifies the rotation of the black hole, and the swirling parameter $j$, which describes the background rotation. The swirling background is characterized by an odd $\mathcal{Z}_2$ symmetry, where the northern and southern hemispheres rotate in opposite directions. The rotation of the black hole embedded into this swirling background leads to non-trivial spin-spin interactions with the background rotation. The spacetime properties in the vicinity of the black hole are significantly influenced by this spin-spin interaction. In order to study the influence on the basic properties of this spacetime, we analyze circular orbits and geometrically thick disks for different spacetime solutions, which are classified by the black hole and swirling spins. We identify stabilizing effects on prograde circular orbits and destabilizing effects on retrograde circular orbits, which originate from the spin-spin interaction and depend mainly on the Kerr rotation. Furthermore, we discovered the emergence of static orbits, which appear due to the background rotation. The symmetry breaking of the spacetime rotation with regard to the equatorial plane highly influences the spatial distribution of circular orbits. This asymmetry causes a concave (convex) distribution of the prograde (retrograde) circular orbits and accordingly, bowl-like deformations of the accretion disk solutions. Moreover, due to the destabilizing effect of the swirling rotation, an outer marginally stable orbit appears, which heavily limits the range of the parameter space in which disk solutions can exist. Due to the possibility of an outer and inner disk cusp, different types of disk solutions are possible. We classify the different types of disk solutions, which differ from each other by the properties of their cusps. Four different scenarios can be identified in which different accretion dynamics could arise.
\end{abstract}

\maketitle

\section{Introduction}

In the last century, the paradigm of gravitational physics has been significantly modified by the publication and development of general relativity. Among its most extraordinary predictions is the existence of black holes, which are believed to be the final stage of a stellar collapse. Obtaining solutions from the field equations of general relativity, i.e., Einstein’s field equations, has shown to be a herculean task. Of particular interest is the existence of exact solutions, since those classes of solutions allow to shed light on the main properties of gravity using analytical techniques. Most prominent are the black hole solutions of the coupled (electro-)vacuum Einstein(-Maxwell) field equations. In general, physically meaningful solutions are found by imposing symmetries on the field equations. Among the most well-known solutions is the Schwarzschild black hole, which describes the exterior gravitational field of a spherical compact object; the rotating counterpart of such solution is the Kerr black hole, uniquely described by the black hole mass and angular momentum. However, the simplest solution possessing rotation must be stationary and axially symmetric. The general formulation to classify solutions with such symmetries was developed by Ernst \cite{Ernst:1967by,Ernst:1967wx}, in which the coupled Einstein-Maxwell field equations are replaced by the \textit{Ernst potentials}. Besides simplifying the study of the field equations under the assumed symmetries, this formalism allows to derive new solutions from old known (\textit{“seed”}) solutions by exploring hidden symmetries within the Ernst potentials, which is often referred to as \textit{Ernst generating techniques} \cite{harrison1968new,griffiths2009exact}. In particular, two transformations are often used to generate new solutions: a) the \textit{Harrison transformation}, which allows the embedding of black holes into a magnetic universe \cite{harrison1968new,Ernst:1976mzr}, and b) the \textit{Ehlers transformations}, which generates the swirling background and the immersion of compact objects into it \cite{Astorino:2022aam}. The latter is a solution that, although it has appeared before \cite{Gibbons:2013yq}, did not receive much attention until it was rediscovered more recently in \cite{Astorino:2022aam}, where the physical properties of this novel family of solutions have been extensively studied. Overall, the swirling background is a non-asymptotically flat solution that describes a rotating background equipped with an odd $\mathcal{Z}_2$ symmetry. The spacetime is characterized by the solution’s frame-dragging, which has an opposite sign in the different hemispheres; thus, the spacetime rotates in opposite directions below and above the equatorial plane. This solution is uniquely characterized by just one parameter, the swirling parameter $j$, which gives rise to the presence of exotic features, such as the appearance of an ergoregion that extends to infinity. A detailed description of the ergoregion and the geodesic motion in the swirling background can be found in \cite{Capobianco:2023kse}. Despite these odd characteristics, this novel solution does not possess other exotic features that often appear in rotating background spacetimes, like for the Gödel universe or for the Taub-NUT spacetime, which contain closed timelike curves \cite{Gödel,TAUB_NUT}. Due to the similarity in the derivation of the swirling family of solutions, it has been taken to as a cousin solution to the magnetic universe. Moreover, it has been pointed out that such rotating spacetime could be considered locally to model cosmic filaments \cite{capobianco2024motion}, which are believed to be the largest structures in our universe. They are made of galaxy clusters arranged in thread-like structures \cite{cosmicfilaments1}. Interestingly, observations reveal that such structures do rotate \cite{cosmicfilaments_spin,cosmicfilaments_spin_2} and can even possess electromagnetic fields \cite{cosmicfilaments_charge}. Thus, the swirling background could potentially be used to model such cosmic filaments. Therefore, the effects of such background on dispersed matter, especially around compact objects immersed in such non-trivial geometry, posts itself as a relevant topic.   

In the vicinity of compact objects, diffuse matter gets attracted by the gravitational pull of the compact object. Over longer time periods, accumulations of matter orbiting the compact object gradually form accretion structures. Since these structures usually have the geometry of a disk or a torus, they are called accretion disks or tori and are characterized by the circular flow of their disk particles. In complex disk interactions, angular momentum gets exchanged from the inner to the outer regions of the disk, which causes matter to migrate from the outside inwards through the disk to smaller and smaller orbits towards the central object. When matter reaches the marginally stable orbit, small perturbations could lead to an inspiral of the matter onto the central object, where it ultimately gets accreted. The energy converted in this process scales with the mass of the central object, and in the case of supermassive black holes, it corresponds to magnitudes that can be found at the upper end of our current energy scales for observed and theorized phenomena in the universe. Accretion disks are therefore particularly important in the study of high-energy physics and in the observation of astrophysical objects, as a large part of the energy is converted into radiation energy, with such high intensities that it can be observed from billions of light years away. This radiation can have a broad frequency spectrum, which mainly depends on the properties of the disk as well as on the properties of the central object. Overall, multiple approaches exist in the modeling of accretion disks, which differ by the dynamical situations they are intended to describe. When it comes to their morphology, they can generally be divided into two classes with regard to their vertical height, namely, thin and thick disks. In this paper, we focus on the latter, whose theoretical foundations have been vastly established in \cite{Abramowicz1971,Abramowicz1974,Fishbone1976,Abramowicz1978,Kozlowski1978,Abramowicz1980,Jarosczynski1980,Paczynsky1980,Paczynski1981,Paczysnki1982}.

The simplest case of a thick disk imposes the \textit{Polish Doughnut} model, in which the disk is modeled as a body composed of a non-selfgravitating perfect fluid, whose properties are solely determined by the spacetime geometry. Due to its simplicity, this approach often {results in analytical solutions that can be directly computed from the metric tensor. Nevertheless, despite its simplicity, such modeling captures the main qualitative properties of possible accretion structures and is often used as an initial condition in more complex general relativistic magnetohydrodynamics simulations \cite{Narayan2012,Igumenshchev2003,McKinney2012}. The Polish Doughnut model has been applied to a variety of alternative gravity theories and compact objects, such as de Sitter black holes, scalarized black holes, NUT spacetimes, and Born-Infeld teleparallel gravity \cite{Rezzolla2003,Stuchlik2009,Chakraborty2015,Jefremov2017,Teodoro2021S,Teodoro2021B,Bahamonde2022, Zhou2024}. Furthermore, the thick disk model has recently been generalized to the swirling family of solutions, more precisely to the Schwarzschild black hole immersed in a swirling background \cite{Chen2024}. The spacetime of a Schwarzschild black hole immersed in a swirling background is fully characterized by the mass of the black hole and the swirling parameter $j$. It was shown that the presence of the rotating background modifies the typical disk structures around the Schwarzschild black hole by deforming the thick disk into a bowl-shaped matter distribution as well as by the appearance of an outer disk cusp, which originates from instabilities of the circular geodesics within the spacetime \cite{Chen2024}.

Here, we will focus on generalizing this work to the Kerr black hole immersed in a swirling background (KBHSB). This spacetime is characterized by three parameters, namely the black hole mass $M$, the Kerr spin parameter $a = J/M$, where $J$ is the angular momentum of the black hole, and the swirling parameter $j$.  In contrast to the swirling motion, the Kerr rotation does not change its sign with respect to the equatorial plane; therefore the black hole is co-rotating with the swirling background in one hemisphere and counter-rotating in the other. This breaking of symmetry and the resulting spin-spin interaction of the Kerr rotation with the swirling rotation has an influence on the main properties of the local spacetime environment. The motion of particles in the local vicinity of the black hole is especially affected by the spin-spin interaction and the symmetry breaking. As such, unique effects arise on the properties of basic geodesic motion around the black hole. In this work, we aim to shed light on these effects by analyzing the influence of the Kerr and swirling rotations on the properties of circular orbits and geometrically thick accretion structures.

This work is organized as follows. In section 2, we introduce the KBHSB solution and briefly discuss its main characteristics. In section 3, an introduction to the theory of geometrically thick accretion disks is presented, and in section 4 we analyze qualitatively the circular orbits and thick disk solutions of the KBHSB spacetime. Throughout this work, we use the Einstein summation convention and geometrized units, where $G = c = 1$, with $G$ as the gravitational constant and $c$ as the speed of light. Note that, in this units, the swirling parameter $j$ has the dimension $[j] = 1/M^2$. The metric signature used is $(-,+,+,+)$ and for all KBHSB solutions considered in this work, the black hole mass is normalized to $M = 1$.

\section{Kerr black hole in a swirling background}

The line element describing a Kerr black hole immersed in a swirling background can be written in Boyer-Lindquist coordinates as \cite{Astorino:2022aam}

\begin{equation}
    {\rm d}s^2 = \frac{1}{\mathcal{F}(r,\theta)} \left( {\rm d}\varphi + \omega {\rm d}t  \right)^2 + \mathcal{F}(r,\theta) \left[ -\varrho^2 {\rm d}t^2 + \Sigma \sin^2\theta \left( \frac{{\rm d}r^2}{\Delta} + {\rm d}\theta^2 \right) \right],
\end{equation}

where the metric functions $\mathcal{F}$ and $\omega$, that encode the swirling parameter $j$, can be expressed in a finite power series in the form of

\begin{equation}
    \mathcal{F} = \frac{\mathcal{F}_0 + j \mathcal{F}_1 + j^2 \mathcal{F}_2}{R^2 \Sigma \sin^2\theta} \ \ \ \ \ ; \ \ \ \ \ 
    \omega = \frac{\omega_0 + j \omega_1 + j^2 \omega_2}{\Sigma},
\end{equation}

with

\begin{eqnarray}
    \mathcal{F}_0 &=& R^4, \ \ \ \quad 
    \mathcal{F}_1 = 4 a m \Xi \cos\theta R^2, \ \ \ \quad 
    \mathcal{F}_2 = 4 a^2 m^2 \Xi^2 \cos^2\theta + \Sigma^2 \sin^4\theta  \nonumber
    \\
    \omega_0 &=& 2 a m r, \ \ \ \quad 
    \omega_1 = 4\cos\theta (m a^4 - r^4(r-2m) - \Delta a^2 r - a (r-m)\Omega), \nonumber \\
    \omega_2 &=& 2m \left[ 3ar^5 - a^5(r+2m) + 2a^3r^2(r+3m) -r^3 (\cos^2\theta -6)\Omega + a^2 \Omega \left( \cos^2\theta(3r-2m) -6(r-m) \right) \right], \nonumber
\end{eqnarray}
where

\begin{eqnarray}
    \Delta &=& r^2-2mr+a^2, \ \ \ \qquad \ \ \ 
    \varrho^2 = \Delta \sin^2\theta, \ \ \ \qquad \ \ \ 
    \Sigma = \left( r^2+a^2 \right)^2 - \Delta a^2 \sin^2\theta, \nonumber 
    \\ 
    \Omega &=& \Delta a \cos^2\theta, \ \ \ \qquad 
    \Xi = r^2 \left( \cos^2\theta -3 \right) -a^2 \left( 1+\cos^2\theta \right), \ \ \ \qquad \ \ \ 
    R^2 = r^2 + a^2 \cos^2\theta. \nonumber
\end{eqnarray}
The pure Kerr solution is recovered for vanishing $j$, the immersion of Schwarzschild in a swirling background is obtained by $a=0$, and the pure swirling background is found for $m=0$ and $a=0$. As noted in \cite{Astorino:2022aam} the spin-spin interaction between the Kerr black hole and the background frame dragging gives rise to a conical singularity on the symmetry axis. The presence of conical singularities in the immersion of different spacetimes into the swirling background was already pointed out in \cite{Astorino_CS}, where the accelerated Kerr solution was considered and the metric could be regularized by fixing the swirling parameter. Nonetheless, this appears not to be possible in the KBHSB case. Furthermore, the KBHSB solution possesses two horizon surfaces, an inner and an outer one, which in these coordinates are found by imposing $g^{rr} = 0$, and thus are defined by
\begin{equation}
    r_\pm = m \pm \sqrt{m^2 - a^2}, 
\end{equation}
which is independent of $j$ and equivalent to the Kerr solution. However, it should be noted that the presence of the background dragging does deform the horizon geometry. Considering the main interest of this paper, we will focus only on the outer horizon here. Taking a slice of constant $t$ leads to the induced metric
\begin{equation}
\label{KBHSU horizon}
    {\rm d}s^2_{hor} = \Sigma(r_{\pm}, \theta) \sin^2\theta \mathcal{F}(r_\pm,\theta) {\rm d}\theta^2 + \frac{{\rm d}\varphi^2}{\mathcal{F}(r_\pm,\theta)}.
\end{equation}
Insights about the shape of the horizon can be taken by considering an isometric embedding of the two-dimensional surface (\ref{KBHSU horizon}) into a three-dimensional Euclidean space. However, a full embedding is not always possible. For the pure Kerr solution, i.e. $j=0$, the event horizon geometry around the poles becomes hyperbolic as the Kerr parameter increases ($a \in (\sqrt{3}/2,1)$), and the embedding of those regions into Euclidean space is not feasible. For this purpose, insights about the geometry can be shown by considering the composition of an Euclidean embedding with the metric $ds^2_{E} = dx^2 + dy^2 + dz^2$ and a pseudo-Euclidean embedding described by the metric $ds^2_{pE} = dx^2 + dy^2 - dz^2$ \cite{Smarr:1973zz,Kleihaus:2015aje}. 

The presence of the swirling background affects the north and south hemispheres differently, which is also directly reflected by the horizon geometry. We note that even small variations of $j$ deform the usual oblate structure of the Kerr horizon. For a positive spin-spin relation ($aj > 0$) the neighborhood of $(r = r_+, \theta = \pi)$ becomes hyperbolic. For a rapidly rotating Kerr black hole, where the neighborhood of the poles is already hyperbolic, the vicinity of $(r = r_+,\theta = 0)$ is dragged into an elliptic geometry. In Fig. \ref{fig:horizons} we show a representation of the outer horizon for a variation of the swirling parameter $j \in \{0, 10^{-4}, 10^{-3}\}$, and the Kerr parameter, $a \in \{0.5, 0.9\}$. For a negative spin-spin relation ($aj < 0$) an analogous behavior would be present, where just the northern and southern hemispheres are interchanged.

When $j$ increases further, the geometry of the event horizon is significantly modified. Some examples are shown in Fig. \ref{horizons_comparison}, where we have chosen the limiting Kerr parameter $a =\sqrt{3}/2$, for which the horizon geometry starts to become hyperbolic, to illustrate the hyperbolic behavior appearing due to the swirling dragging.

Despite the significant changes in geometry, the horizon area $\mathcal{A}$ is not affected by the presence of the swirling parameter $j$,

\begin{equation}
    \mathcal{A} = \int_{0}^{2\pi} \int_{0}^{\pi} \sqrt{g_{\theta\theta} g_{\varphi\varphi}} {\rm d}\theta{\rm d}\varphi = 
    \int_{0}^{2\pi} \int_{0}^{\pi} \sqrt{\sin^2\theta \Sigma} {\rm d}\theta{\rm d}\varphi \left. \right|_{r=r_\pm} = 8m\pi r_{\pm},
\end{equation}

which has already been seen for the Schwarzschild black hole in a swirling background \cite{Astorino:2022aam, Moreira:2024sjq}. The ergosurfaces for the KBHSB solution are rather complicated and change significantly depending on $j$. Since it appears not to be possible to find simple analytical relations, as it is possible for pure Kerr and the swirling background, we perform a qualitative study on the boundaries of the ergoregions, which are defined by the condition $g_{tt} = 0$. The pure Kerr solution has two of these ergosurfaces, an inner and an outer closed surface that touches the respective horizons on the symmetry axis, thus

\begin{equation}
    r_{E_-}(m,a,\theta) \leq r_{-} < r_{+} \leq r_{E_+}(m,a,\theta),
\label{eq:Horizons_Ergo}
\end{equation}

where $r_{E_\pm}$ is the radial coordinate of the inner and outer ergosurface, respectively. These surfaces are also present in the KBHSB solution and the constraints given in Eq. \ref{eq:Horizons_Ergo} still hold for the KBHSB case. Furthermore, two additional ergosurfaces appear, which, for small values of $j$, are disconnected patches in the southern and northern hemispheres extending to infinity. These surfaces are a property of the swirling background. They also appear for the Schwarzschild black hole immersed in a swirling background. A detailed discussion of these ergosurfaces can be found in \cite{Astorino:2022aam,Capobianco:2023kse,Moreira:2024sjq}. For the main interest of this paper, we will restrict the discussion of the ergoregions to the outer area, i.e. $r > r_+$.

The ergoregions of the KBHSB are intrinsically related to the spin-spin interaction between the black hole and swirling background. For smaller variations of $j$, the ergosurfaces resemble roughly a superposition of the intrinsic ergoregions from each spacetime (Kerr and swirling background) described above. This is illustrated in Fig. \ref{fig:ergosurfaces}. In Fig. \ref{fig:ergosurface_3D}, we show a representation of these surfaces in a three-dimensional space for a small swirling parameter value of $j = 0.01$.

The KBHSB ergosurface, corresponding to the outer Kerr surface, remains qualitatively similar for small values of $j$ (Fig. \ref{fig:ergosurfaces} (a)). The ergosurfaces originating from the swirling background show a minor deviation for smaller $j$, but their general structure is similar. However, as $j$ increases the ergosurfaces deviate more and more from the Kerr and swirling background surfaces and are no longer resembling a superposition of the two limiting solutions. The non-linear spin-spin interaction causes greater symmetry breaking for larger $j$, and the ergosurfaces become more asymmetrical with respect to the equatorial plane.

Generally, we conclude that the spin-spin interaction between the black hole and the swirling background enhances the symmetry breaking with regard to the major spacetime properties, as showcased for the horizons and ergosurfaces. This symmetry breaking also has an influence on the properties of circular orbits and thick disks, as it will be demonstrated in the following sections.

\begin{figure}[H]
\centering
\begin{subfigure}{.31\textwidth}
  \centering
  \includegraphics[width=\linewidth]{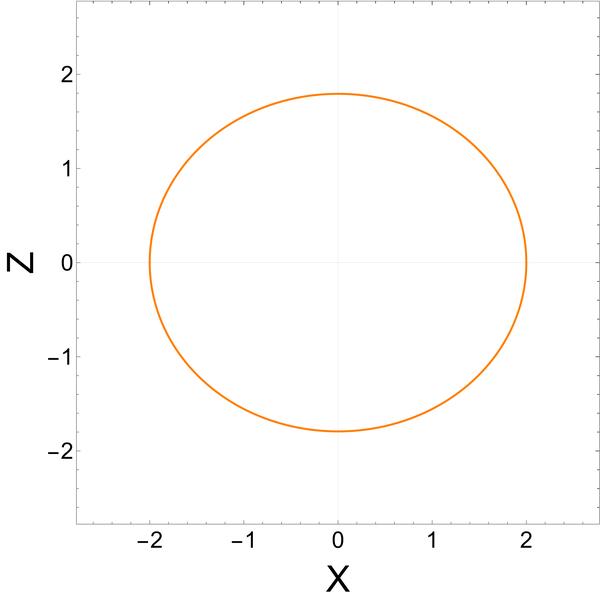}
  \caption{$a = 0.5$, $j = 0$ }
\end{subfigure}
\begin{subfigure}{.31\textwidth}
  \centering
  \includegraphics[width=\linewidth]{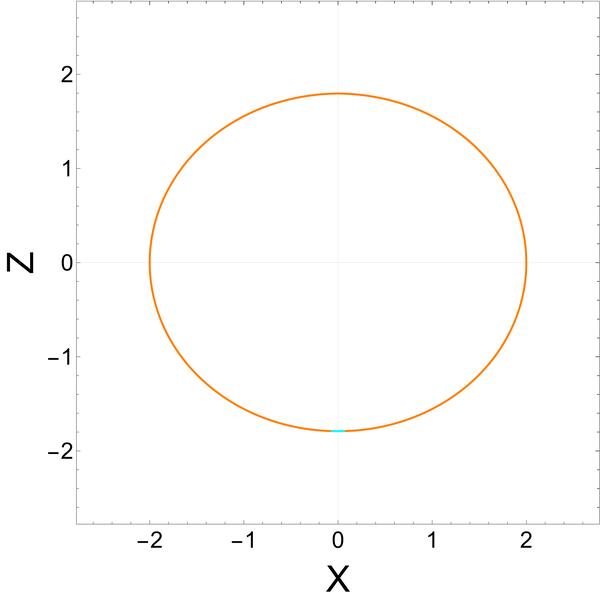}
  \caption{$a = 0.5$, $j=10^{-4}$}
\end{subfigure}
\begin{subfigure}{.31\textwidth}
  \centering
  \includegraphics[width=\linewidth]{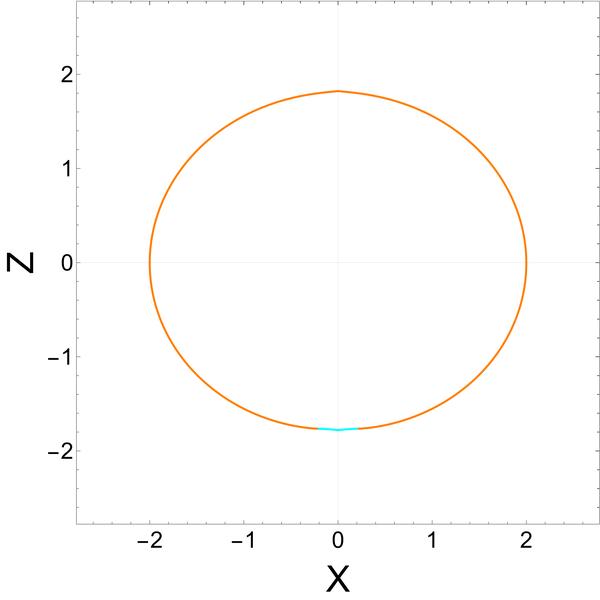}
  \caption{$a = 0.5$, $j=10^{-3}$}
\end{subfigure}

\begin{subfigure}{.31\textwidth}
  \centering
  \includegraphics[width=\linewidth]{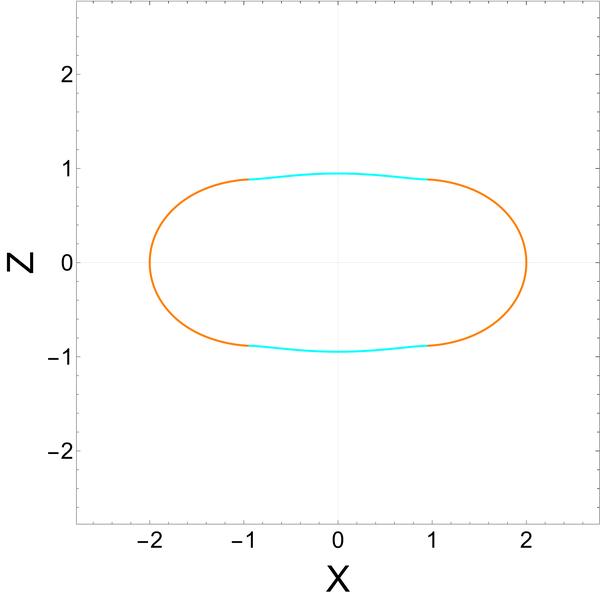}
  \caption{$a = 0.9$, $j=0$}
\end{subfigure}
\begin{subfigure}{.31\textwidth}
  \centering
  \includegraphics[width=\linewidth]{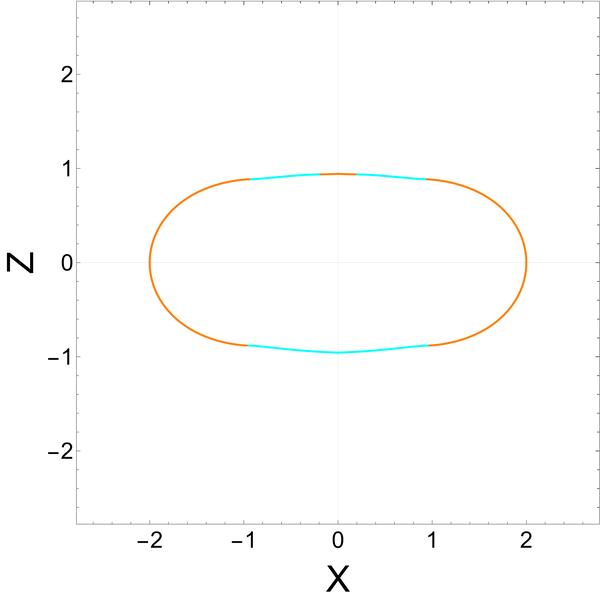}
  \caption{$a = 0.9$, $j=10^{-4}$}
\end{subfigure}
\begin{subfigure}{.31\textwidth}
  \centering
  \includegraphics[width=\linewidth]{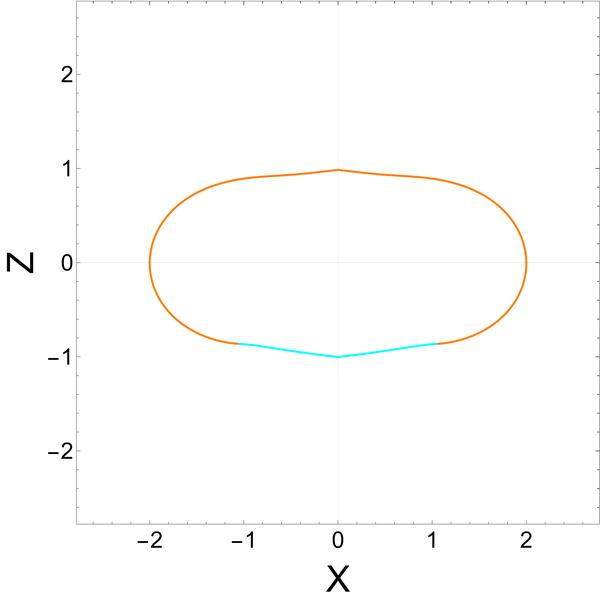}
  \caption{$a = 0.9$, $j=10^{-3}$}
\end{subfigure}

\begin{subfigure}{.325\textwidth}
  \centering
  \includegraphics[width=\linewidth]{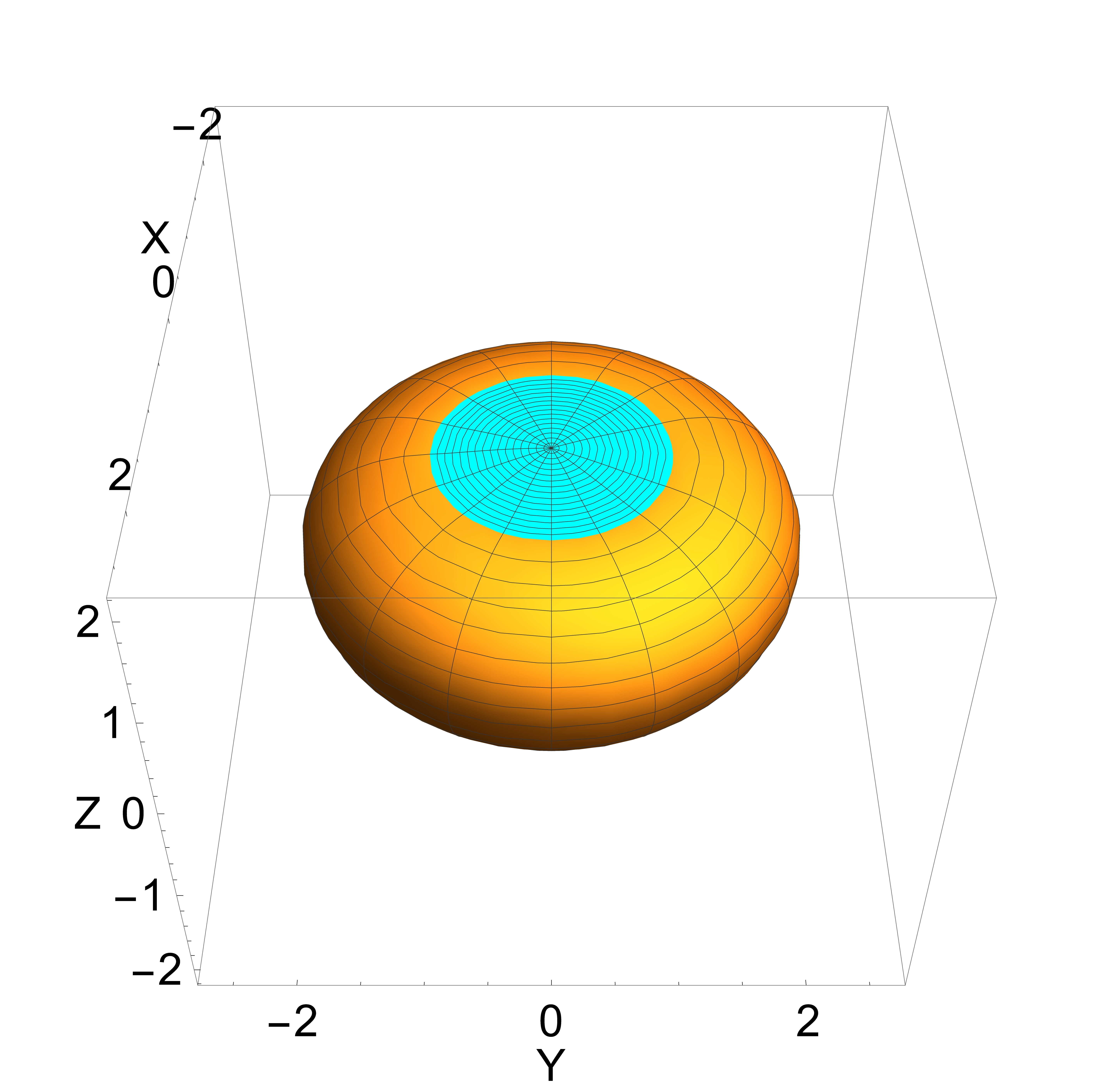}
  \caption{$a = 0.9$, $j=0$}
\end{subfigure}
\begin{subfigure}{.325\textwidth}
  \centering
  \includegraphics[width=\linewidth]{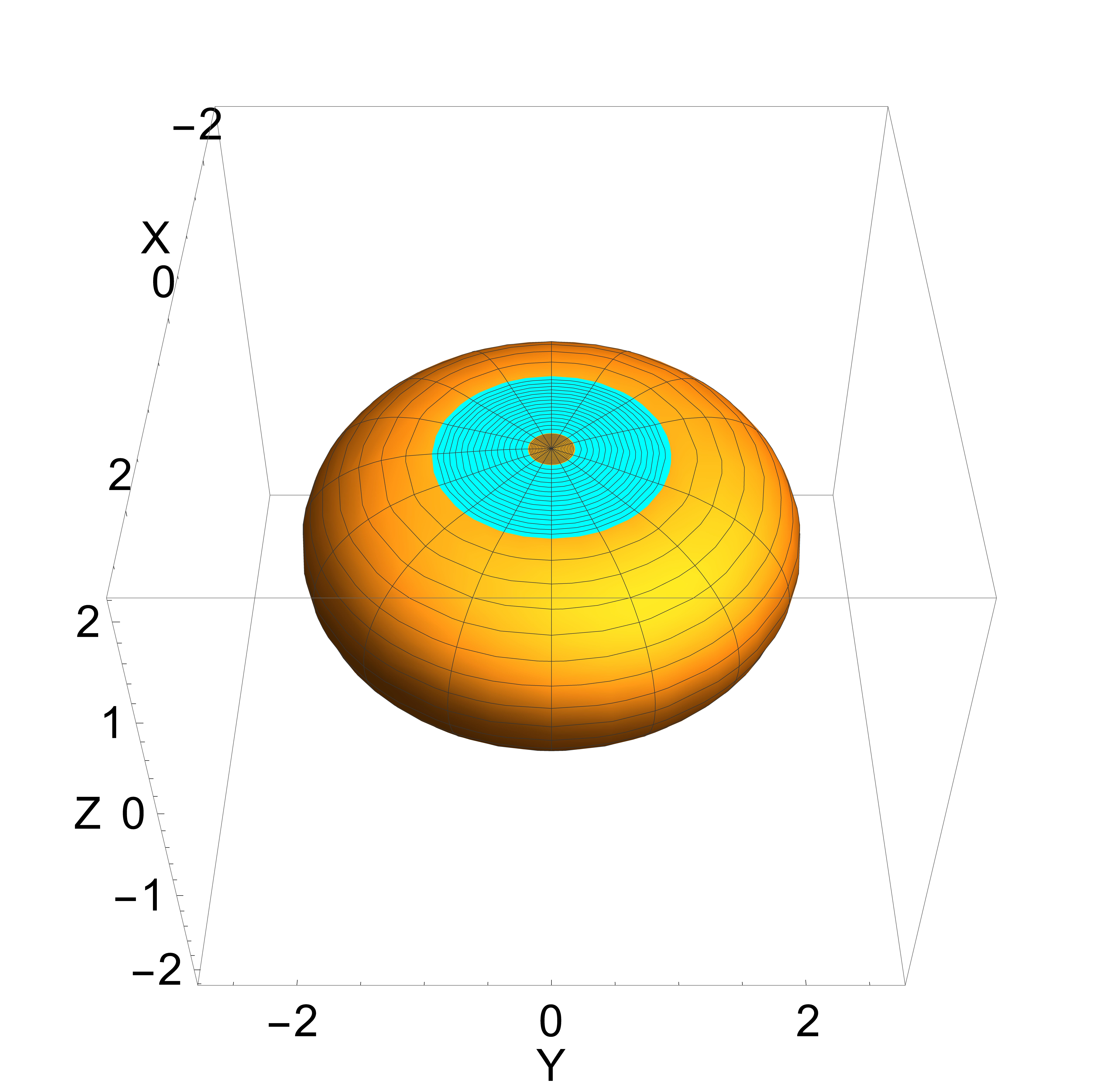}
  \caption{$a = 0.9$, $j=10^{-4}$}
\end{subfigure}
\begin{subfigure}{.325\textwidth}
  \centering
  \includegraphics[width=\linewidth]{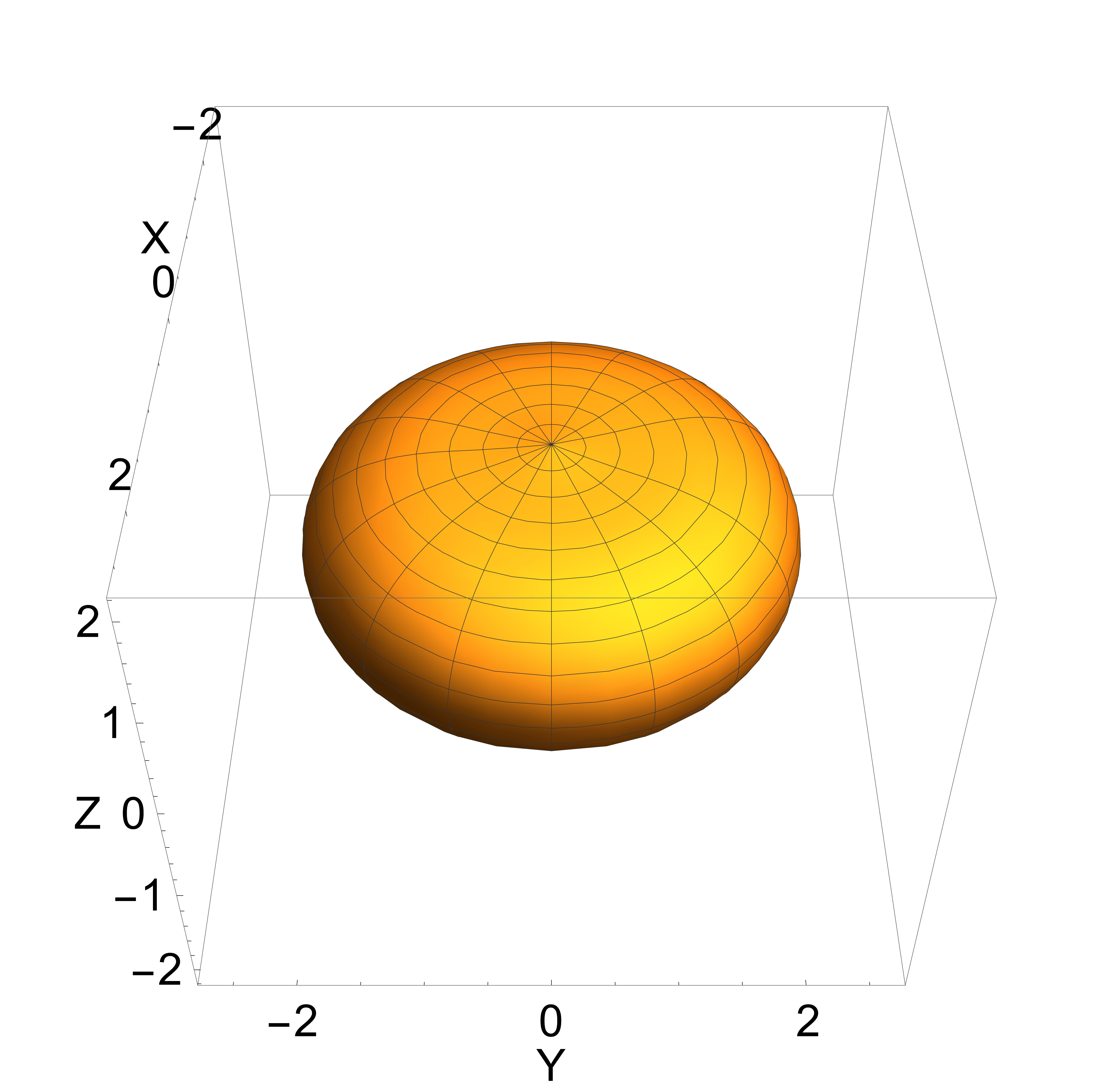}
  \caption{$a = 0.9$, $j=10^{-3}$}
\end{subfigure}
\caption{Embeddings of the outer horizon geometry for different KBHSB solutions and small variations of the swirling parameter $j$. The orange curves and surfaces represent an Euclidean embedding, the cyan curves and surfaces represent a pseudo-Euclidean embedding. The top row shows the horizon geometry for a KBHSB with Kerr parameter $a = 0.5$, where the vicinity of the southern pole becomes hyperbolic due to the swirling dragging. This effect increases with $j$. The middle row shows the geometry for a rapidly rotating KBHSB with $a = 0.9$. Here, the regions close to the northern and southern poles are already hyperbolic and with increasing $j$ the horizon region close to the northern pole becomes elliptic. The bottom row presents a 3D representation of the embedding for the rapidly rotating case with $a = 0.9$. The equatorial plane has been set to be $Z(\pi/2)=0$ in all plots.}
\label{fig:horizons}
\end{figure}

\begin{figure}[H]
\centering
\begin{subfigure}{.325\textwidth}
  \centering
  \includegraphics[width=\linewidth]{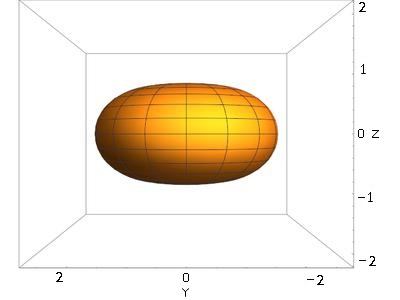}
  \caption{$j = 0$}
\end{subfigure}
\begin{subfigure}{.325\textwidth}
  \centering
  \includegraphics[width=\linewidth]{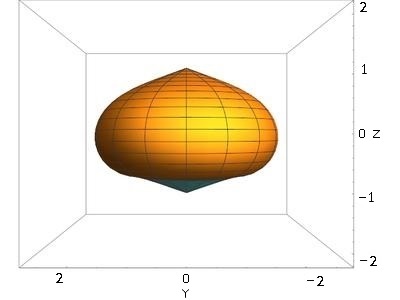}
  \caption{$j = 0.01$}
\end{subfigure}
\begin{subfigure}{.325\textwidth}
  \centering
  \includegraphics[width=\linewidth]{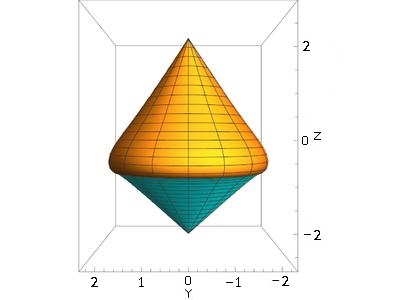}
  \caption{$j = 0.1$}
\end{subfigure}
\caption{Embedding of the outer horizon geometry for a fixed value of $a = \sqrt{3}/2$ and three values of the swirling parameter, $j \in \{ 0, 0.01, 0.1 \}$. The equatorial plane is set to $Z(\pi/2) = 0$. With increasing $j$ the horizon geometry becomes more conical and the symmetry breaking between the northern and southern hemispheres is more accentuated. A large part of the horizon geometry in the southern hemisphere becomes hyperbolic for larger values of $j$.}
\label{horizons_comparison}
\end{figure}

\begin{figure}[H]
\centering
\begin{subfigure}{.325\textwidth}
  \centering
  \includegraphics[width=\linewidth]{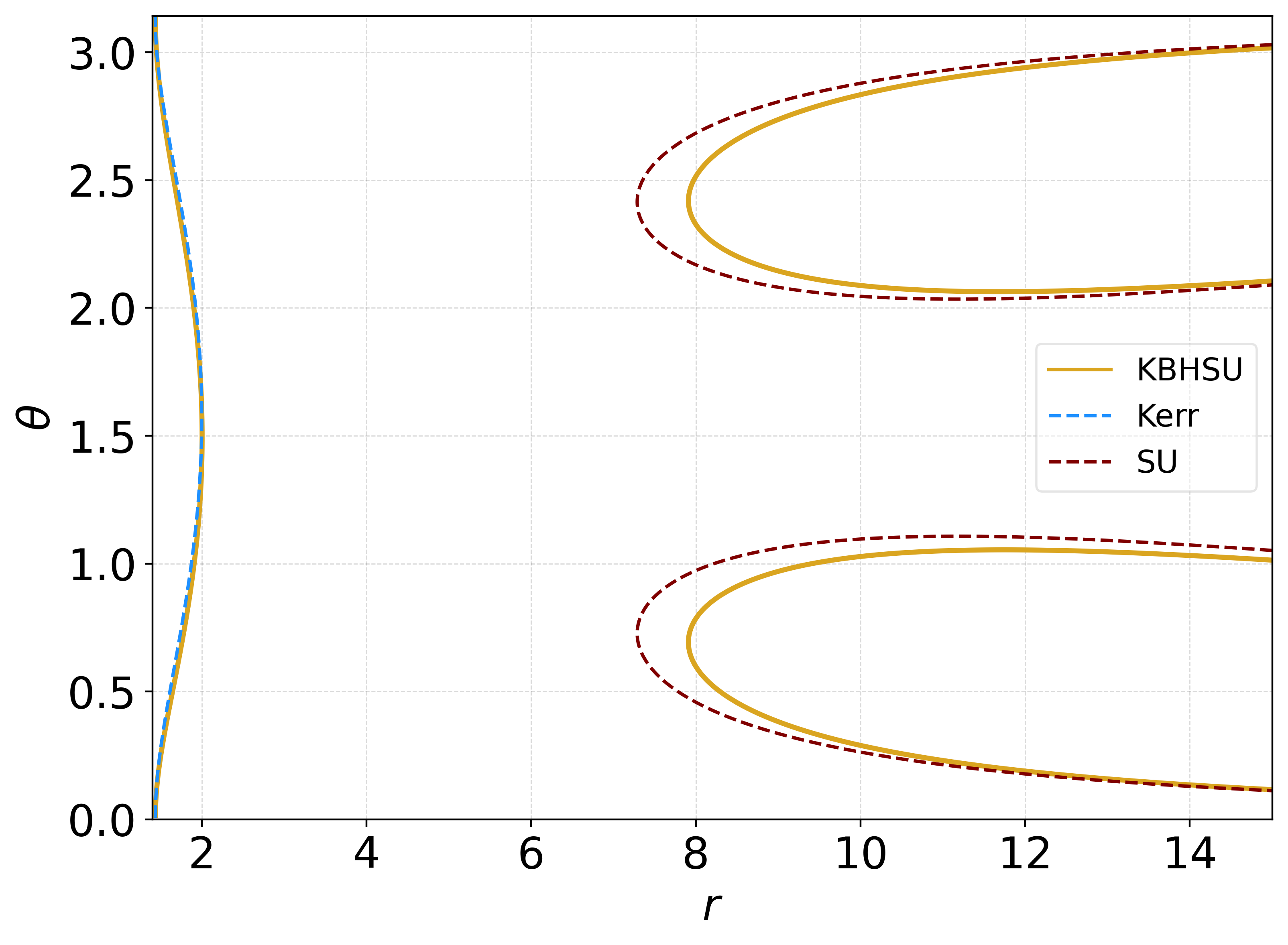}
  \caption{$j = 0.01$}
\end{subfigure}
\begin{subfigure}{.325\textwidth}
  \centering
  \includegraphics[width=\linewidth]{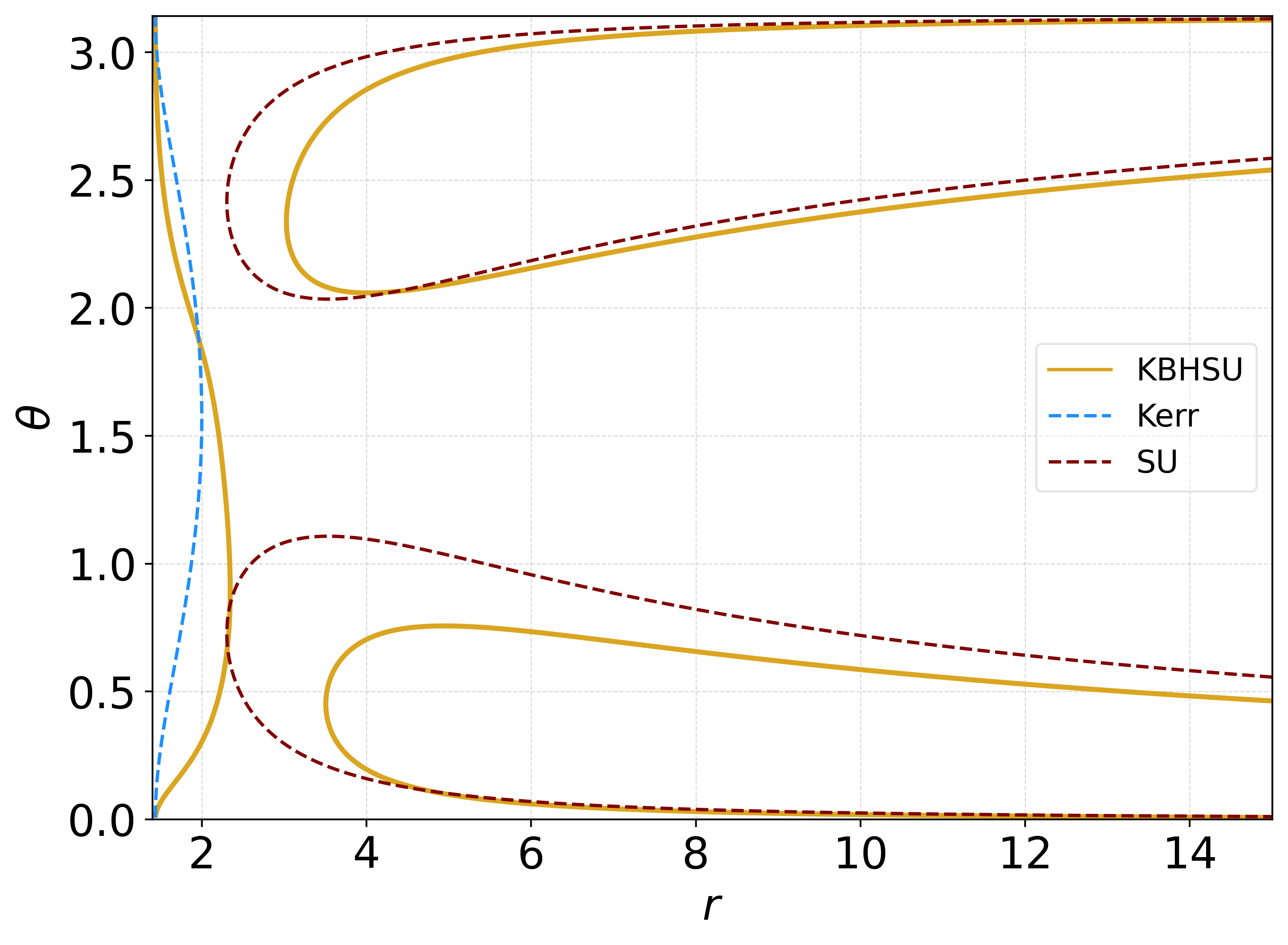}
  \caption{$j = 0.1$}
\end{subfigure}
\caption{Comparison of the ergosurfaces for KBHSB, Kerr, and the swirling background in the $r-\theta$ plane for two different values of the swirling parameter. The orange curves show the outer KBHSB ergosurfaces, the blue-dashed curve shows the outer Kerr ergosurface and the red-dashed curves show the swirling background ergosurfaces. The Kerr parameter is set to $a = 0.9$ in both plots.}
\label{fig:ergosurfaces}
\end{figure}

\begin{figure}[H]
    \centering
    \includegraphics[width=0.35\linewidth]{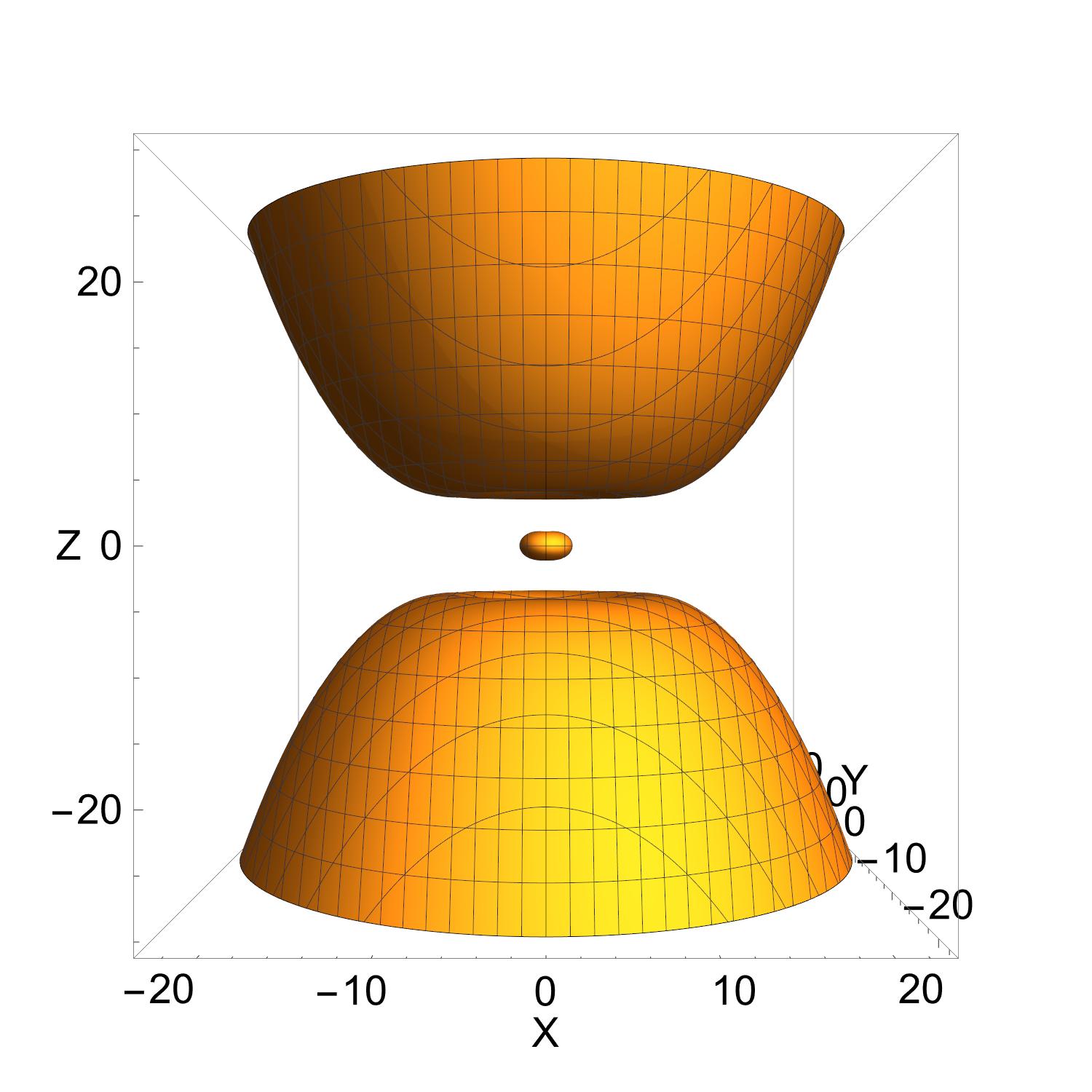}
    \caption{Representation of the outer KBHSB surfaces in a three-dimensional space with $Z(\pi/2) = 0$. The Kerr parameter is set to $a = 0.9$ and the swirling parameter to $j = 0.01$. The inner structure showcases the ergosurface originating from the outer Kerr surface, and the structures in the northern and southern hemispheres showcase the ergosurfaces from the swirling background. This representation is obtained by conducting a rotation around the symmetry axis.}
    \label{fig:ergosurface_3D}
\end{figure}

\section{Geometrically Thick Disks}
In axisymmetric and stationary spacetimes, geometrically thick disks can be modeled as a body of fluids, where the fluid particles are moving on circular orbits around a central compact object. The stress-energy tensor of the disk, $T^{\mu \nu}$, can be written as a sum of the different contributing terms, i.e. the fluid, viscosity, radiation, and magnetic terms. In general, the degree of complexity of the model can increase significantly if the various terms are taken into account. In the Polish doughnut model, the viscosity, radiation, and magnetic terms are neglected and only the fluid part is taken into account. The accretion disk is therefore modeled as a perfect fluid and due to its simplicity, it offers a suitable model to investigate the qualitative properties of geometrically thick accretion disks. The stress-energy tensor is then given by
\begin{align}
    T^{\mu \nu} = T^{\mu \nu}_{Fluid} = \rho h u^\mu u^\nu + p g^{\mu \nu},
\end{align}
where $\rho$ is the rest mass density, $h$ is the specific enthalpy, $u^\mu = (u^t, 0, 0, u^\varphi)$ is the four-velocity of the fluid particles, $p$ is the pressure and $g^{\mu \nu}$ is the contra-variant metric tensor. To construct disk solutions, the fundamental conservation laws, i.e. the conservation of energy-momentum and the continuity equation need to be solved,
\begin{align}
    \nabla_\mu T^{\mu \nu} &= 0 
    \label{eq:energy-momentum} \\
    \nabla_\mu (\rho u^\mu) &= 0.
    \label{eq:continuity}
\end{align}
The continuity equation (\ref{eq:continuity}) is always trivially satisfied since $u^r = u^\theta = 0$ for circular motion and $\partial_t = \partial_\varphi = 0$ due to the axisymmetry and stationarity of the spacetime. By applying the orthogonal projection tensor $h^\alpha_\beta = \delta^\alpha_\beta + u^\alpha u_\beta$ to eq. (\ref{eq:energy-momentum}) and by rewriting it in terms of the specific angular momentum $\ell = - \frac{u_\varphi}{u_t}$ and the angular velocity $\Omega = \frac{u^\varphi}{u^t}$ of the particle, the relativistic Euler equations can be derived,
\begin{align}
    \partial_\mu (\ln|u_t|) - \frac{\Omega}{1 - \Omega \ell} \partial_\mu \ell = - \frac{1}{\rho h} \partial_\mu p,
    \label{eq:Euler}
\end{align}
where $-u_t$ is the mass-normalized energy. As a necessary integrability condition the von Zeipel theorem must be presumed, which states that the angular velocity can be written as a function of the specific angular momentum, $\Omega = \Omega(\ell)$. By assuming a polytropic equation of state, where the pressure is solely a function of the density, such as $p = K \rho^\Gamma$, with $K$ and $\Gamma$ as constants, the von Zeipel theorem is always satisfied. Hence, the eq. (\ref{eq:Euler}) becomes integrable and can be written as
\begin{align}
    \ln|u_t| - \ln|(u_t)_{in}| - \int_{\ell_{in}}^{\ell} \frac{\Omega}{1 - \Omega \ell'} {\rm d}\ell'= - \int_{p_{in}}^p \frac{{\rm d}p'}{\rho h} = - \int_{h_{in}}^h \frac{ {\rm d}h'}{h'} = - \ln h,
    \label{eq:Euler_integrated}
\end{align}
where the subscript $in$ refers to the inner edge of the accretion disk. Assuming a uniform specific angular momentum distribution of the disk, the integral term regarding the specific angular momentum vanishes and the right-hand side can be defined as an effective potential, $\mathcal{W} = \ln|u_t|$ and $\mathcal{W}_{in} = \ln|(u_t)|_{in}$, which corresponds to the combined gravitational and centrifugal potential of a disk particle. The effective potential at the inner edge of the disk $\mathcal{W}_{in}$ is taken as a free parameter which should not be lower than $\mathcal{W}_{center}$ and not higher than 0, since particles with $\mathcal{W} > 0$ could escape to infinity. In order to represent all physically feasible solutions (composed of particles on bound orbits), we will set $\mathcal{W}_{in} = 0$ in all further calculations. Rewriting the specific enthalpy in eq. (\ref{eq:Euler_integrated}) in terms of the rest mass density is leading to,
\begin{align}
    \mathcal{W} - \mathcal{W}_{in} = \ln \left( \left| 1 + \frac{K \Gamma \rho^{\Gamma - 1}}{\Gamma - 1} \right| \right).
\end{align}
The rest mass density can therefore be expressed as a function of the effective potential $\mathcal{W}$. As a consequence, the equi-density and equi-pressure surfaces coincide with the equi-potential surfaces. The four-acceleration $a_\mu$ can be expressed through the partial derivatives of the effective potential, $a_\mu = \partial_\mu \mathcal{W}$. Thus, the extrema of the effective potential correspond to geodesic motion. Minima of $\mathcal{W}$ correspond to stable geodesic motion, where the pressure and rest mass density are maximal. The minima are defined as the accretion disk center. Maxima of $\mathcal{W}$ correspond to unstable geodesic motion and correspond to minima of the pressure and rest mass density, they mark the disk cusps. Cusps are characterized by a self-intersection of the equi-potential surface. Due to the unstable flow at their radial location, small perturbations could trigger accretion processes. By expressing $u_t$ through the metric components, the effective potential $\mathcal{W}$ is solely determined by the spacetime geometry and the specific angular momentum of the disk particles,
\begin{align}
    \mathcal{W} = \frac{1}{2} \ln \left( \frac{g_{t\varphi}^2 - g_{tt} g_{\varphi \varphi}}{\ell^2 g_{tt} + 2 g_{t \varphi} \ell + g_{\varphi \varphi}}\right).
\end{align}
Due to the odd $\mathcal{Z}_2$ spacetime symmetry in the KBHSB solution, the local extrema of the potential are not located in the equatorial plane for $j \neq 0$. In order to determine the locations of the extrema, one needs to solve the system of equations composed of $\partial_r \mathcal{W} = 0$ and $\partial_\theta \mathcal{W} = 0$ for the specific angular momentum $\ell$. The resulting distribution of $\ell$, which solves this system of equations, represents therefore the Keplerian specific angular momentum $\ell_K$ of the circular geodesics for that spacetime. For a chosen specific angular momentum $\ell_0$ of the disk, the location of the disk center is then given by $(r_{center}, \theta_{center}) = \{(r, \theta): \ell_K(r,\theta) = \ell_0 \wedge \partial_r |\ell_K(r,\theta)| > 0\}$ and of a disk cusp by $(r_{cusp}, \theta_{cusp}) = \{(r, \theta): \ell_K(r,\theta) = \ell_0 \wedge \partial_r |\ell_K(r,\theta)| < 0\}$.

\section{Circular Orbits and Disk solutions}
In order to conduct an exemplary study of circular orbits and geometrically thick disks around KBHSB, we have analyzed different solutions, which are classified by their Kerr spin parameter $a$ and the swirling parameter $j$. For a representative analysis regarding the Kerr spin parameter range, we selected solutions in the lower, middle, and upper parameter range, namely $a = 0.2$, $a = 0.5$, and $a = 0.9$. To investigate the influence of the swirling parameter $j$ on the solutions, we have varied $j$ for the mentioned Kerr spin parameter and analyzed the resulting spacetime and disk properties. The swirling parameter $j$ was thereby set to $j \in \{0, \ 10^{-5}, \ 10^{-4}, \ 2 \cdot 10^{-4}, \ 5 \cdot 10^{-4}\}$. Fig. \ref{fig:ell_K} illustrates the specific angular momentum of circular orbits for the different Kerr spin parameters and varying swirling parameters.

\begin{figure}[H]
\centering
\begin{subfigure}{.325\textwidth}
  \centering
  \includegraphics[width=\linewidth]{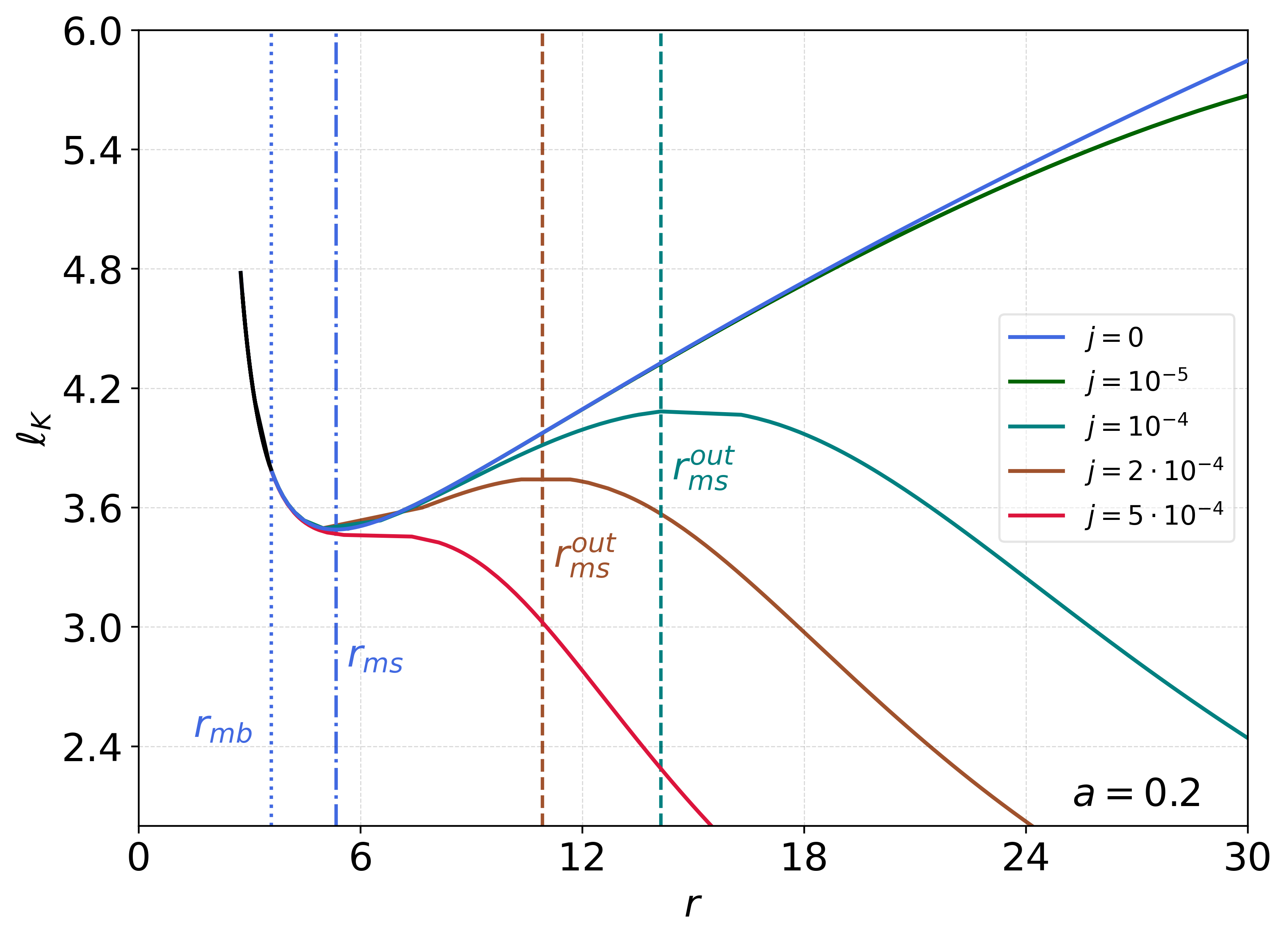}
  \caption{$a = 0.2$, prograde}
\end{subfigure}
\begin{subfigure}{.325\textwidth}
  \centering
  \includegraphics[width=\linewidth]{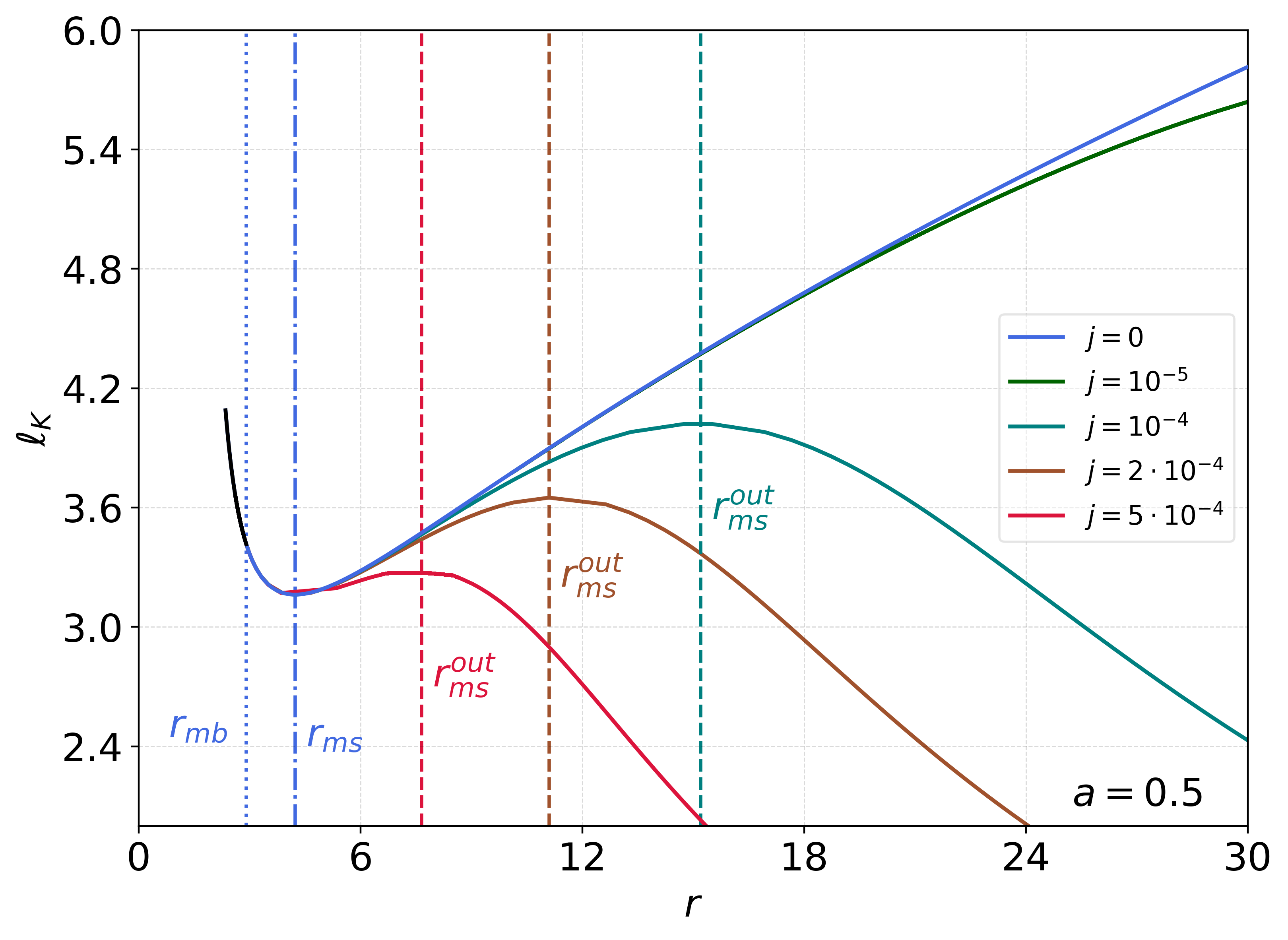}
  \caption{$a = 0.5$, prograde}
\end{subfigure}
\begin{subfigure}{.325\textwidth}
  \centering
  \includegraphics[width=\linewidth]{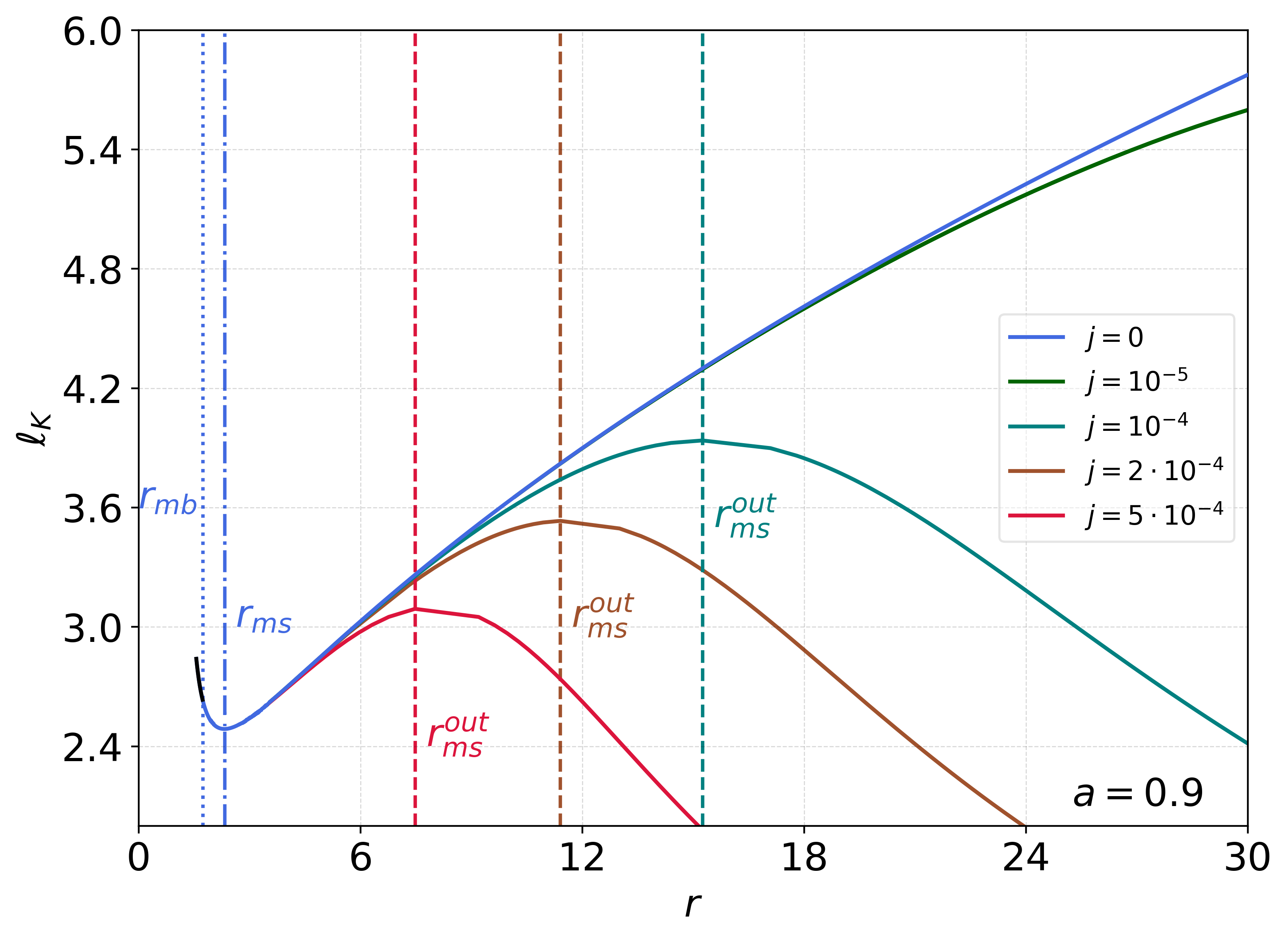}
  \caption{$a = 0.9$, prograde}
\end{subfigure}

\begin{subfigure}{.325\textwidth}
  \centering
  \includegraphics[width=\linewidth]{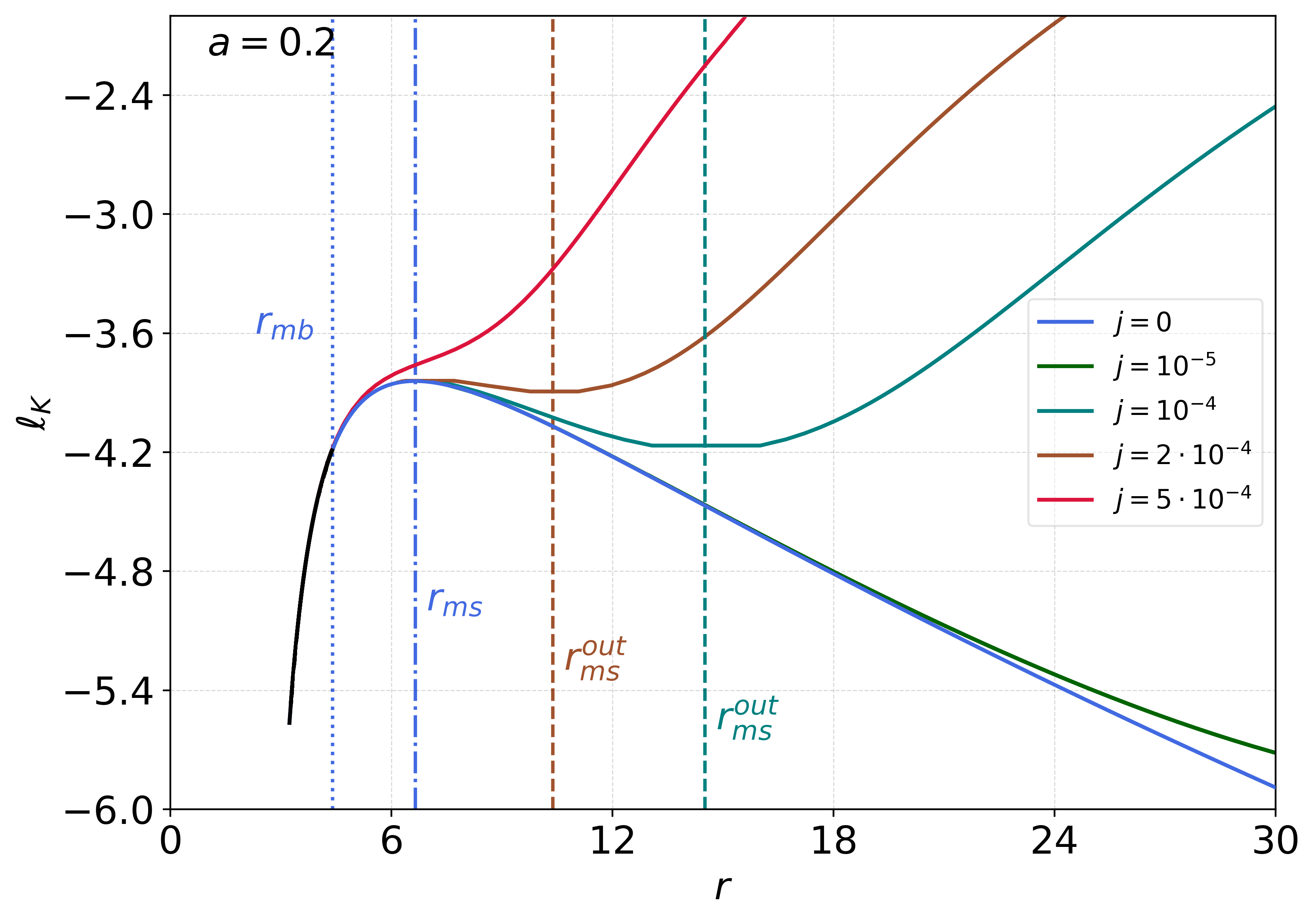}
  \caption{$a = 0.2$, retrograde}
\end{subfigure}
\begin{subfigure}{.325\textwidth}
  \centering
  \includegraphics[width=\linewidth]{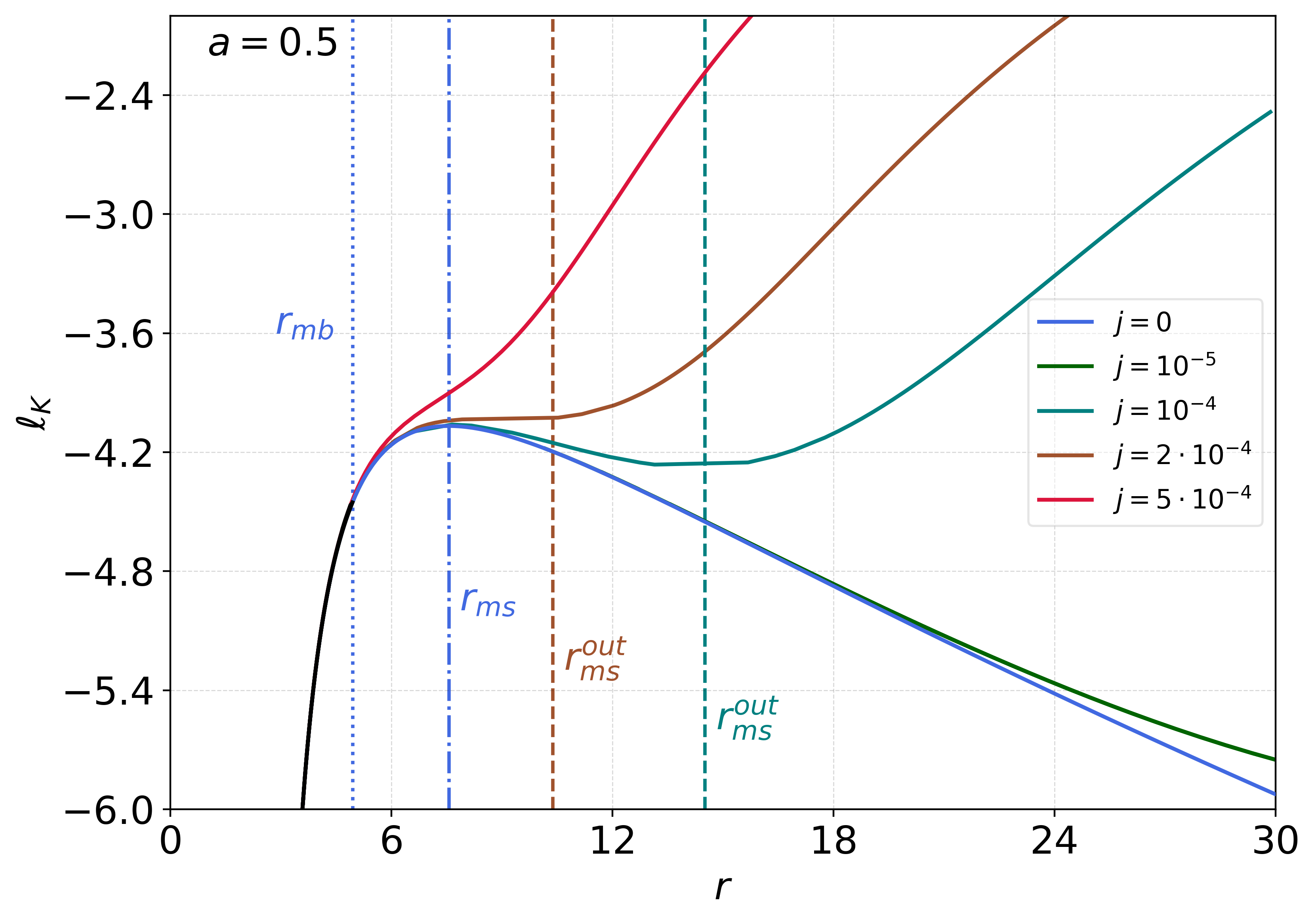}
  \caption{$a = 0.5$, retrograde}
\end{subfigure}
\begin{subfigure}{.325\textwidth}
  \centering
  \includegraphics[width=\linewidth]{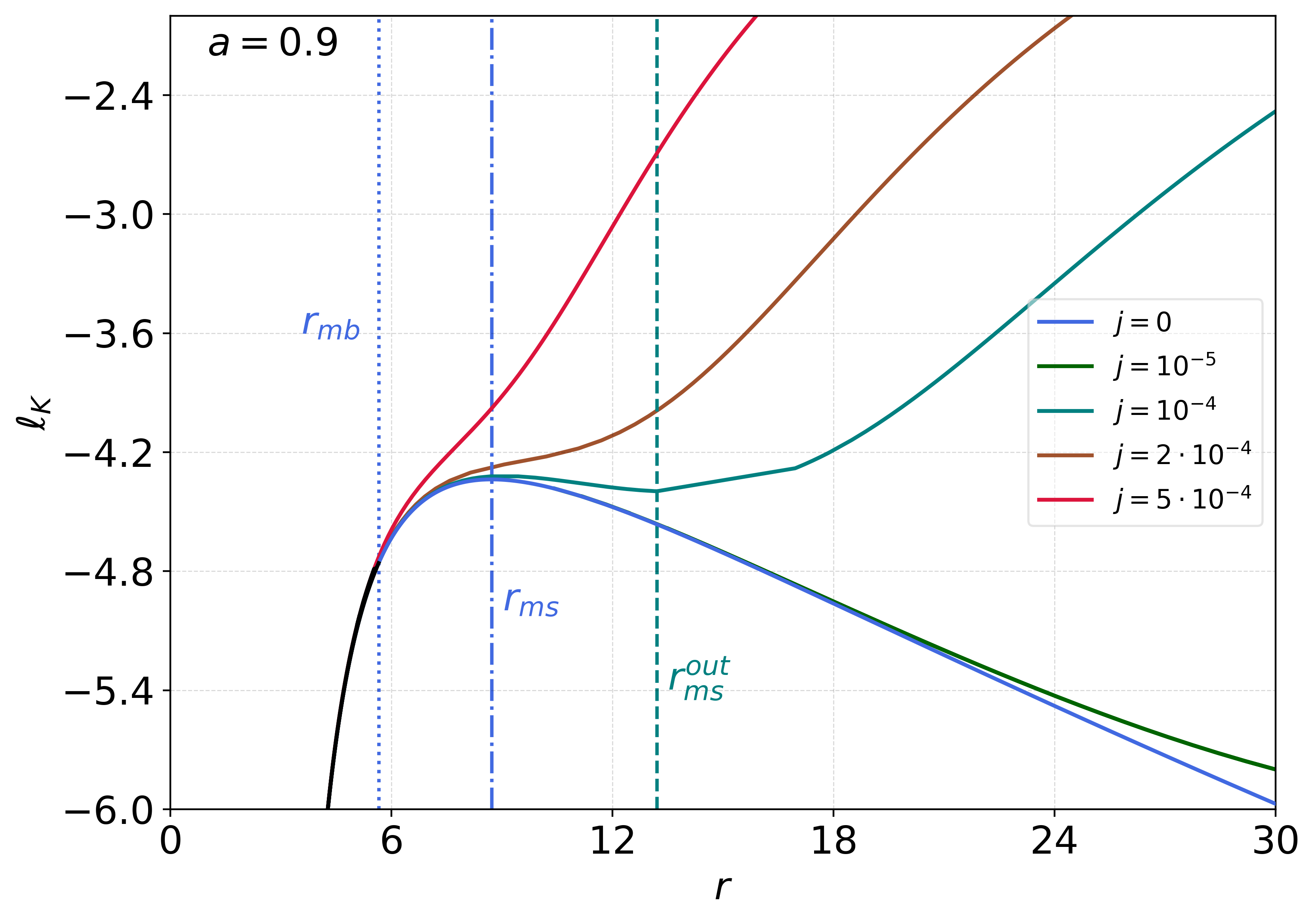}
  \caption{$a = 0.9$, retrograde}
\end{subfigure}
\caption{Specific angular momentum $\ell_K$ of circular orbits for the different Kerr parameter solutions with varying swirling parameter $j$. The upper row presents $\ell_K$ for prograde orbits, and the lower row presents $\ell_K$ for retrograde orbits. The terms prograde and retrograde refer here to the rotation direction of the KBHSB, which is defined by the Kerr spin parameter $a$. The dotted vertical line in each plot marks the marginally bound orbit $r_{mb}$ and the dashed-dotted vertical line marks the marginally stable orbit $r_{ms}$ of the Kerr solution ($j=0$). Dashed vertical lines mark the outer marginally stable orbits $r_{ms}^{out}$, which appear for the KBHSB solutions ($j \neq 0$). Black curve sections mark unbound orbits.}
\label{fig:ell_K}
\end{figure}

The specific angular momentum of circular orbits changes significantly for the KBHSB solutions compared to the Kerr solutions. Up to a critical value of the swirling parameter, $j_c$, the specific angular momentum develops an outer extremum. For prograde orbits, this extremum is a maximum; for retrograde orbits, it is a minimum (Fig. \ref{fig:ell_K} upper and lower row, respectively). For radial values beyond the outer extremum, the specific angular momentum converges asymptotically to 0 in both cases. The outer extremum corresponds to the emergence of an outer marginally stable orbit for KBHSB solutions. Stable circular orbits are therefore only possible in a specific spacetime region, which is lower bounded by the inner marginally stable orbit and upper bounded by the outer marginally stable orbit. The inner marginally stable orbits of the KBHSB solutions are closely located to the marginally stable orbit of the Kerr solution. A greater deviation of $\ell_K$ from the Kerr solution is only noticeable from the inner extremum outwards. For solutions with a swirling parameter greater than the critical swirling parameter, $j > j_c$, the extrema vanish and the specific angular momentum becomes monotonic; all circular orbits are therefore unstable. For prograde orbits, the critical value $j_c$ increases with increasing Kerr spin parameter $a$, thus stable circular orbits are possible for a larger range of the swirling parameter $j$ (Fig. \ref{fig:ell_K} upper row). For retrograde orbits, the value of $j_c$ decreases with increasing Kerr spin parameter $a$; stable circular orbits for fast rotating KBHSB occur therefore for a smaller range of the swirling parameter (Fig. \ref{fig:ell_K} lower row). Furthermore, the swirling rotation causes the emergence of static orbits for prograde as well as retrograde motion. An illustration of the rest specific angular momentum, which is used to determine the static orbits can be found in Fig. \ref{fig:static_orbits}.

\begin{figure}[H]
\centering
\begin{subfigure}{.325\textwidth}
  \centering
  \includegraphics[width=\linewidth]{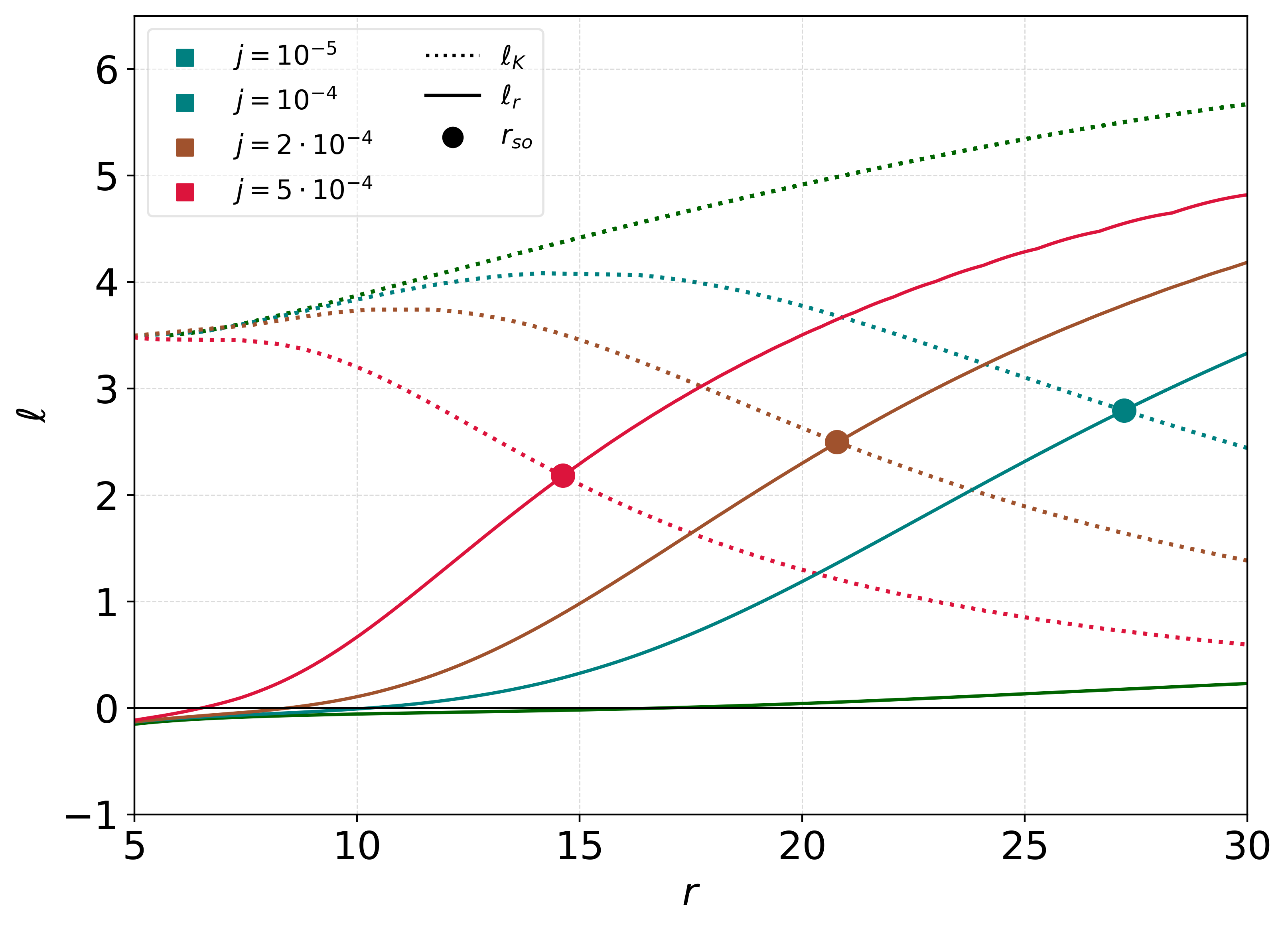}
  \caption{$a = 0.2$, prograde}
\end{subfigure}
\begin{subfigure}{.325\textwidth}
  \centering
  \includegraphics[width=\linewidth]{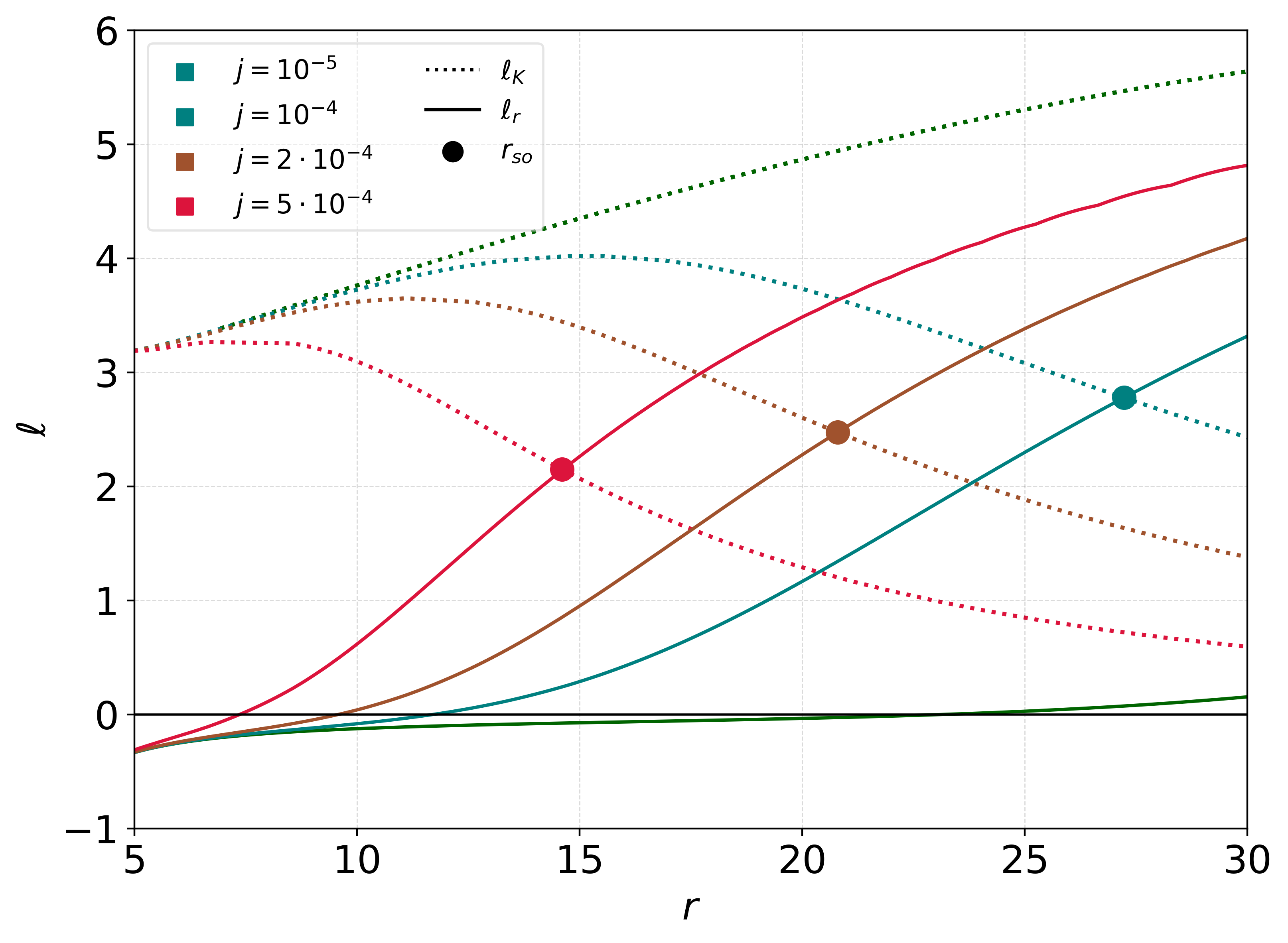}
  \caption{$a = 0.5$, prograde}
\end{subfigure}
\begin{subfigure}{.325\textwidth}
  \centering
  \includegraphics[width=\linewidth]{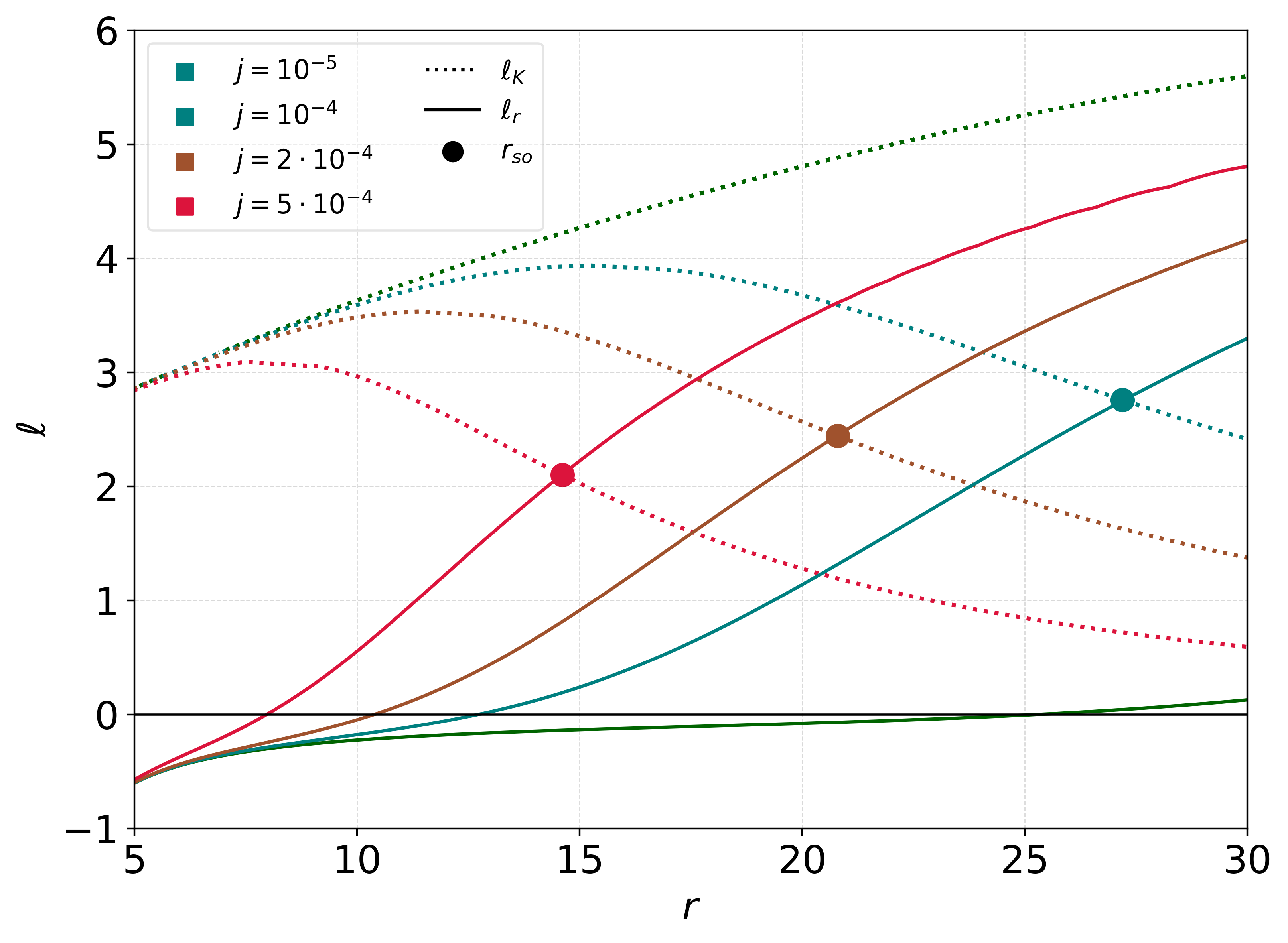}
  \caption{$a = 0.9$, prograde}
\end{subfigure}

\begin{subfigure}{.325\textwidth}
  \centering
  \includegraphics[width=\linewidth]{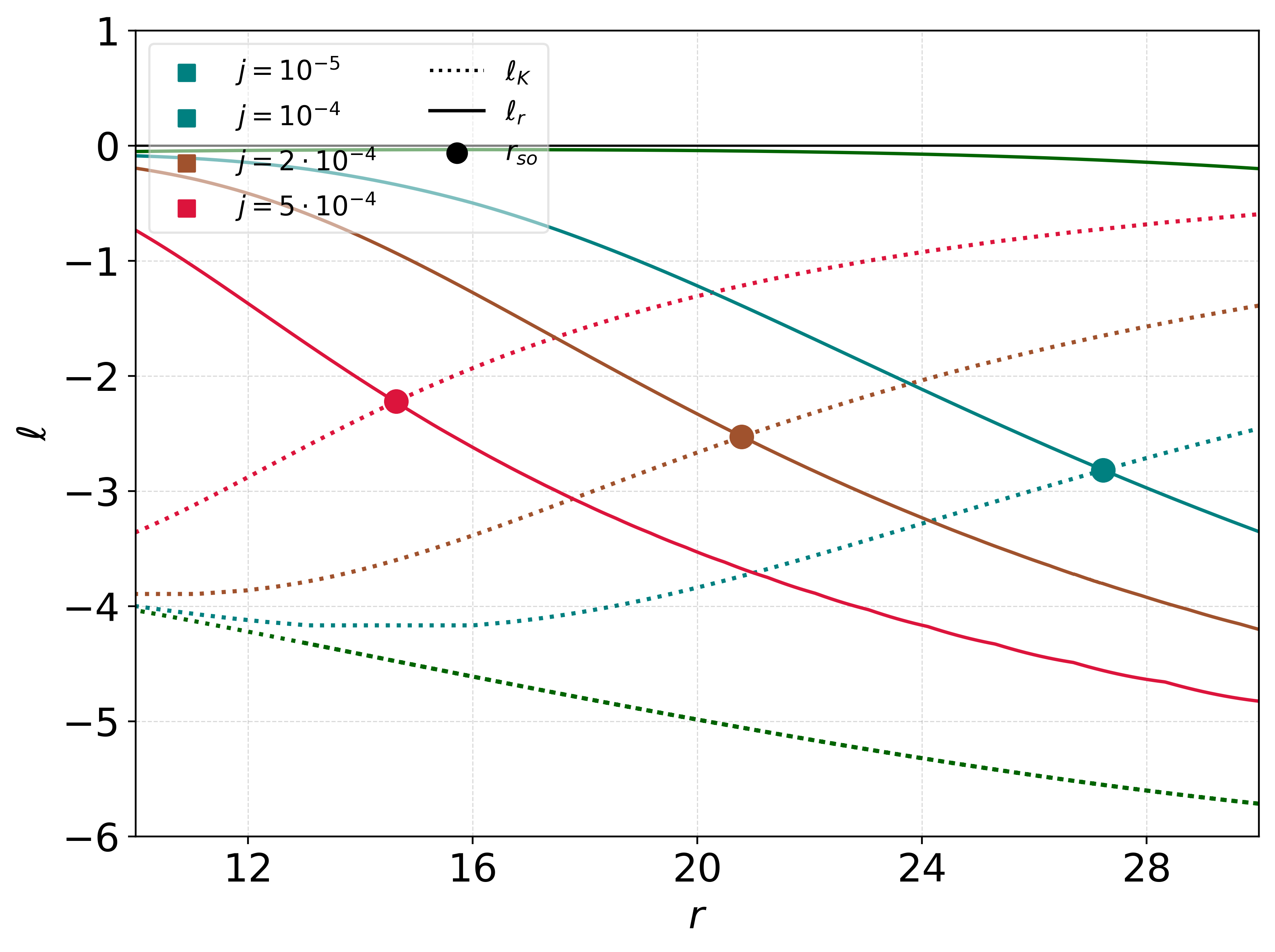}
  \caption{$a = 0.2$, retrograde}
\end{subfigure}
\begin{subfigure}{.325\textwidth}
  \centering
  \includegraphics[width=\linewidth]{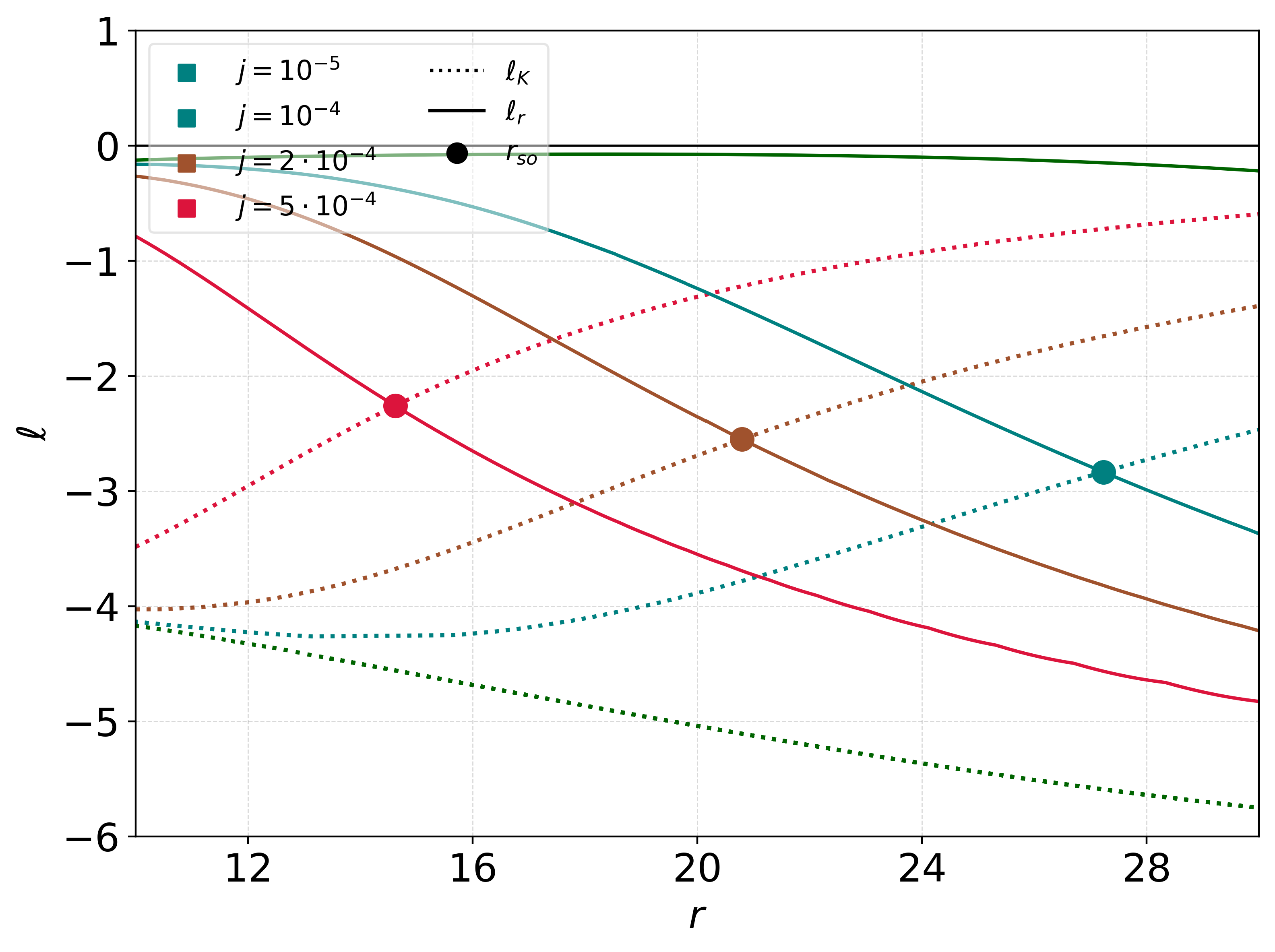}
  \caption{$a = 0.5$, retrograde}
\end{subfigure}
\begin{subfigure}{.325\textwidth}
  \centering
  \includegraphics[width=\linewidth]{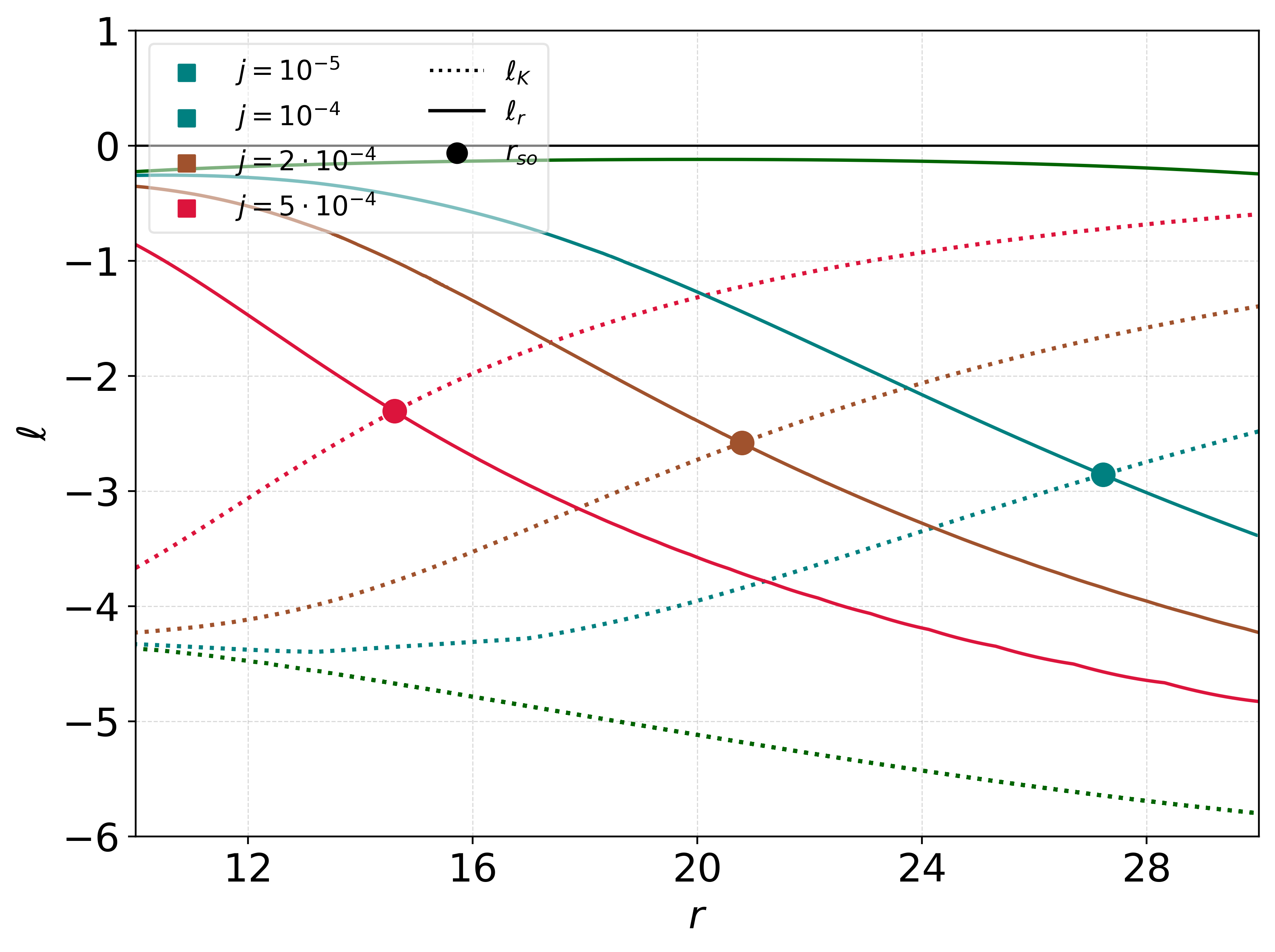}
  \caption{$a = 0.9$, retrograde}
\end{subfigure}
\caption{Rest specific angular momentum, $\ell_r = -\frac{g_{t\varphi}}{g_{tt}}$, plotted alongside the prograde and retrograde circular orbits for the various KBHSB solutions. The previously shown $\ell_K$ distributions are plotted as dotted curves for comparison. The circular marker symbols indicate the radial location of the static orbits $r_{so}$. At these radial locations $\ell_K$ is identical to $\ell_r$, as a consequence the corresponding circular orbit is a static orbit.}
\label{fig:static_orbits}
\end{figure}

The rest specific angular momentum $\ell_r$ alongside the circular orbits has an intersection with $\ell_K$ for the analyzed KBHSB solutions. The specific angular momentum of the circular orbit corresponding to this intersection is equal to the specific angular momentum a particle at rest would have in the co-moving reference frame. Therefore, a test particle, which is initially at rest in the co-moving reference frame, would always stay at rest in this reference frame in the absence of disturbances. However, it should be noted that all static orbits are unstable since the intersection point for each solution is outwards of $r_{ms}^{out}$. Furthermore, for $r > r_{so}$, the absolute value of $\ell_K$ is smaller than $\ell_r$ for prograde as well as retrograde motion. Thus, all circular orbits outwards of the static orbit are counter-rotating in the co-moving reference frame of the spacetime background. With increasing $j$, the radial location $r_{so}$ of the static orbits is moving inwards, regardless of the Kerr parameter $a$. The differences between the various Kerr parameters are in general negligible, thus $r_{so}$ is located for each solution roughly at the same radial location for a fixed $j$, regardless of the rotational direction. Therefore, we conclude that the Kerr parameter and rotation direction have a lesser impact on the properties of static orbits, which are mostly determined by $j$.

As mentioned in section 3 the circular geodesics are not located in the equatorial plane, in contrast to the Kerr solution. To each point of the presented curves for the $\ell_K$ and $\ell_r$ distributions, there is a corresponding varying $\theta$ coordinate. For a better understanding regarding the spatial distribution of the circular orbits, a cross-section plot is shown in Fig. \ref{fig:ell_K_2D} for the various analyzed KBHSB solutions.

\begin{figure}[H]
\centering
\begin{subfigure}{.325\textwidth}
  \centering
  \includegraphics[width=\linewidth]{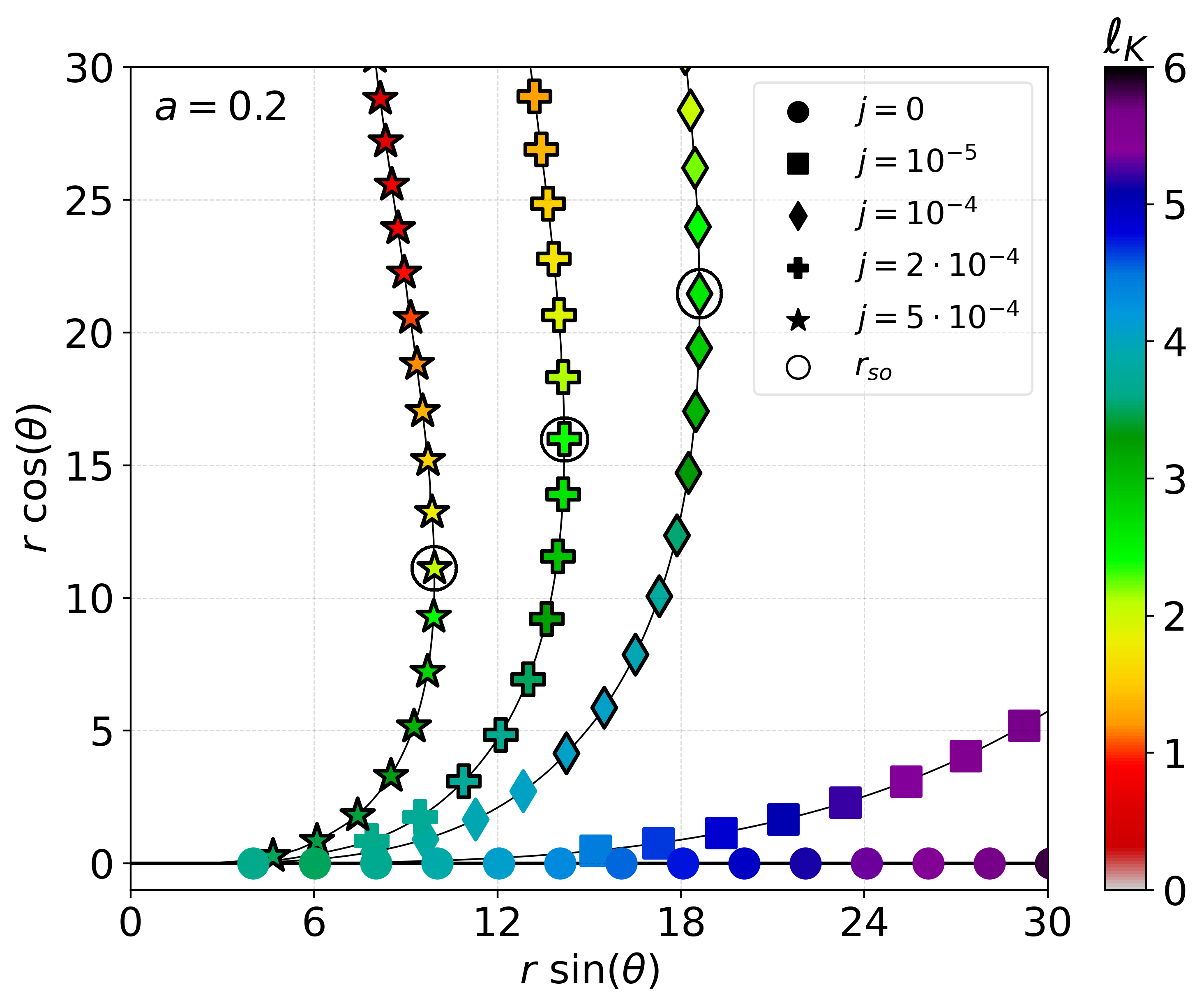}
  \caption{$a = 0.2$, prograde}
\end{subfigure}
\begin{subfigure}{.325\textwidth}
  \centering
  \includegraphics[width=\linewidth]{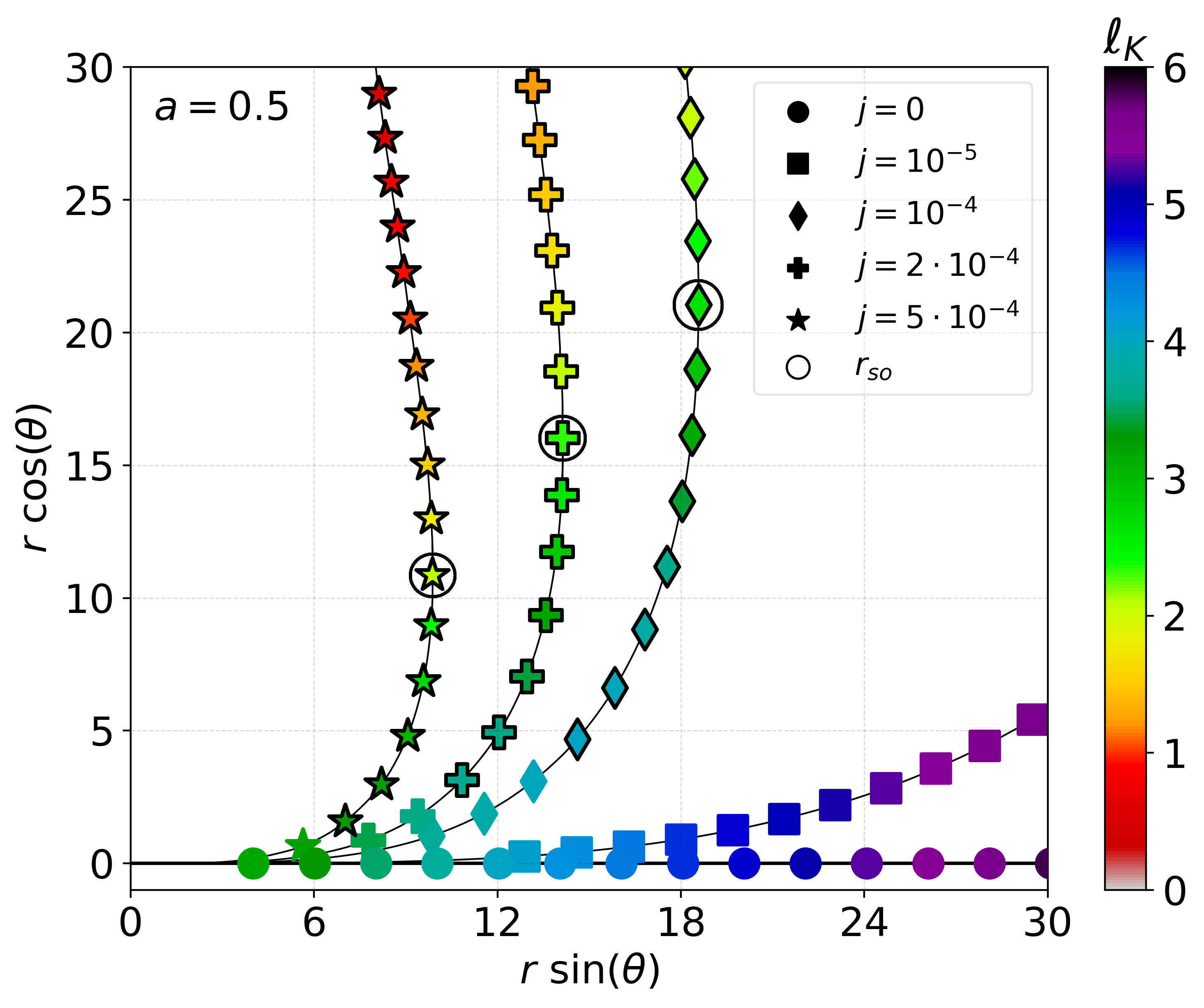}
  \caption{$a = 0.5$, prograde}
\end{subfigure}
\begin{subfigure}{.325\textwidth}
  \centering
  \includegraphics[width=\linewidth]{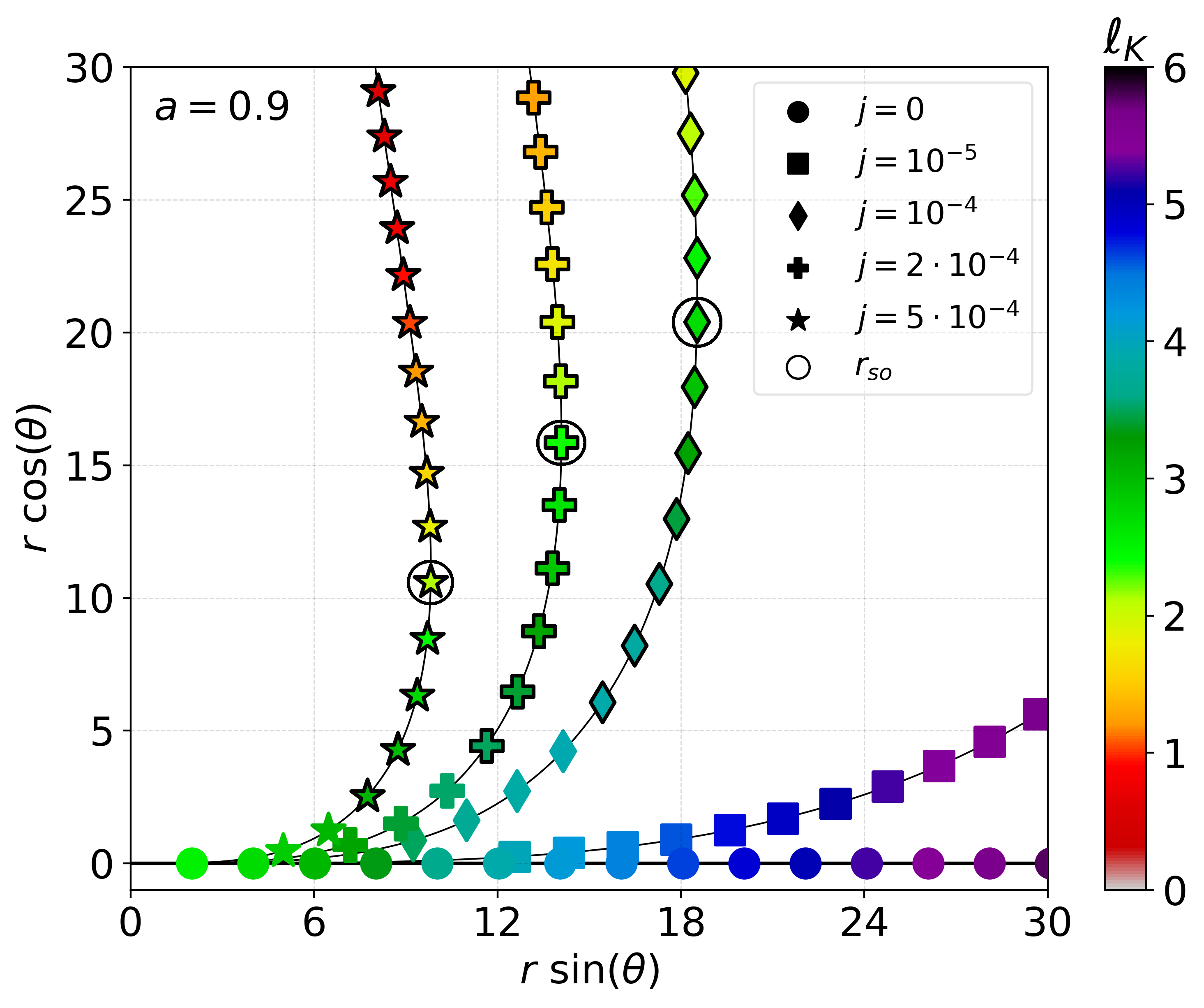}
  \caption{$a = 0.9$, prograde}
\end{subfigure}

\begin{subfigure}{.325\textwidth}
  \centering
  \includegraphics[width=\linewidth]{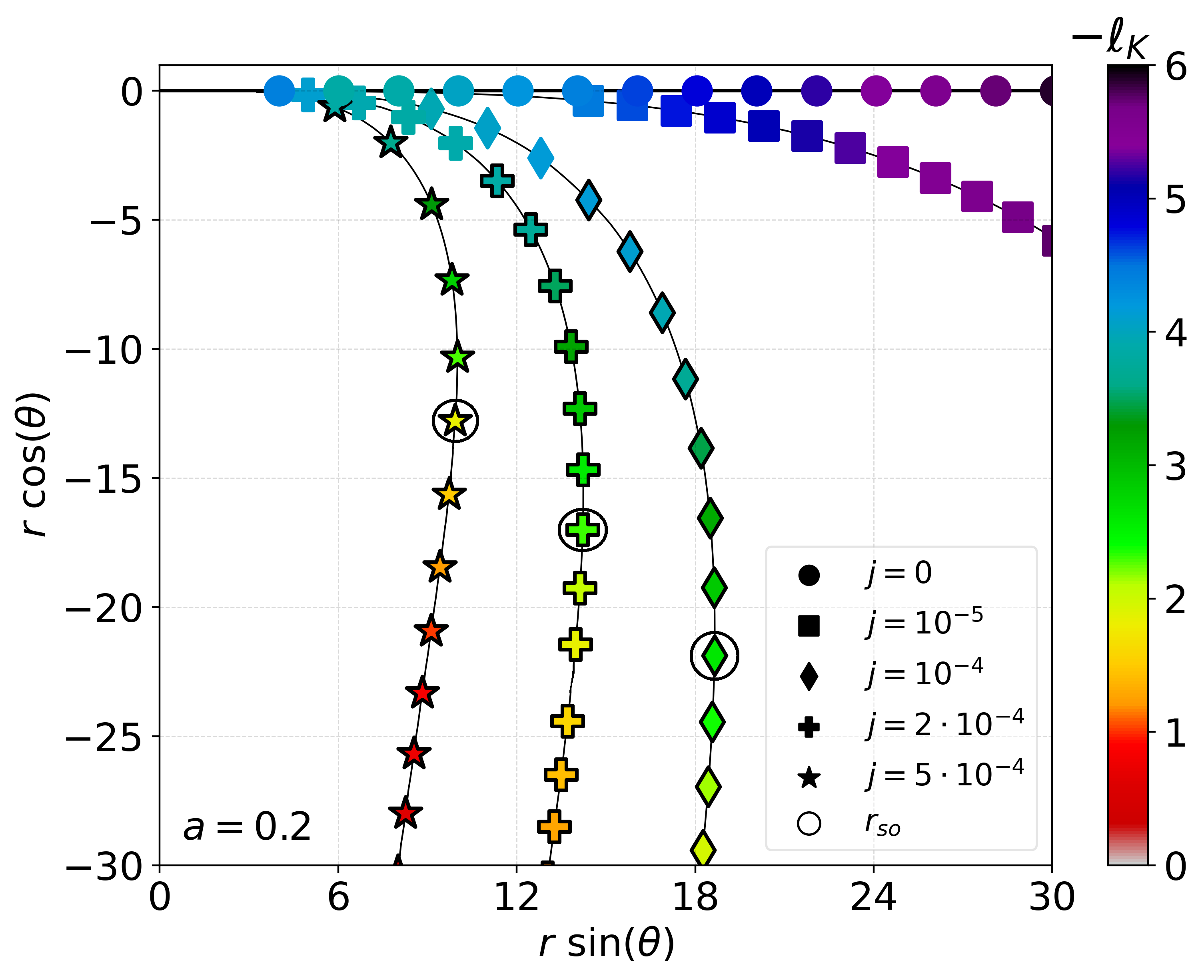}
  \caption{$a = 0.2$, retrograde}
\end{subfigure}
\begin{subfigure}{.325\textwidth}
  \centering
  \includegraphics[width=\linewidth]{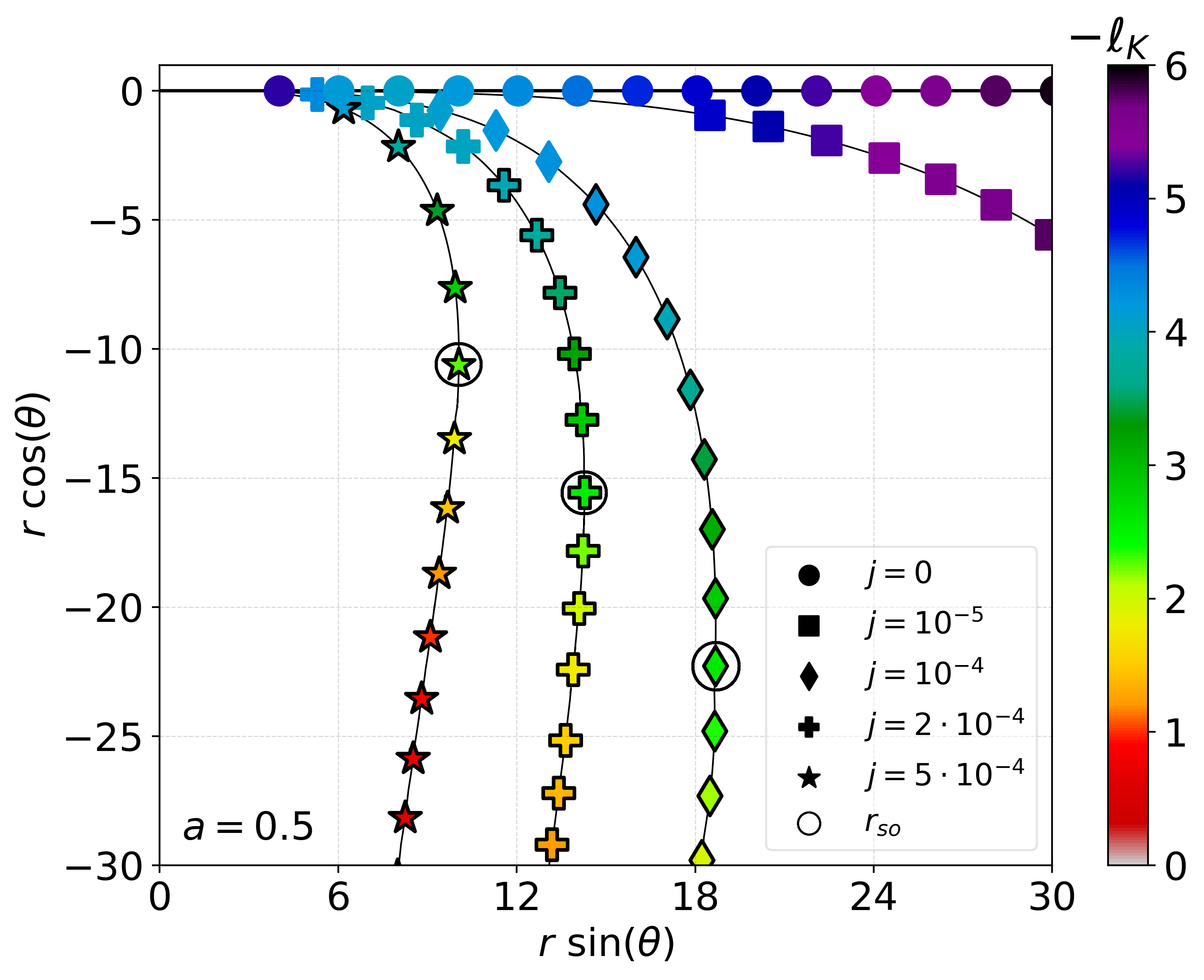}
  \caption{$a = 0.5$, retrograde}
\end{subfigure}
\begin{subfigure}{.325\textwidth}
  \centering
  \includegraphics[width=\linewidth]{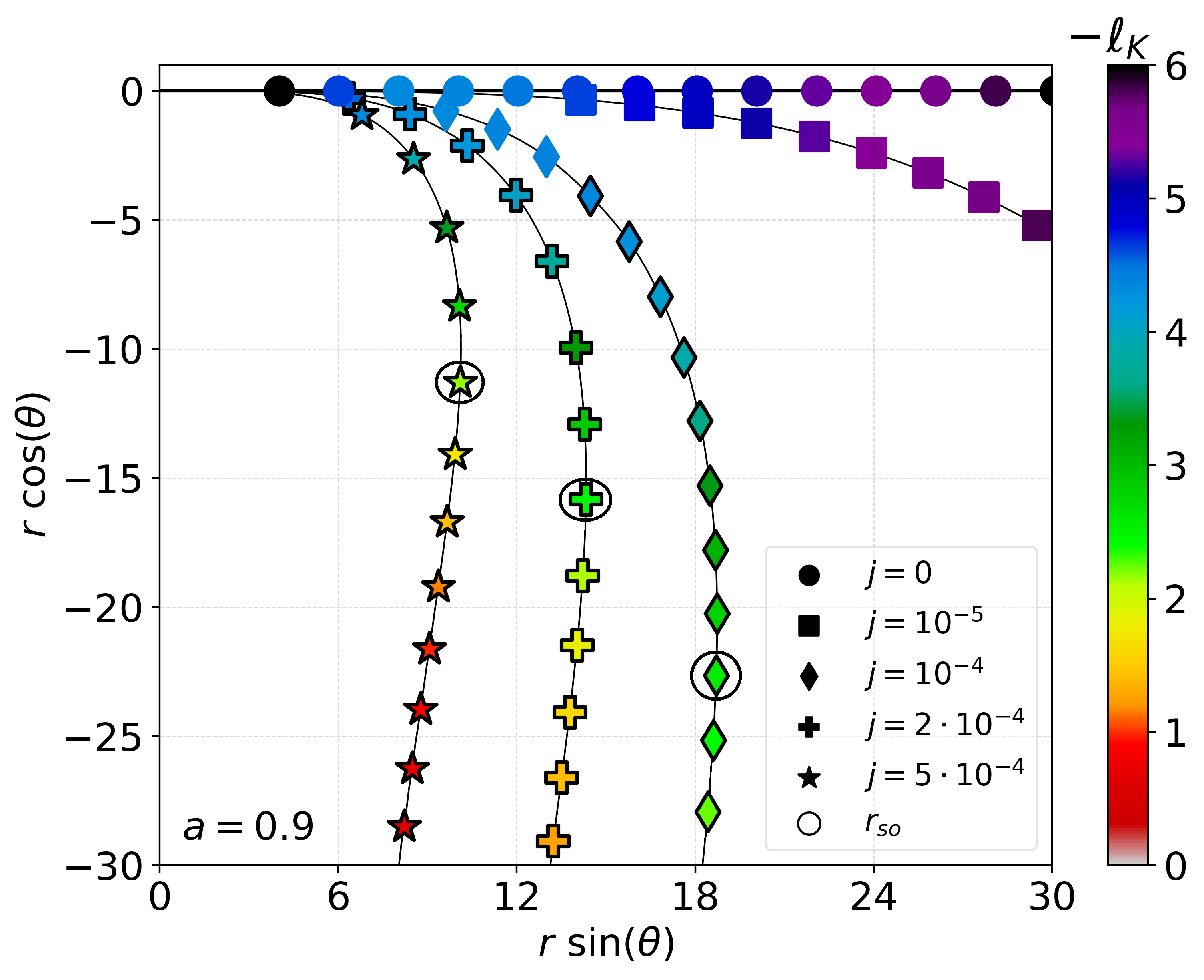}
  \caption{$a = 0.9$, retrograde}
\end{subfigure}
\caption{Cross section plots of circular orbits and their corresponding specific angular momentum $\ell_K$ for the different Kerr parameter solutions with varying swirling parameter $j$. The upper row presents $\ell_K$ of prograde orbits and the lower row presents $\ell_K$ for retrograde orbits. The terms prograde and retrograde refer here to the rotation direction of the KBHSB, which is defined by the Kerr spin parameter $a$. The different symbols mark selected solutions for circular orbits for varying $j$. They are coloured according to their specific angular momentum, which is described by the scale on the right of each plot. Unstable orbits are marked by a black edge around the symbol, and the static orbits are marked by a circle.}
\label{fig:ell_K_2D}
\end{figure}

The spatial location of KBHSB circular orbits is highly affected by the swirling parameter, it deviates from the equatorial plane and is located in planes orthogonal to the rotational axis in the upper and lower hemispheres. With increasing $j$ the verticality of the circular orbit distribution increases, and for $\ell_K \rightarrow 0$ the height of the corresponding circular orbit above or below the equatorial plane diverges. The location of static orbits is largely unaffected by $a$ and for increasing $j$ it moves closer to the equatorial plane. Stable orbits are only possible close to the equatorial plane. In the case of prograde motion, the orbits are solely located in the upper hemisphere (Fig. \ref{fig:ell_K_2D} upper row); they are therefore co-moving with the swirling background. For retrograde motion, the orbits are solely located in the lower hemisphere (Fig. \ref{fig:ell_K_2D} lower row). Since the frame-dragging direction of the swirling background changes its sign for the lower hemisphere, a counter-rotating test particle regarding the Kerr rotation is co-moving with the swirling background in the lower hemisphere. All circular orbits are therefore co-moving with their surrounding spacetime background reference frame. The spatial distribution of circular orbits throughout the Kerr parameter range is very similar, thus the influence of the Kerr parameter is negligible. We conclude that the general properties of circular orbits for KBHSB are largely determined by the swirling parameter, with the Kerr parameter playing a minor role. However, the Kerr parameter has a significant influence on the values of the angular momentum distribution and the stability of the circular orbits. The rotation of the KBHSB has a stabilizing effect on the co-rotating orbits and a destabilizing effect on the counter-rotating orbits. This can be attributed to the rotation of the swirling background. Orbits located in the lower hemisphere are co-rotating with the spacetime background but counter-rotating with the black hole, which is inducing instabilities due to the opposite spin direction of the swirling and Kerr rotation. The magnitude of this effect grows with increasing swirling and Kerr parameters $j$ and $a$. 

As a consequence, accretion structures can exist only for a specific range of $\ell_0$, which is lower bounded by $\ell_{ms}^{in}$ and upper bounded by $\ell_{ms}^{out}$, since only for $|\ell_0| \in [|\ell_{ms}^{in}|, |\ell_{ms}^{out}|]$ stable circular orbits exist. Thus, disk solutions with static orbits and static surfaces do not exist. The disk center can only be located between $r_{ms}^{in}$ and $r_{ms}^{out}$ and with increasing $j$ the solution space for prograde accretion structures gets smaller, as the solution range of stable orbits decreases. Due to the stabilizing effect of the KBHSB rotation, the solution space for prograde (retrograde) accretion structures is greater (smaller) for faster rotating KBHSBs compared to slower rotating ones. This is depicted exemplary in Fig. \ref{fig:W_j4} for $j = 10^{-4}$ and varying values for $\ell_0$.

\begin{figure}[H]
\centering
\begin{subfigure}{.325\textwidth}
  \centering
  \includegraphics[width=\linewidth]{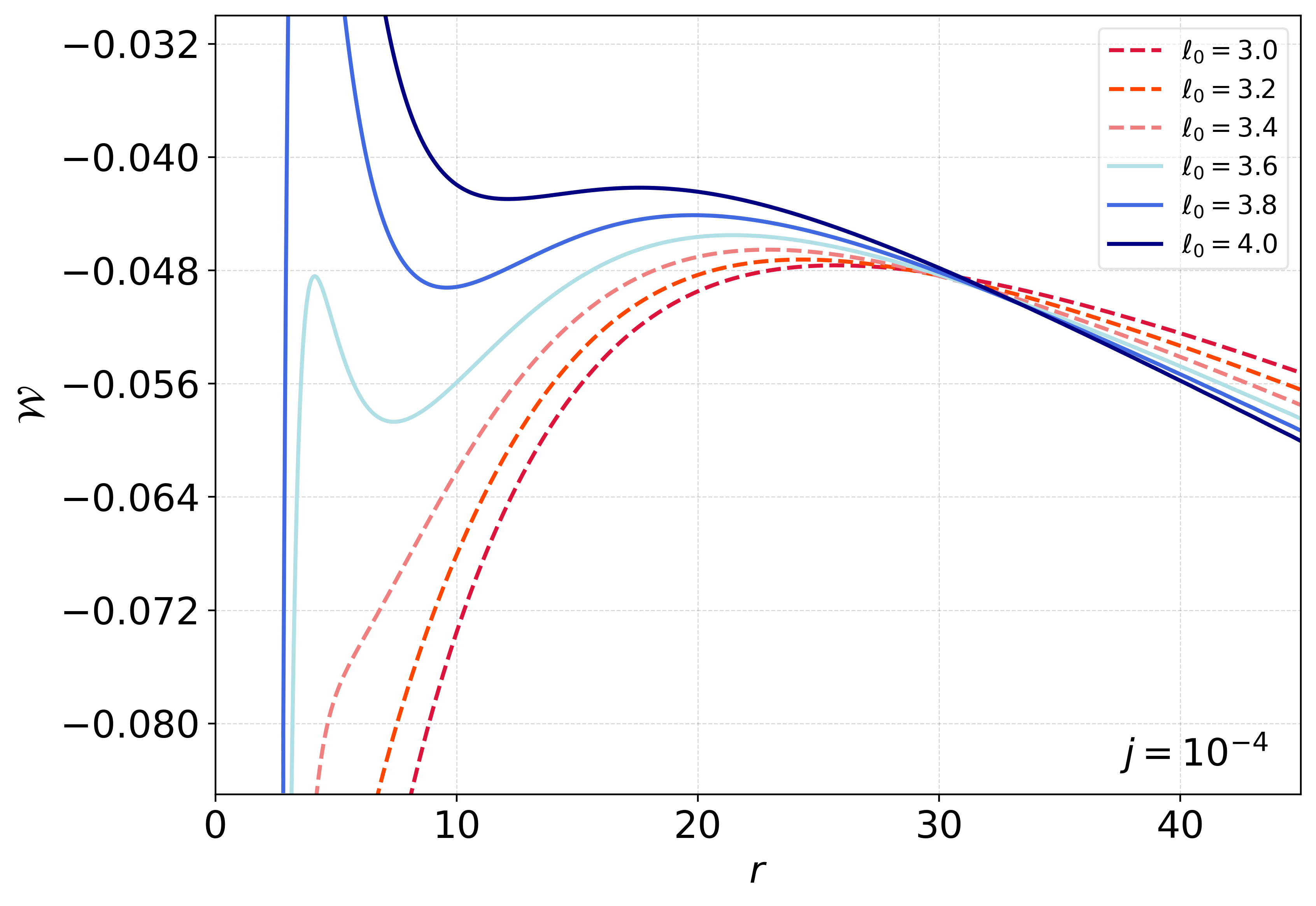}
  \caption{$a = 0.2$, prograde}
\end{subfigure}
\begin{subfigure}{.325\textwidth}
  \centering
  \includegraphics[width=\linewidth]{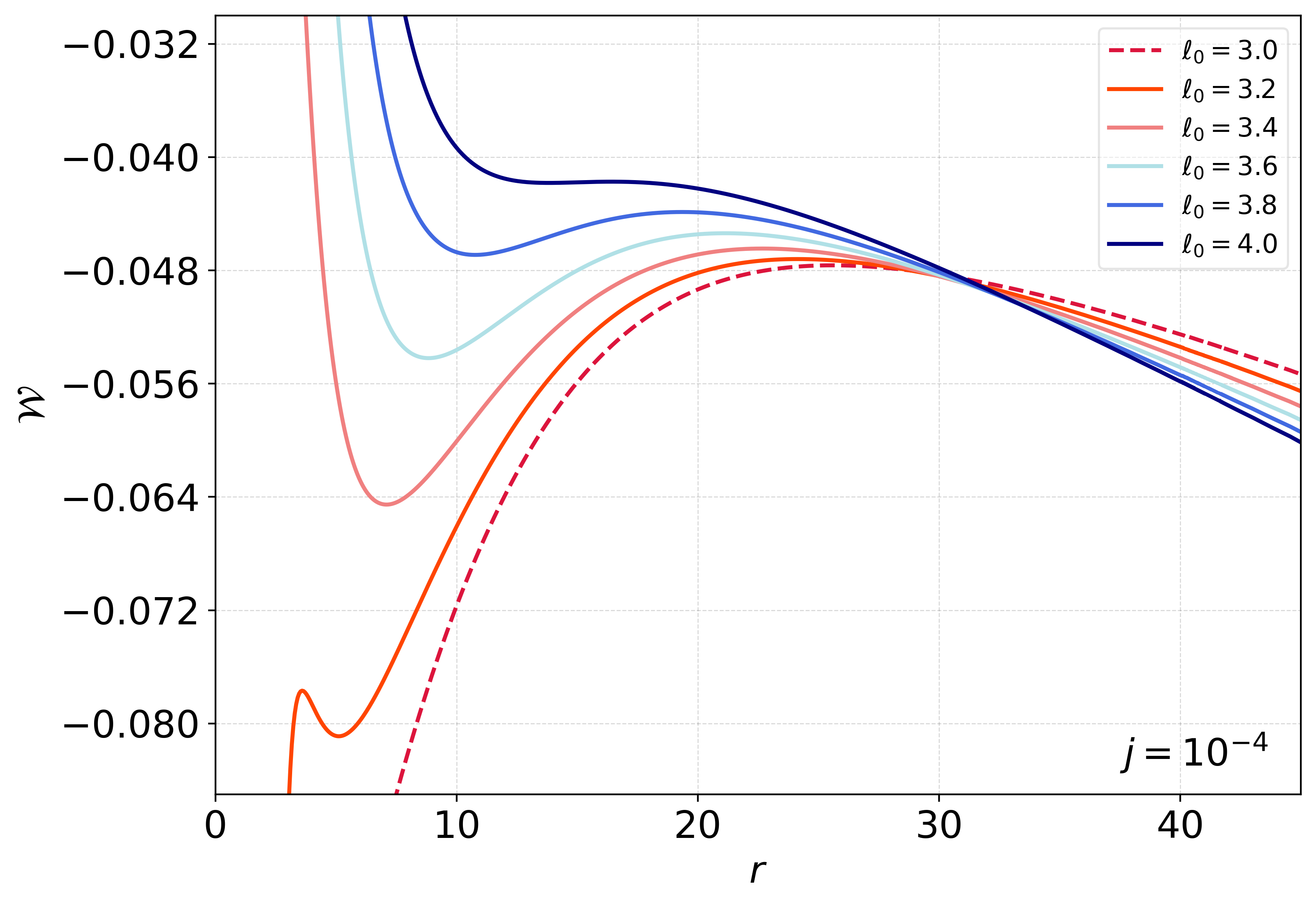}
  \caption{$a = 0.5$, prograde}
\end{subfigure}
\begin{subfigure}{.325\textwidth}
  \centering
  \includegraphics[width=\linewidth]{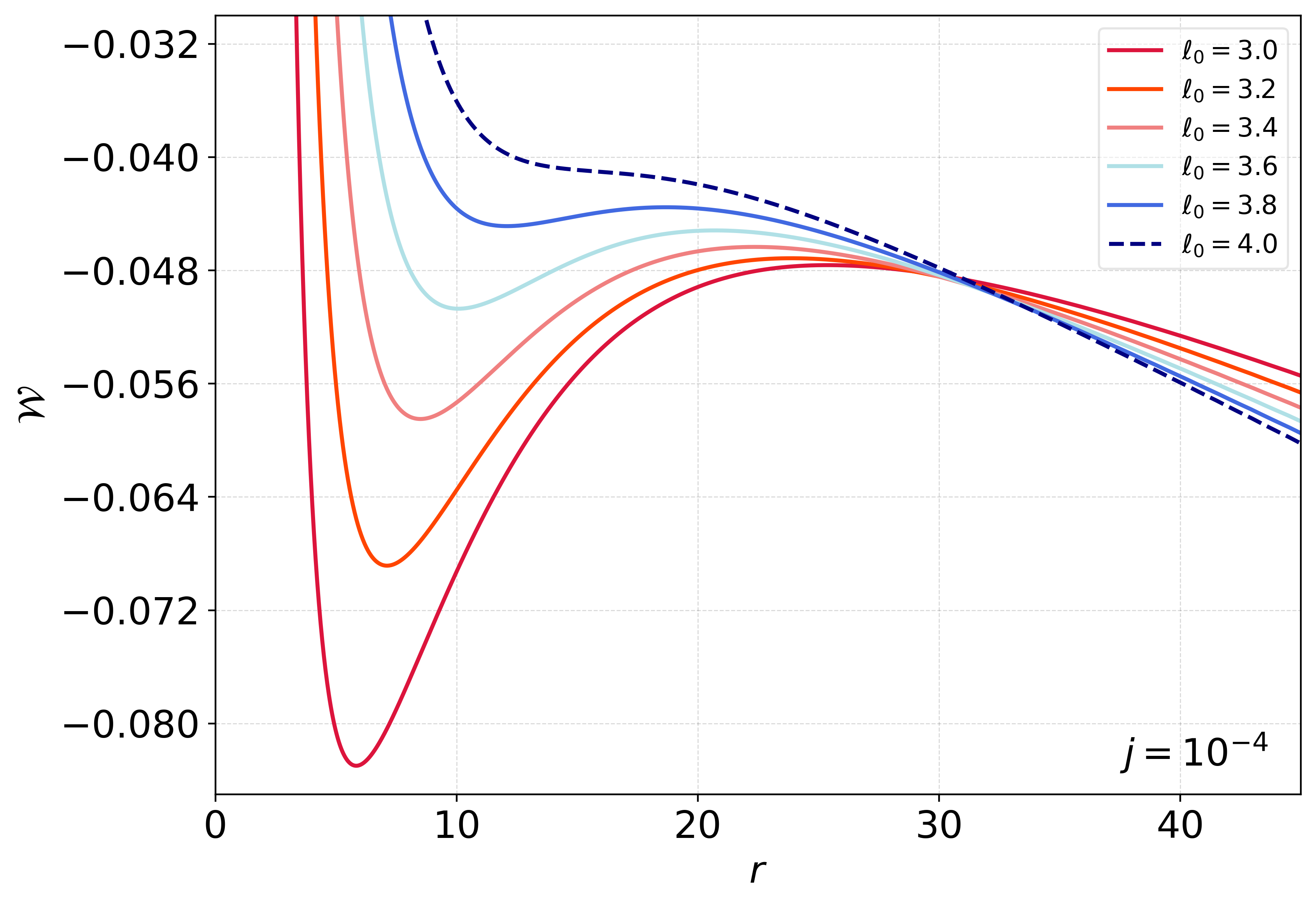}
  \caption{$a = 0.9$, prograde}
\end{subfigure}

\begin{subfigure}{.325\textwidth}
  \centering
  \includegraphics[width=\linewidth]{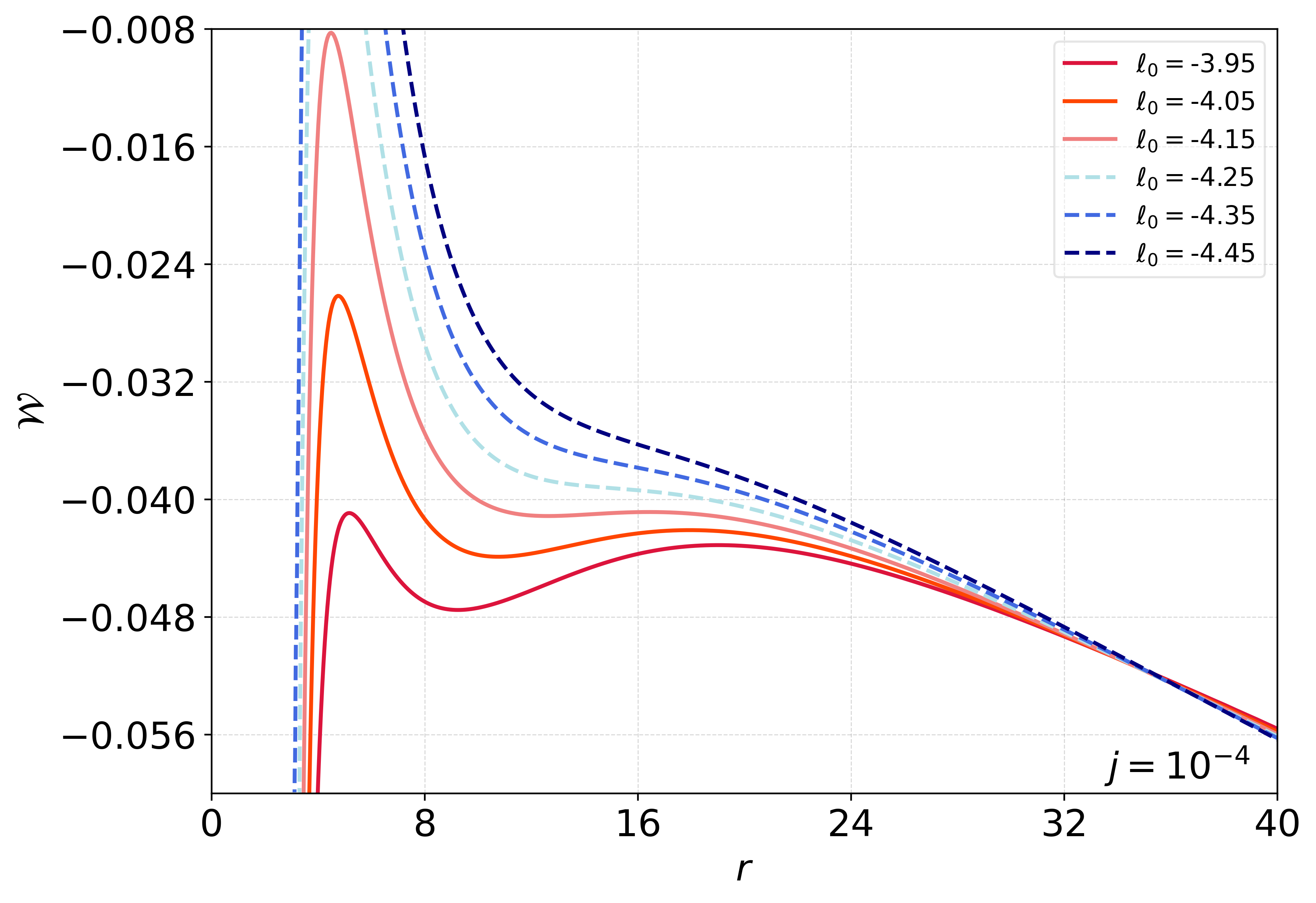}
  \caption{$a = 0.2$, retrograde}
\end{subfigure}
\begin{subfigure}{.325\textwidth}
  \centering
  \includegraphics[width=\linewidth]{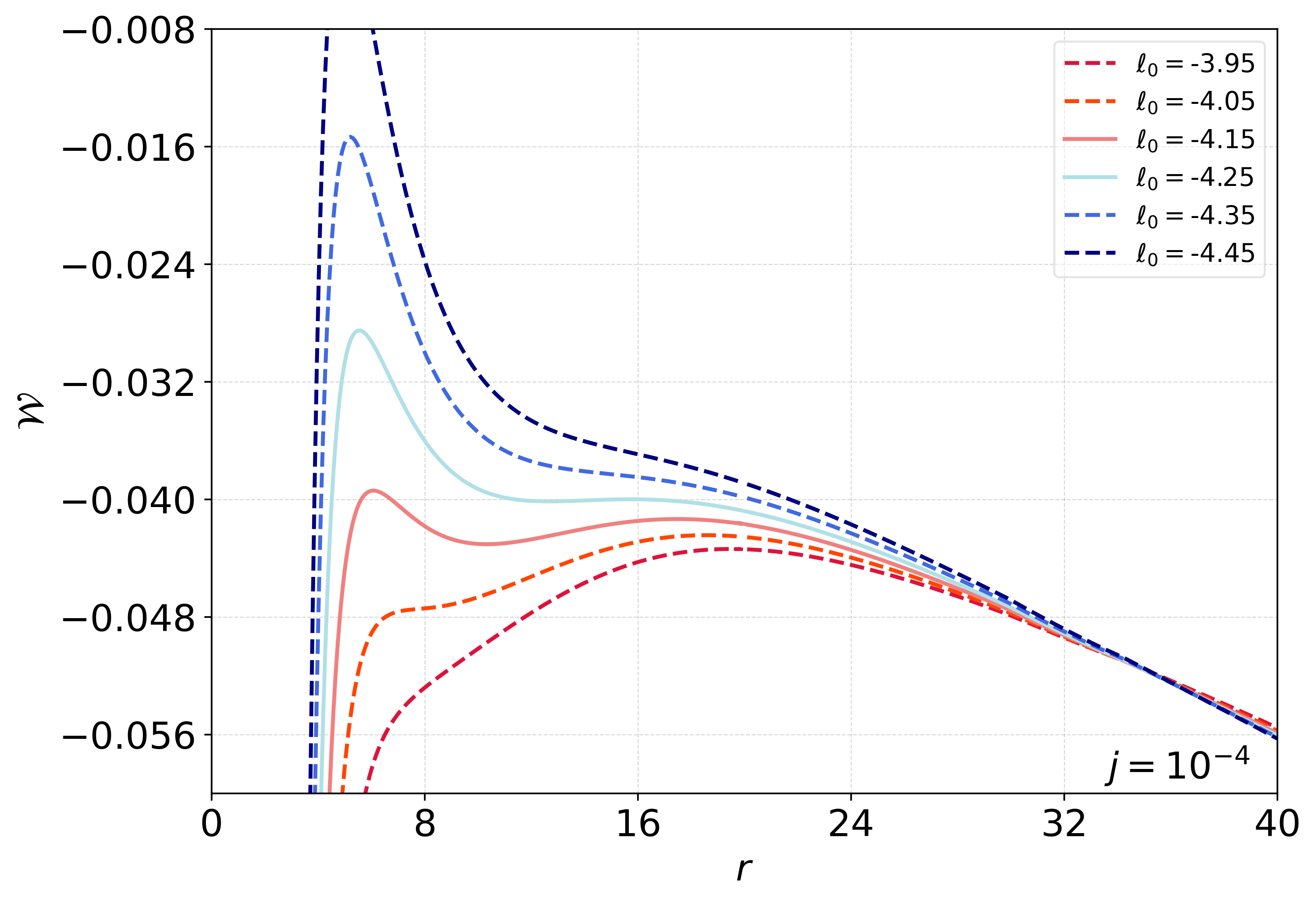}
  \caption{$a = 0.5$, retrograde}
\end{subfigure}
\begin{subfigure}{.325\textwidth}
  \centering
  \includegraphics[width=\linewidth]{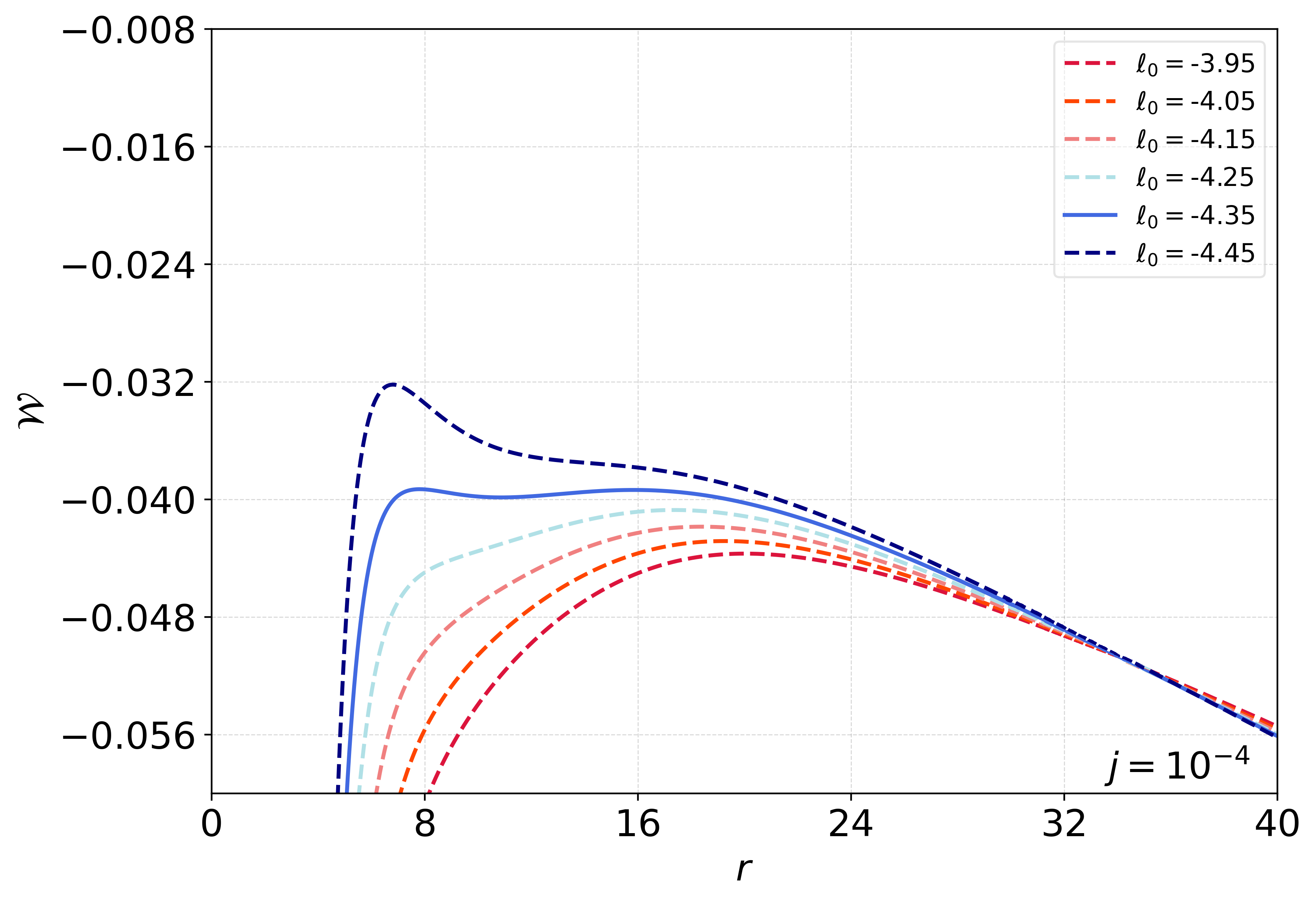}
  \caption{$a = 0.9$, retrograde}
\end{subfigure}
\caption{Effective potential $\mathcal{W}$ alongside circular geodesics for the different Kerr spin parameters and various $\ell_0$. The swirling parameter is set to $j = 10^{-4}$ in all solutions. The upper row presents $\mathcal{W}$ for prograde motion and the lower row for retrograde motion (referring to the rotation direction defined by the Kerr parameter $a$). Dashed curves indicate solutions for which no closed disk surfaces can exist due to the absence of a local minimum.}
\label{fig:W_j4}
\end{figure}

With increasing $a$ prograde disk exists for a greater range of $\ell_0$ for the same $j$ (Fig. \ref{fig:W_j4} upper row). For retrograde disks the solution space narrows down for the same $j$ and $\ell_0$ parameter range (Fig. \ref{fig:W_j4} lower row). All possible disk solutions can be classified into two main types: Type 1 solutions are composed of an inner cusp and an outer cusp, between which the disk center is located. The inner cusp corresponds to an inner local maximum, the disk center to the local minimum, and the outer cusp to the outer local maximum of the effective potential $\mathcal{W}$. These solutions can be further classified into three subtypes: Type 1a disks, where the inner maximum is greater than the outer maximum (Fig. \ref{fig:W_j4} (d)). Type 1b disks, where the outer maximum is greater than the inner maximum (Fig. \ref{fig:W_j4} (b)). And the special case Type 1c, where the inner and outer potential maxima have the same value. It can be viewed as the limiting case of a Type 1a or 1b solution. In contrast to Type 1 solutions, Type 2 solutions are only composed of a disk center and an outer cusp, the effective potential has one minimum and only one maximum, which is located outwards to the minimum (Fig. \ref{fig:W_j4} (c)). With increasing $a$ the disk solution space moves from mostly Type 1 solutions to Type 2 solutions for prograde disks. For retrograde disks all solutions are of Type I. Exemplary cross-section plots of the different disk types are presented in Fig. \ref{fig:disk_types} for prograde disks. In the case of retrograde disks, the defined disk types are analogous; besides that only solutions of Type 1 exist as mentioned.

\begin{figure}[H]
\centering
$\begin{array}{cc}
\begin{subfigure}{.325\textwidth}
  \centering
  \includegraphics[width=\linewidth]{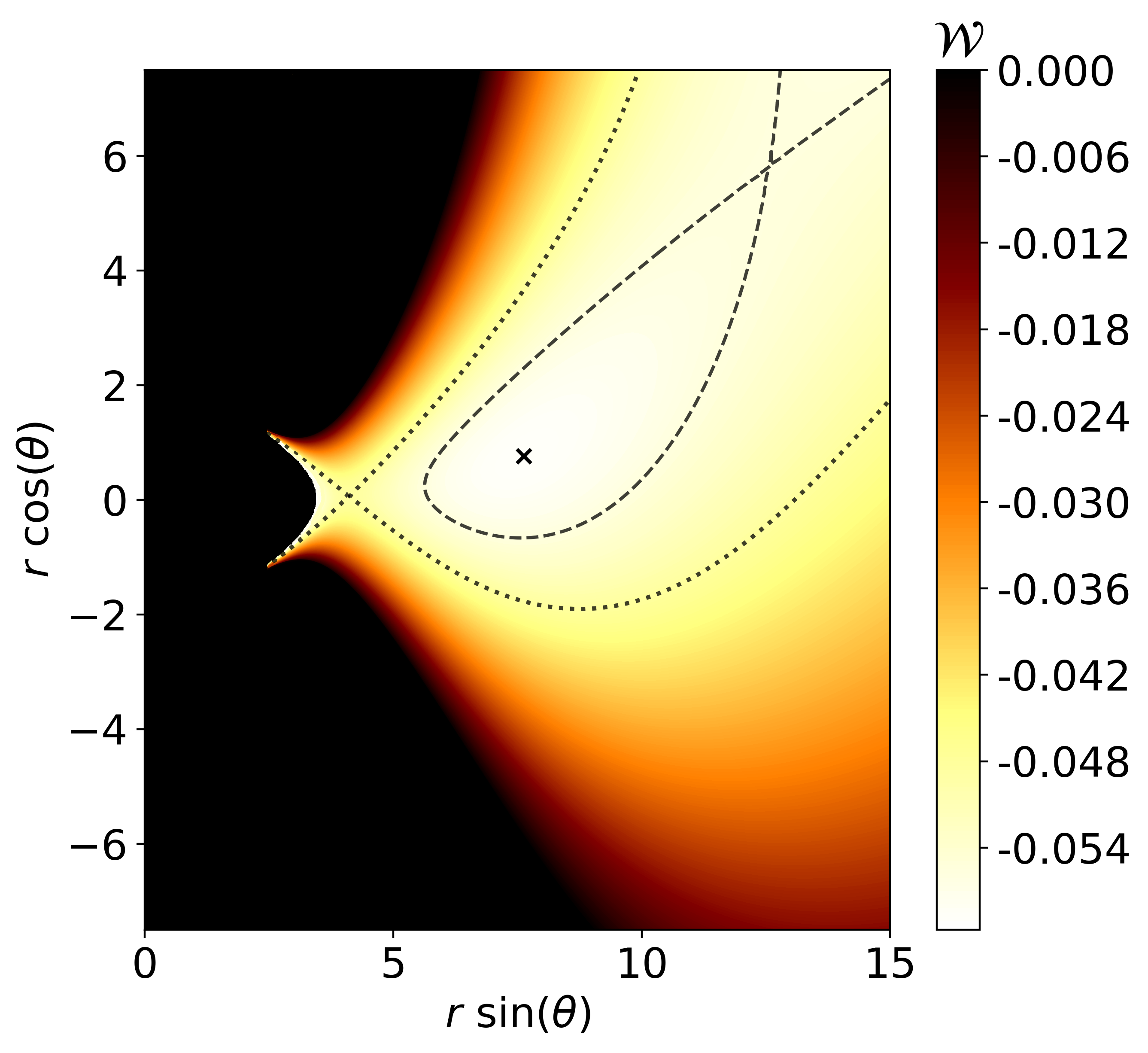}
  \caption{Type 1a}
\end{subfigure} &
\begin{subfigure}{.325\textwidth}
  \centering
  \includegraphics[width=\linewidth]{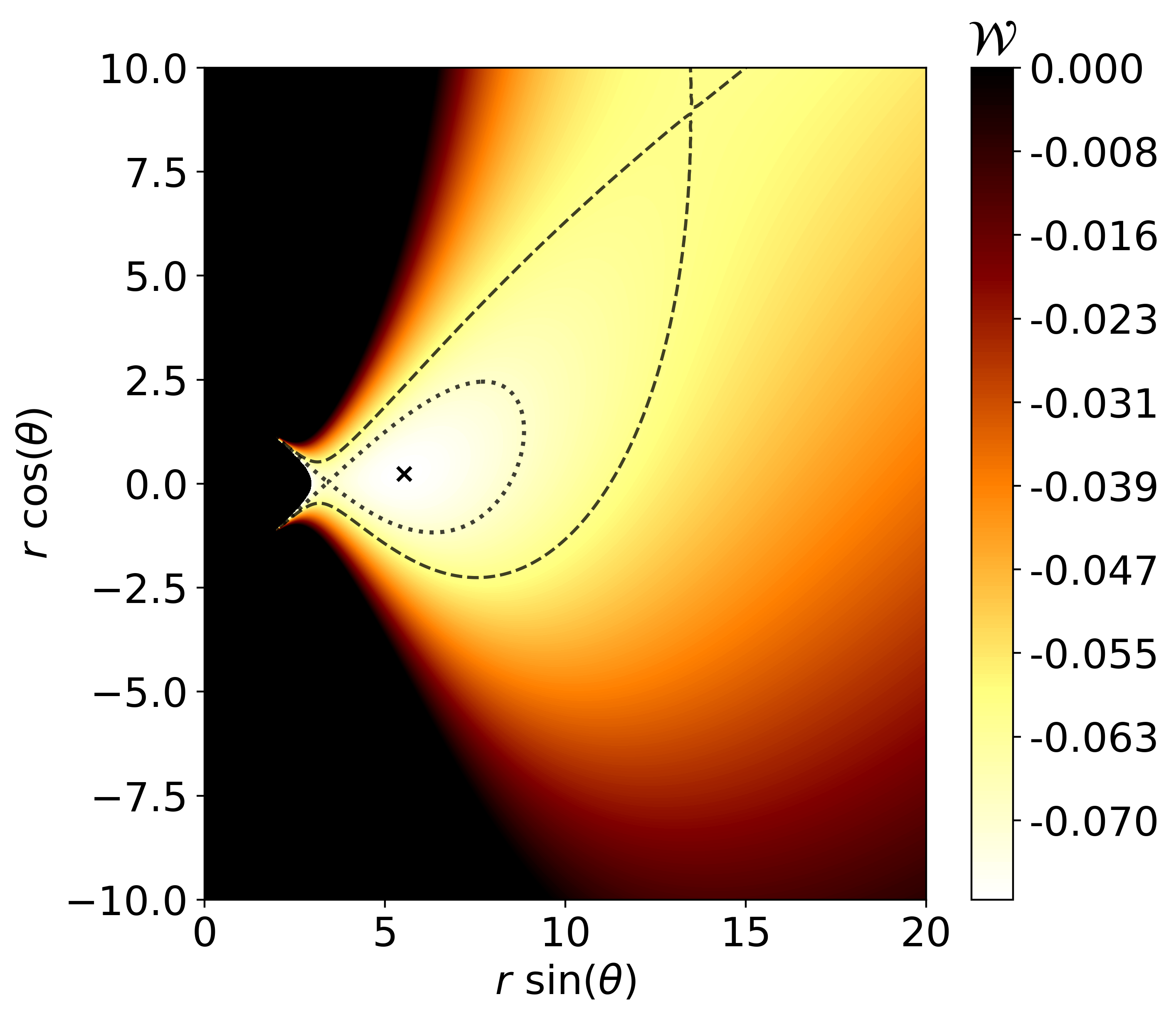}
  \caption{Type 1b}
\end{subfigure}
\end{array}$
\begin{subfigure}{.325\textwidth}
  \centering
  \includegraphics[width=\linewidth]{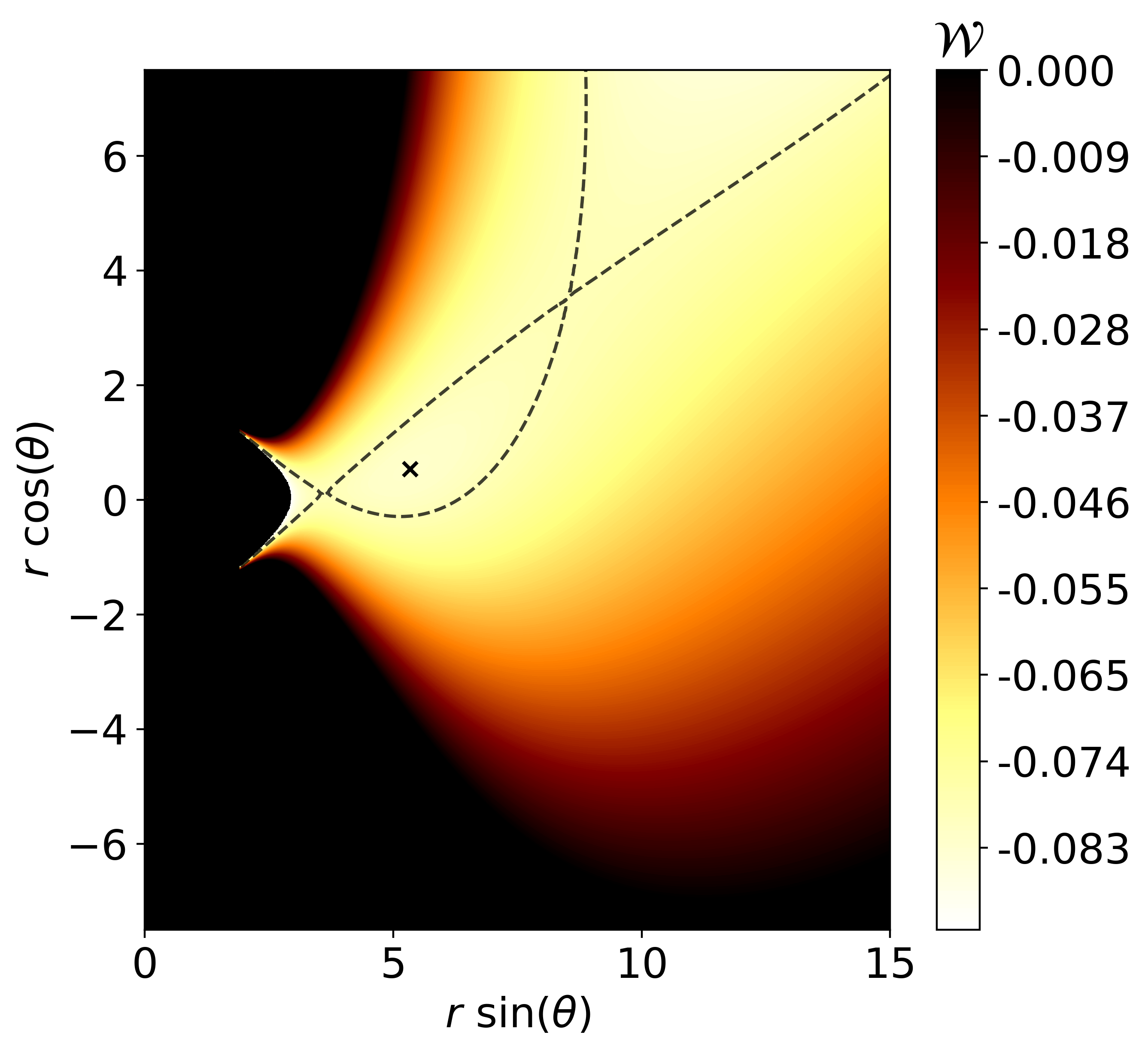}
  \caption{Type 1c}
\end{subfigure}
\begin{subfigure}{.325\textwidth}
  \centering
  \includegraphics[width=\linewidth]{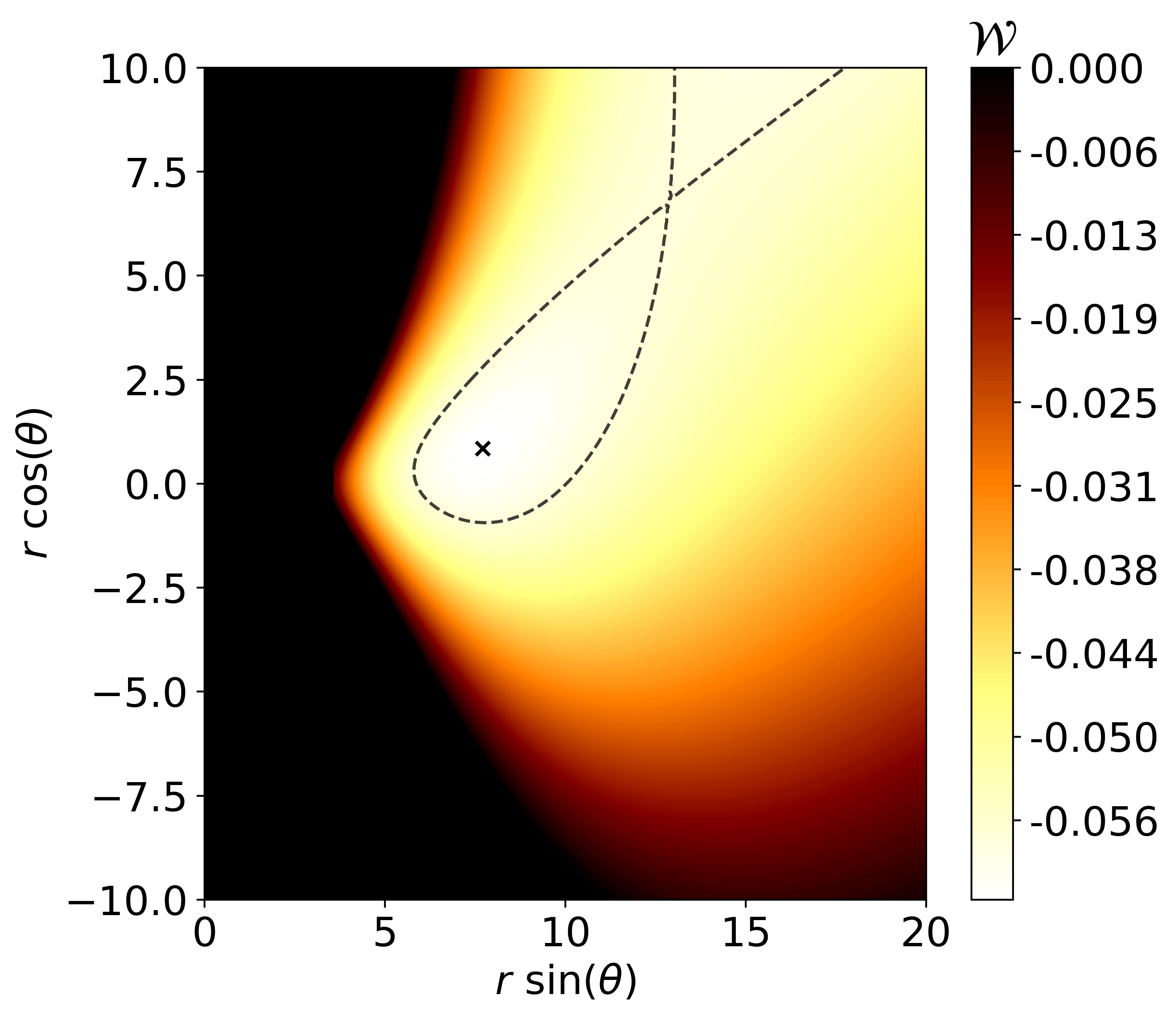}
  \caption{Type 2}
\end{subfigure}
\caption{Meridional cross section of $\mathcal{W}$ for the different disk types. As an example for a Type 1a disk, the parameters $a = 0.2$, $j = 2 \cdot 10^{-4}$ and $\ell_0 = 3.6$ where used. For Type 1b the parameters $a = 0.5$, $j = 2 \cdot 10^{-4}$ and $\ell_0 = 3.23$. For Type 1c the parameters $a = 0.5$, $j = 5 \cdot 10^{4}$ and $\ell_0 = 3.1964$ and for Type 2 the parameters $a = 0.5$, $j = 2 \cdot 10^{-4}$ and $\ell_0 = 3.45$. The dotted curve corresponds to the equi-potential surface of the inner cusp, the dashed curve corresponds to the equi-potential surface of the outer cusp. The $\times$ symbol marks the location of the disk center.}
\label{fig:disk_types}
\end{figure}

For Type 1a disks, the equi-potential surface of the inner cusp is not closed, it is open towards the KBHSB center and expands from the cusp outwards to infinity. The equi-potential surface of the outer cusp is semi-closed, with the closed region containing the disk center and the open region expanding from the outer cusp towards infinity. This enclosed region with the disk center is fully surrounded by the equi-potential surface of the inner cusp. Depending on the initial conditions various disk evolutions seem possible, as the complexity of disk dynamics could increase due to the presence of two cusps. Matter could be accreted through the inner cusp by flowing past the center region, without having a major effect on the dynamics and stability of the center region, as it is fully enclosed by the equi-potential surface of the outer cusp. On a different note, the whole disk could become unstable, since matter from the disk center could disperse through the outer cusp to unstable orbits, leading to a diffusion of the whole inner disk region. In the case of Type 1b disks, the equi-potential surface of the inner cusp is semi-closed, it fully encloses the disk center and is open towards the KBHSB center. The equi-potential surface of the outer cusp is open towards the KBHSB center and from the outer cusp outwards to infinity. It surrounds the inner cusp potential surface and the disk center. Accretion processes through both cusps are conceivable. Matter from outside could accrete through the outer cusp towards the KBHSB center, flowing past the center region. Accretion from the disk center through the inner cusp towards the KBHSB could also take place. Since the center region is fully surrounded by the outer cusp potential surface, an indirect accretion process could take place, where matter flows through the outer cusp towards the inner potential surface, feeding the central region of the accretion structure, which could trigger an accretion process through the inner cusp. In the limiting case Type 1c, the effective potential at the inner and outer cusp is identical; the corresponding equi-potential surface has two intersections, it is open towards the KBHSB center and open outwards of the outer cusp. Between the inner and outer cusp it fully encloses the disk center region. An accretion process may happen through the outer cusp towards the disk center and through the inner cusp towards the KBHSB center. The disk could also disperse if matter from the center region flows through the outer cusp to the outer regions. For Type 2 disks only an outer cusp exists, with the equi-potential surface of the cusp being closed towards the KBHSB and open outwards of the cusp towards infinity. As for Type 1a disks it fully encloses the disk center. Due to the absence of an inner cusp, no accretion processes without additional inner-physical disk effects are possible. Nevertheless, the discussed runaway instability for Type 1 disks could also apply to Type 2 disks, where matter dispersing through the outer cusp towards unstable orbits could lead to a diffusion of the center region.

In order to investigate the influence of the Kerr parameter on prograde disk properties, different exemplary solutions for the same parameters and varying Kerr parameter are presented in Fig. \ref{fig:disks_l3.7}.

\begin{figure}[H]
\centering
\begin{subfigure}{.325\textwidth}
  \centering
  \includegraphics[width=\linewidth]{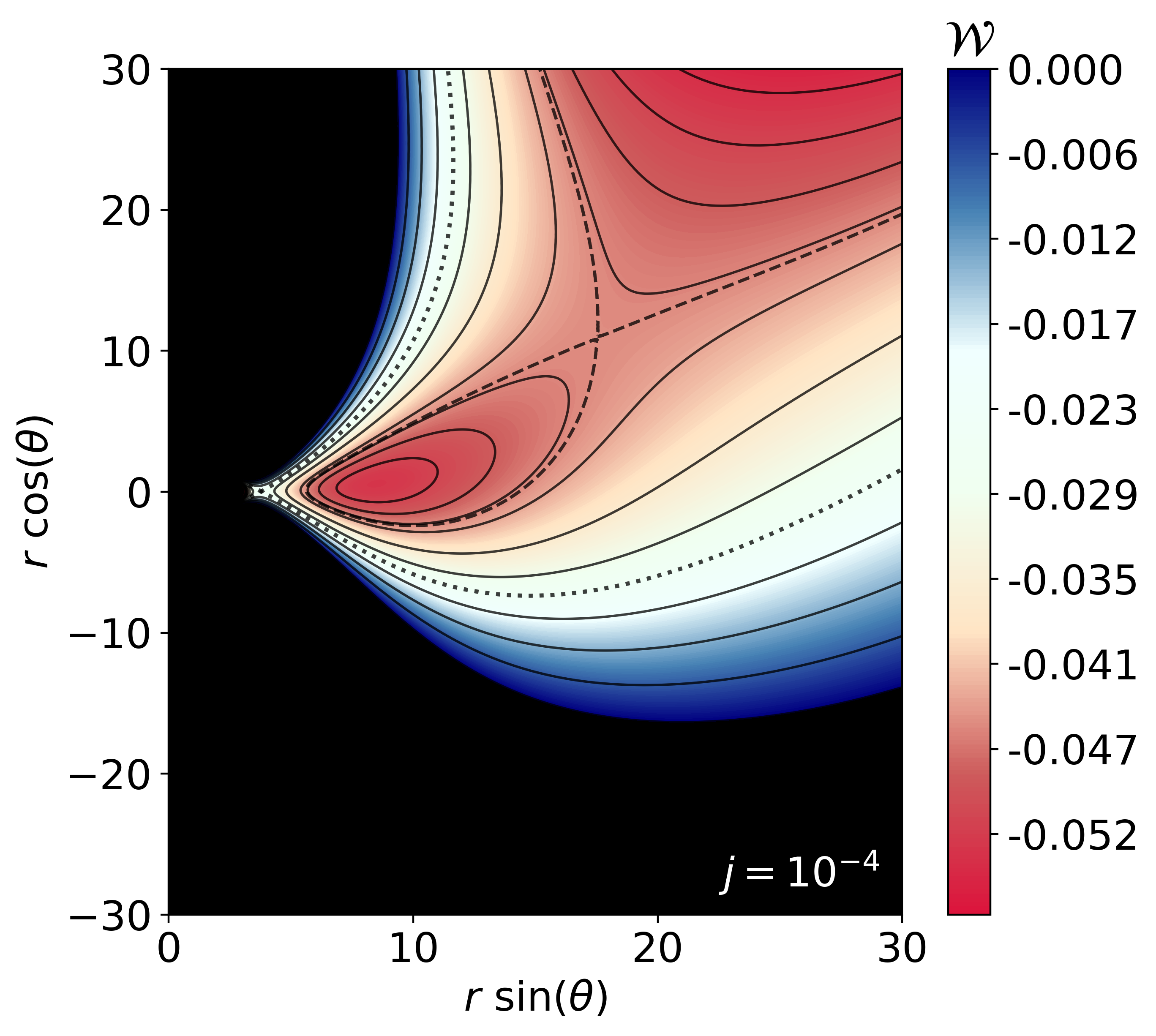}
  \caption{$a = 0.2$, $\ell_0 = 3.7$}
\end{subfigure}
\begin{subfigure}{.325\textwidth}
  \centering
  \includegraphics[width=\linewidth]{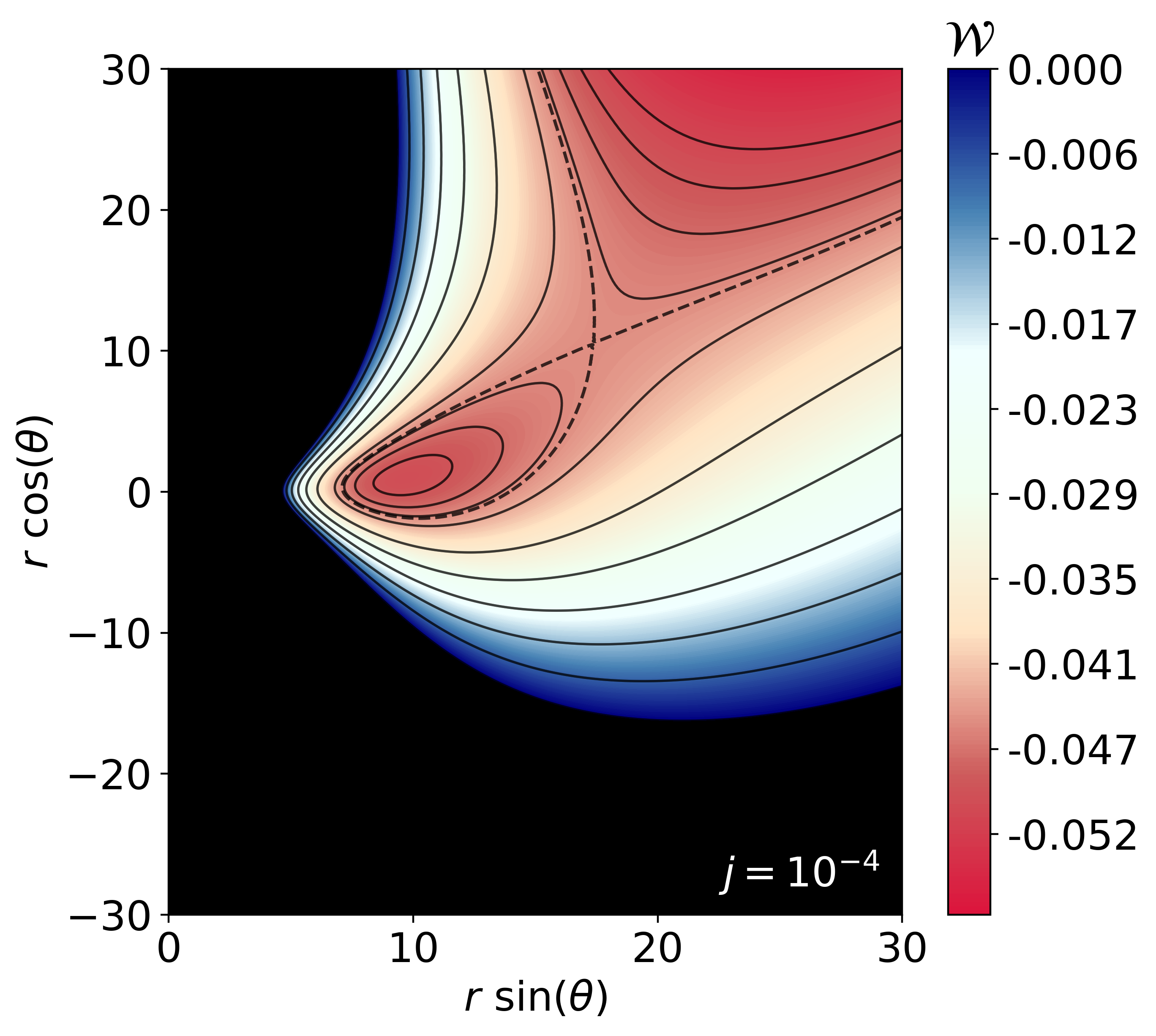}
  \caption{$a = 0.5$, $\ell_0 = 3.7$}
\end{subfigure}
\begin{subfigure}{.325\textwidth}
  \centering
  \includegraphics[width=\linewidth]{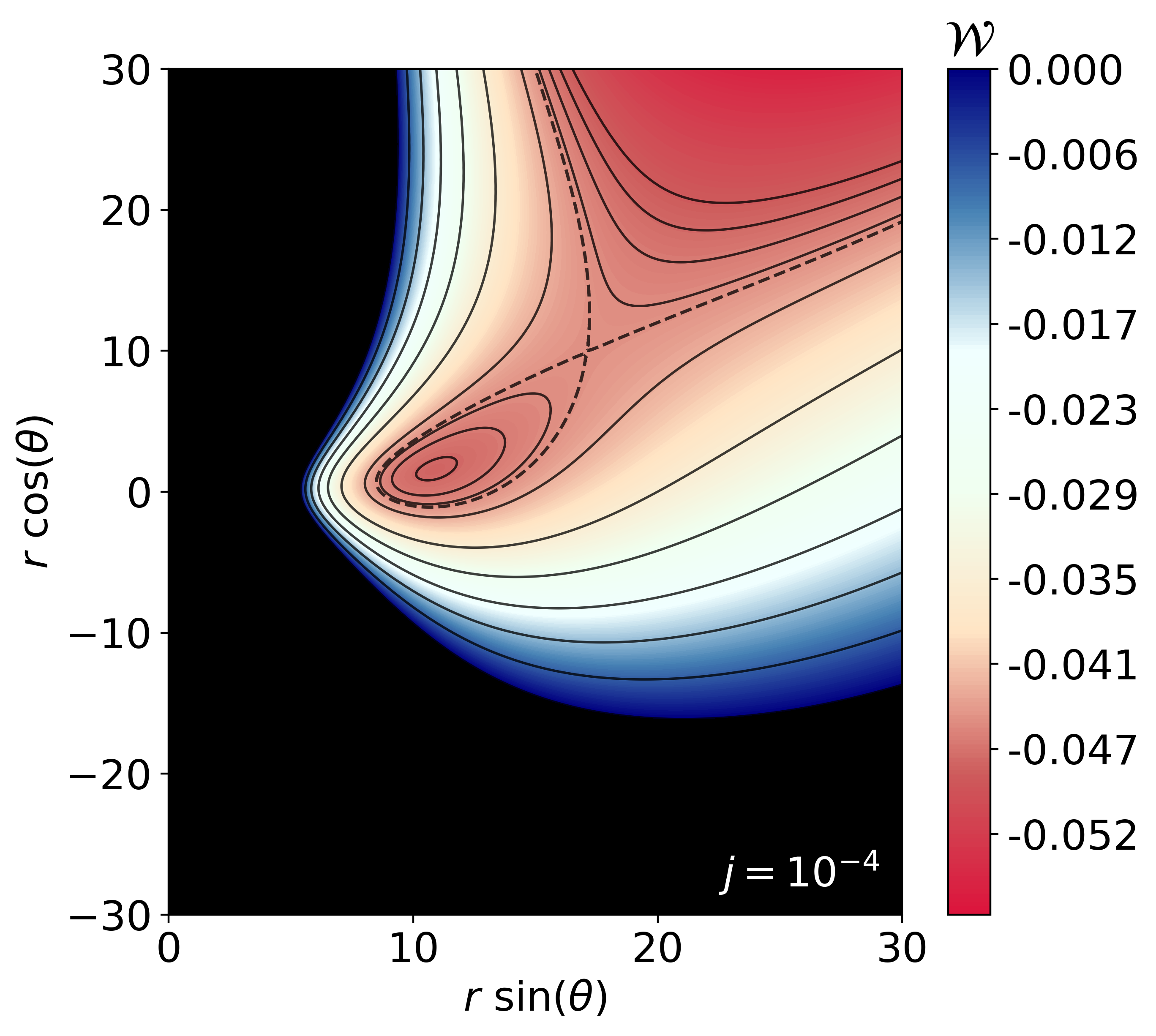}
  \caption{$a = 0.9$, $\ell_0 = 3.7$}
\end{subfigure}
\caption{Meridional cross section of the effective potential $\mathcal{W}$ for different disk solutions with the same
swirling parameter $j = 10^{-4}$ and different Kerr parameter $a$. The specific angular momentum of the disks is set to $\ell_0 = 3.7$ in all solutions. The lower boundary of the scale is set in all solutions to $\mathcal{W} = -0.055$ for better comparison. Black curves represent equi-potential surfaces. The dotted curve in (a) corresponds to the equi-potential surface going through the inner cusp. Dashed curves represent the equi-potential surface going through the outer cusp.}
\label{fig:disks_l3.7}
\end{figure}

Using the same disk parameters besides the Kerr parameter, reveals that prograde disk solutions are in their general morphology largely unaffected by the Kerr parameter. All disk solutions are asymmetrical to the equatorial plane, and their spatial distribution is similar, regardless of the Kerr parameter. They extend mainly into the upper hemisphere with only a small region extending into the lower hemisphere. This effect originates from the $\mathcal{Z}_2$ symmetry of the swirling background and increases with increasing $j$. We follow, that even in the vicinity of the KBHSB, the Kerr rotation gets dominated by the swirling rotation, albeit $j$ being small compared to $a$. However, differences regarding the physical disk properties are noticeable. For the same disk parameters, the effective potential for the slower Kerr rotation has a deeper minimum and therefore a higher density at the center of the disk compared to the faster rotating black hole. The disk solution for $a = 0.2$ is of Type 1a, where the disk center is enclosed by the equi-potential surface of the outer cusp, which itself is surrounded by the equi-potential surface of the inner cusp (Fig. \ref{fig:disks_l3.7} (a)). The $a = 0.5$ and $a = 0.9$ disk solutions are of Type 2, with only an outer cusp. Since they do not possess an inner cusp, no accretion process without further inner-physical disk effects is possible. For faster rotating KBHSBs, prograde disks solutions can exist for higher $j$ and smaller values of lower $\ell_0$, leading to more compact disks, as shown in Fig. \ref{fig:disk_a0.9_l2.55}.

\begin{figure}[H]
\centering
\begin{subfigure}{.325\textwidth}
  \centering
  \includegraphics[width=\linewidth]{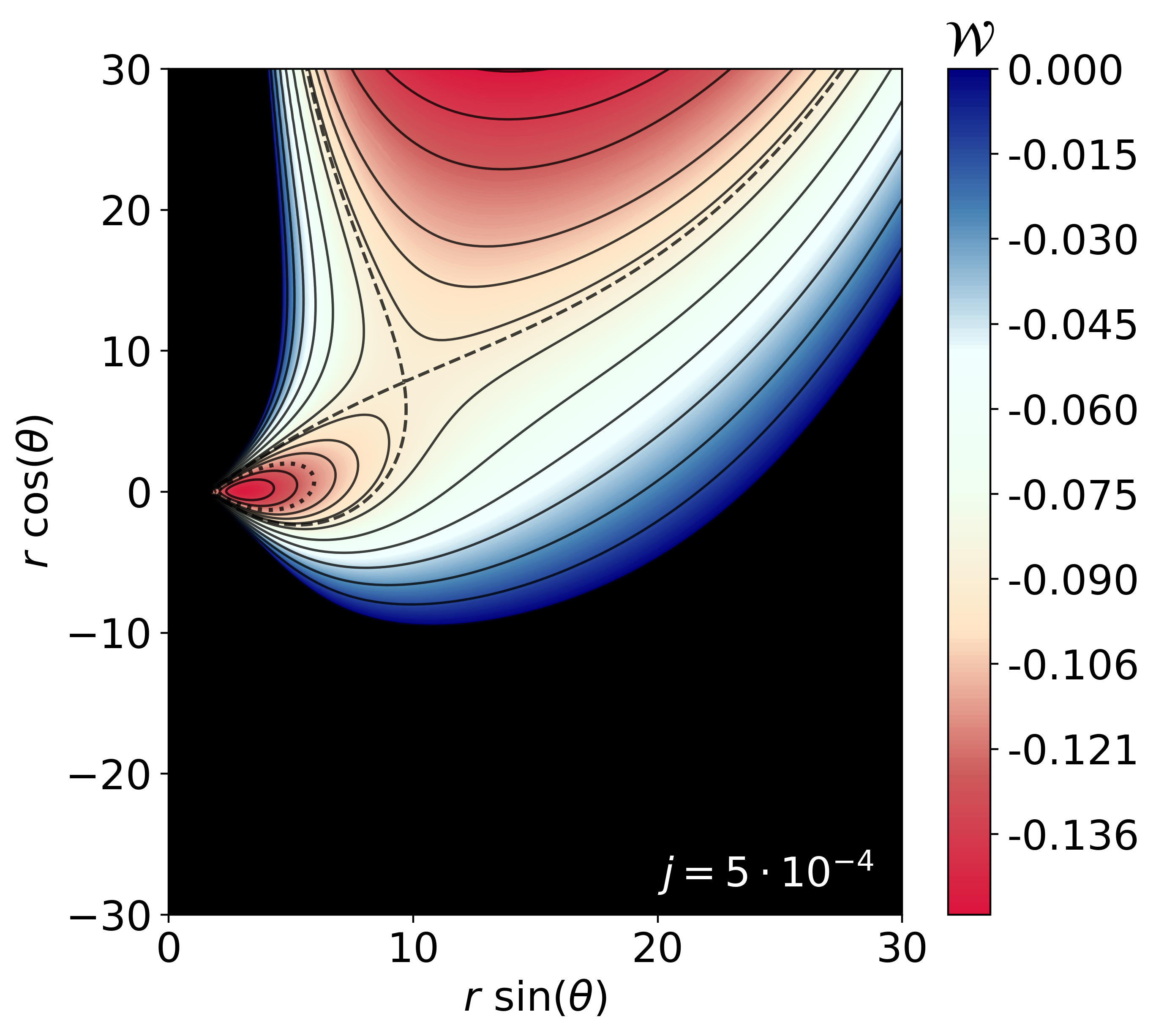}
  \caption{}
\end{subfigure}
\begin{subfigure}{.325\textwidth}
  \centering
  \includegraphics[width=\linewidth]{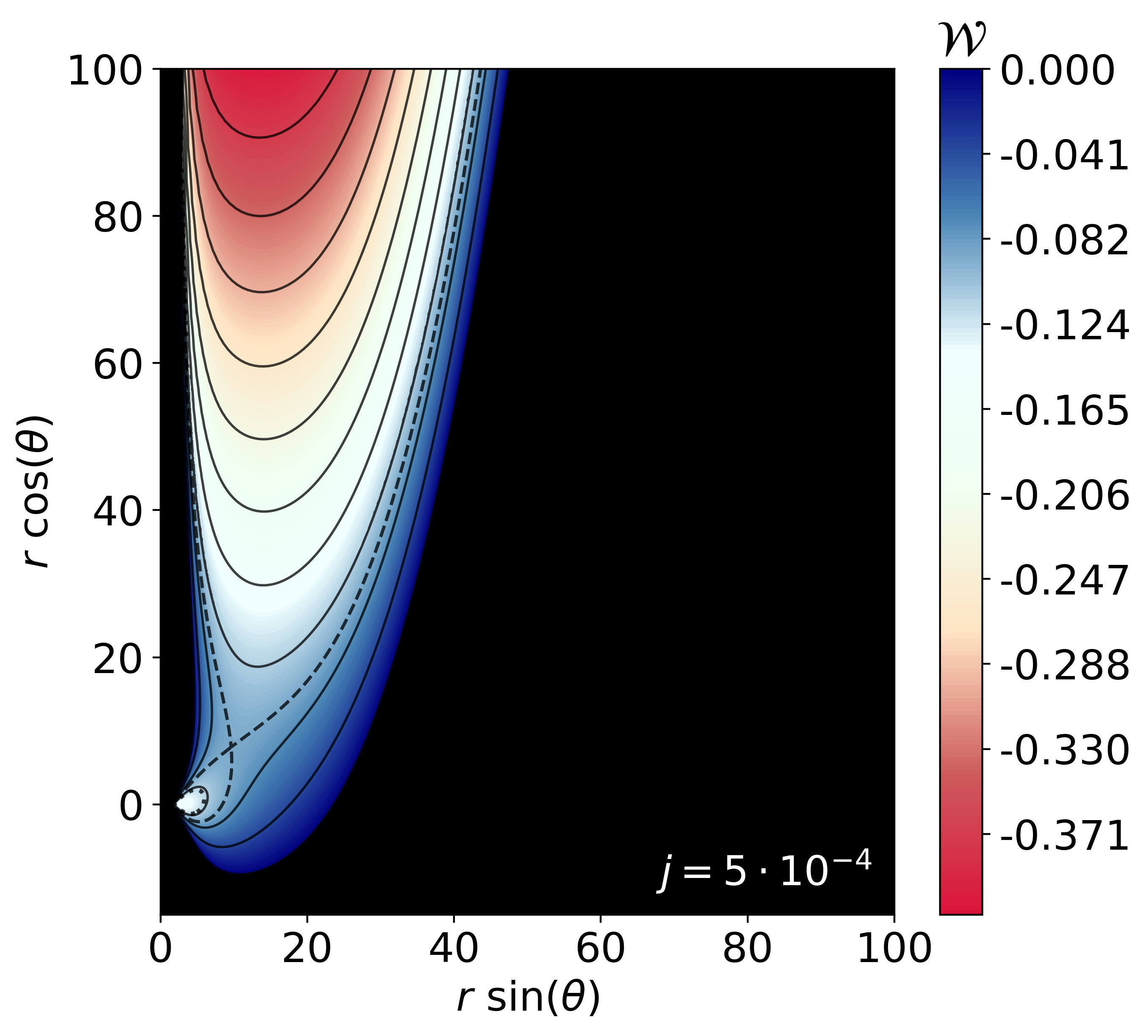}
  \caption{}
\end{subfigure}
\caption{Meridional cross section of the effective potential $\mathcal{W}$ for a Type 1b disk with $a = 0.9$, $j = 5 \cdot 10^{-4}$ and $\ell_0 = 2.55$. (a) presents a more detailed view of the center region, with the dotted curve corresponding to the equi-potential surface of the inner cusp. (b) presents a broader view of the vertical structure. The dashed curves represent the equi-potential surface going through the outer cusp in both plots.}
\label{fig:disk_a0.9_l2.55}
\end{figure}

In this Type 1b solution above, the center region is highly compact, most of the matter is located close the density maximum (minimum of $\mathcal{W}$) and the disk center is located closely to the KBHSB center. The effective potential decreases monotonically outwards the outer cusp, also on a greater scale. We follow that with increasing $a$ and $j$ and decreasing $\ell_0$ disk solutions are more localized and compact. The verticality of the accretion structure is mainly affected by the swirling parameter. Since with increasing $a$ disk solutions are possible for larger values of $j$, solutions with a greater verticality are possible for faster rotating KBHSB.

In the following Fig. \ref{fig:disks_retro}, examples of retrograde disk solutions are presented for various values of $a$, $j$ and $\ell_0$. Due to the more limited parameter space for retrograde accretion structures, solutions with the same $\ell_0$ and $j$ value do not exist for the picked Kerr parameter ($a \in \{0.2,0.5,0.9\}$). A comparison for different Kerr parameter but same swirling and disk parameters, like done for the prograde disks, is therefore not possible. Thus, we picked exemplary solutions with different parameters to showcase the main properties of retrograde disks.

\begin{figure}[H]
\centering
\begin{subfigure}{.325\textwidth}
  \centering
  \includegraphics[width=\linewidth]{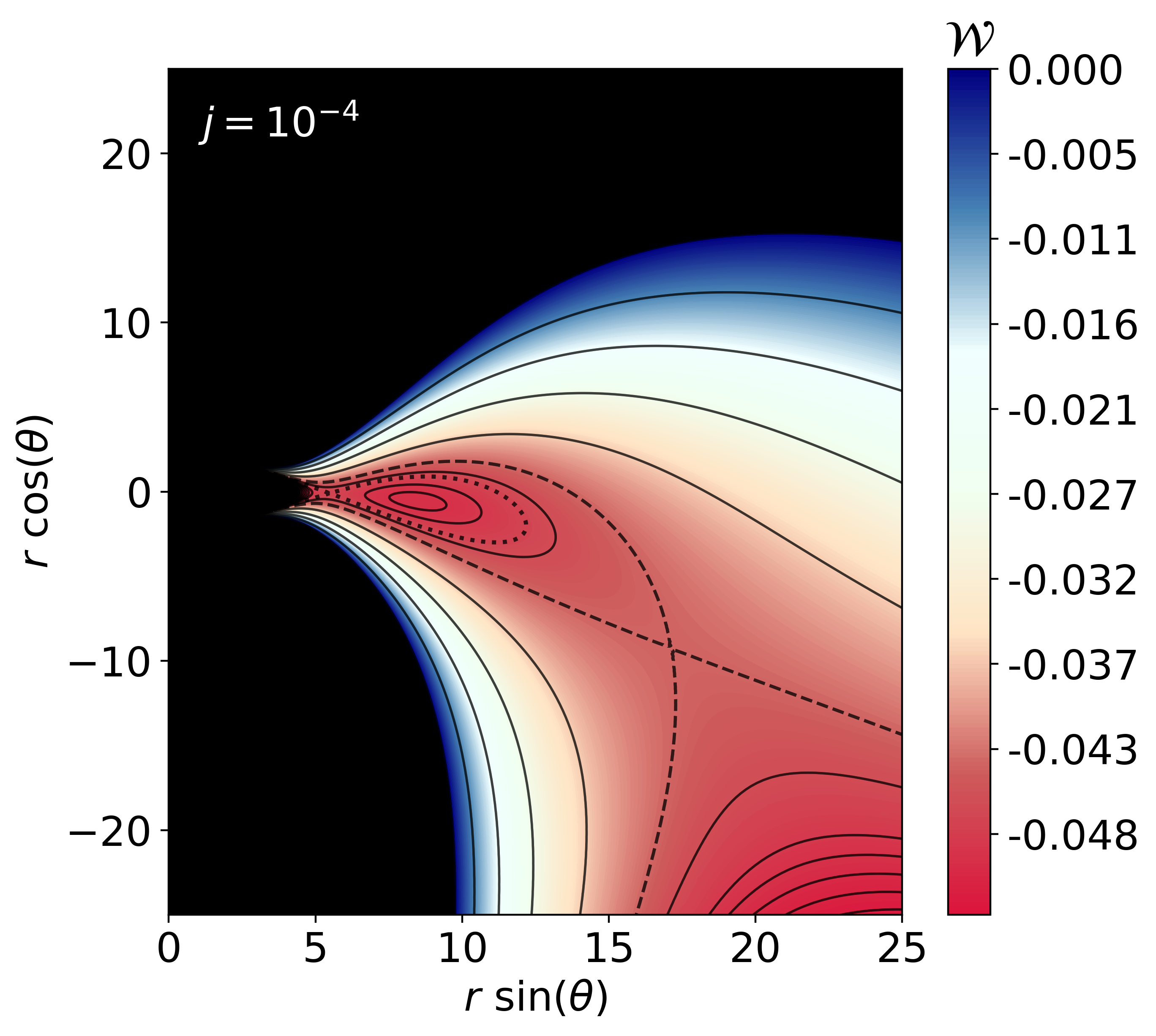}
  \caption{$a = 0.2$, $\ell_0 = -3.7$}
\end{subfigure}
\begin{subfigure}{.325\textwidth}
  \centering
  \includegraphics[width=\linewidth]{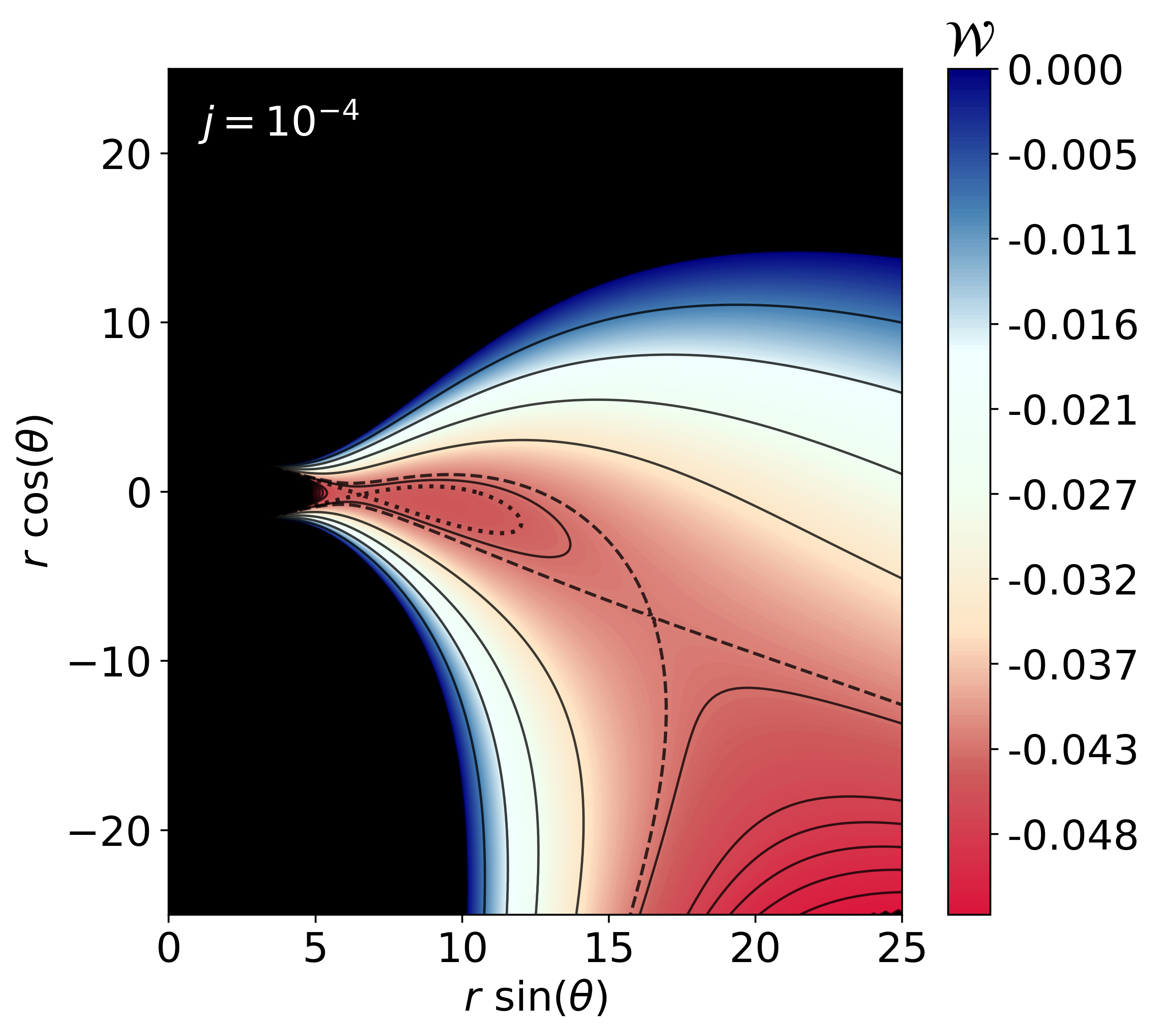}
  \caption{$a = 0.5$, $\ell_0 = -4.1$}
\end{subfigure}
\begin{subfigure}{.325\textwidth}
  \centering
  \includegraphics[width=\linewidth]{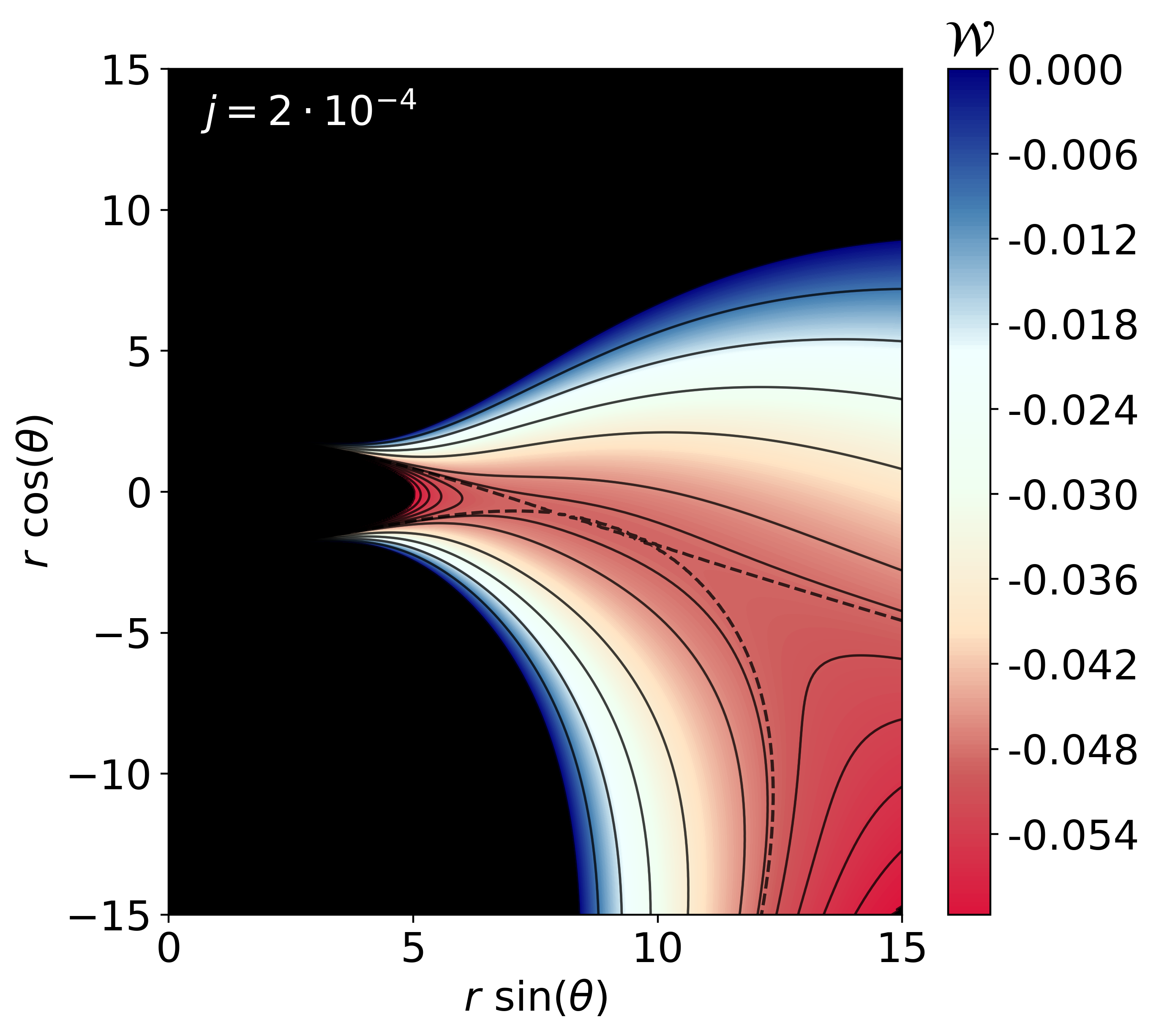}
  \caption{$a = 0.5$, $\ell_0 = -4.0335$}
\end{subfigure}
\caption{Meridional cross section of the effective potential $\mathcal{W}$ for different retrograde disk solutions. Black curves represent equi-potential surfaces. Dotted curves correspond to the equi-potential surface of the inner cusp. Dashed curves represent the equi-potential surface of the outer cusp.}
\label{fig:disks_retro}
\end{figure}

Retrograde disks are similar to prograde disks regarding their asymmetry with respect to the equatorial plane. In contrast to prograde disks, they extend mainly into the lower hemisphere with only a small part extending into the upper hemisphere. This corresponds to the spatial distribution of the retrograde circular orbits, which originates from the change of the swirling rotation in the lower hemisphere. With increasing $j$, the asymmetry and verticality of the accretion structure increases. The Figs. \ref{fig:disks_retro} (a) and (b) present retrograde Type 1b disks, for which the same properties apply as explained for the prograde case. The Fig. \ref{fig:disks_retro} (c) illustrates a retrograde Type 1c disk, where the effective potential at the inner and outer cusp has the same value. This value is also closely located to the potential minimum. The potential difference between the disk center and inner/outer cusp is minuscule, which causes the enclosed region between inner and outer cusp to appear as vanishing. The only stable region of the structure lies inside this small narrow region close to the disk center. Regarding disk dynamics this solution could therefore be highly unstable. The following Fig. \ref{fig:disk_a9_retro} presents a retrograde disk solution for the fast rotating KBHSB with $a = 0.9$.

\begin{figure}[H]
\centering
\begin{subfigure}{.325\textwidth}
  \centering
  \includegraphics[width=\linewidth]{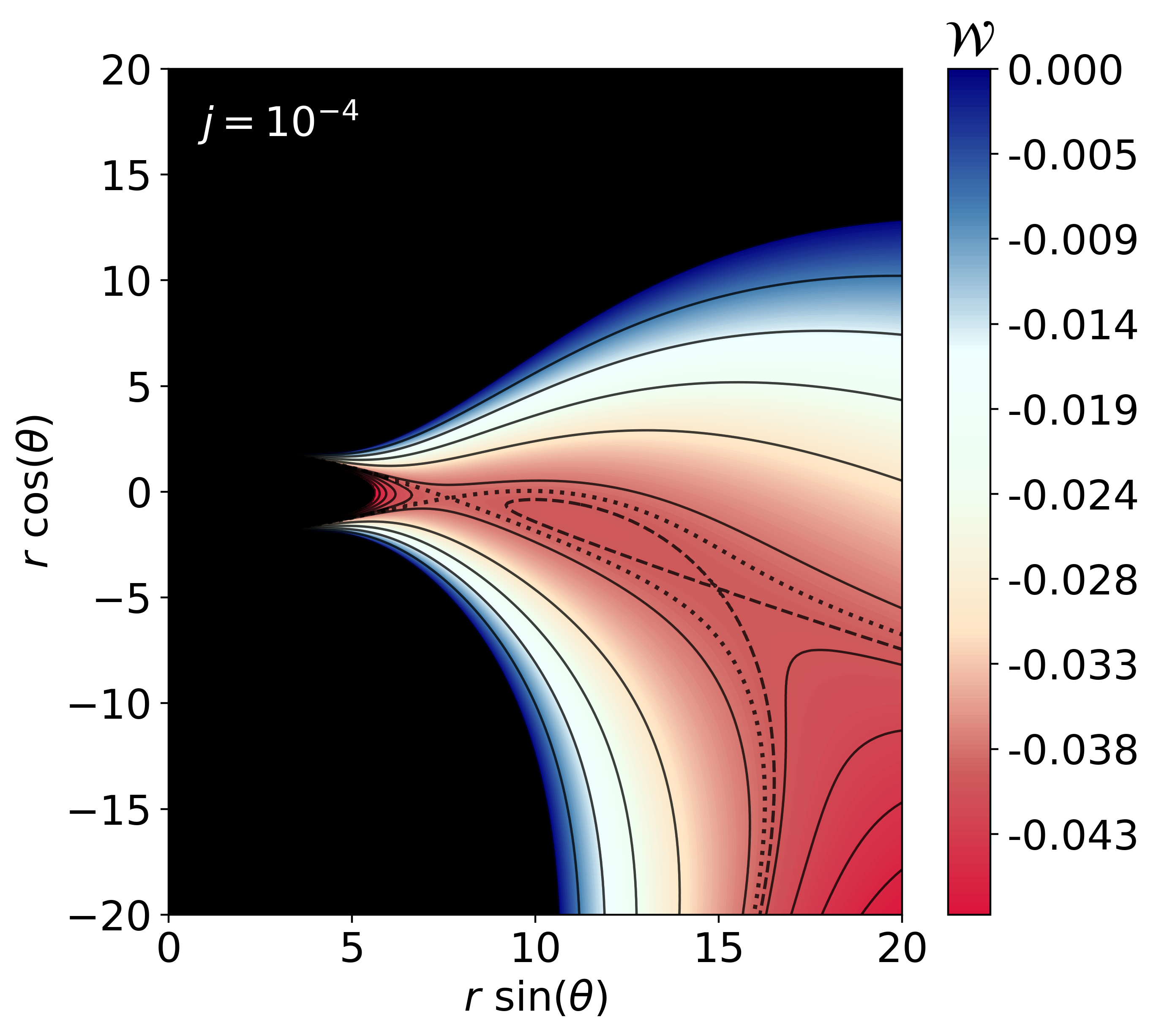}
  \caption{}
\end{subfigure}
\begin{subfigure}{.325\textwidth}
  \centering
  \includegraphics[width=\linewidth]{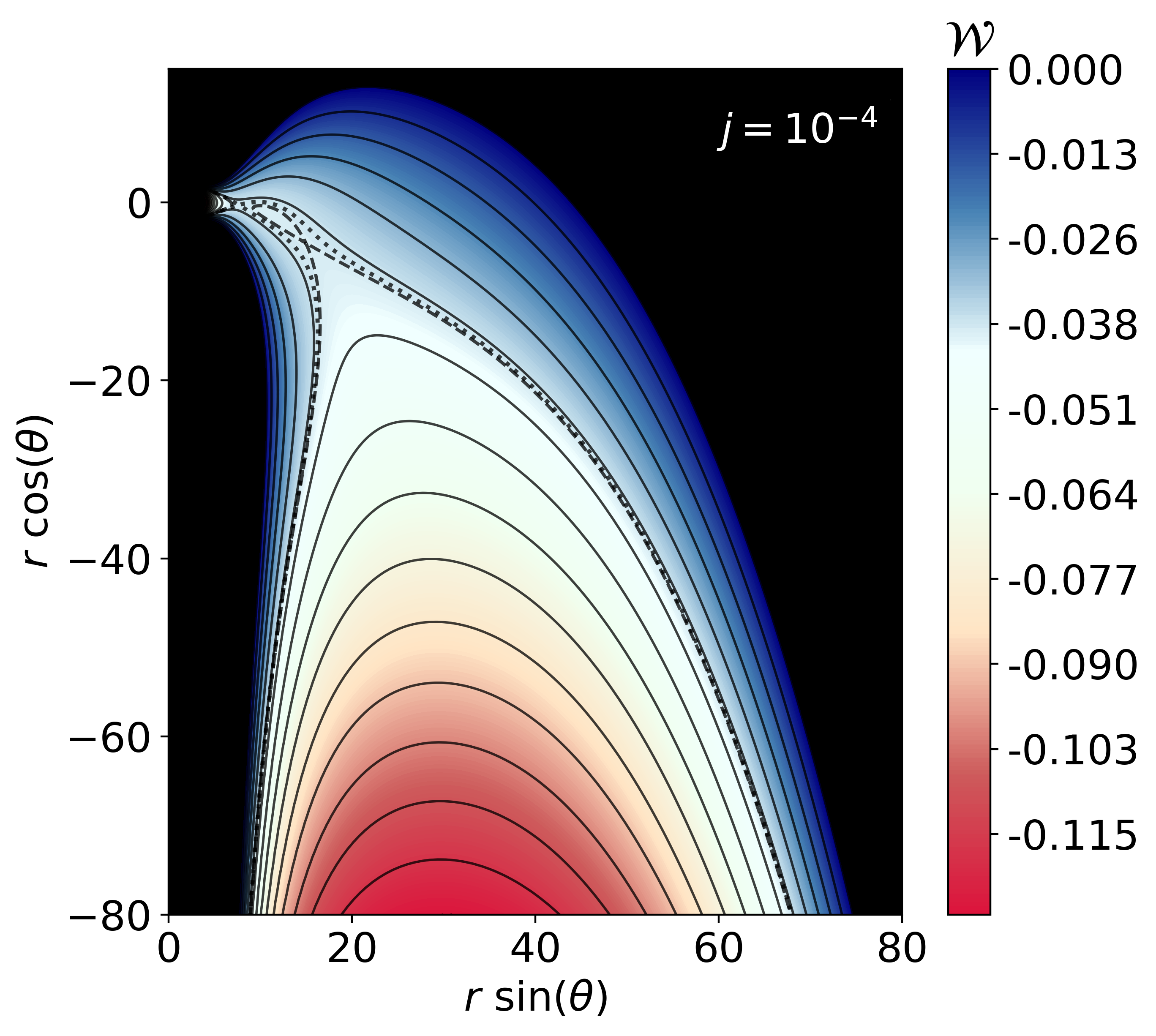}
  \caption{}
\end{subfigure}
\caption{Meridional cross section of the effective potential $\mathcal{W}$ for a retrograde Type 1a disk with $a = 0.9$, $j = 10^{-4}$ and $\ell_0 = -4.36$. (a) presents a more detailed view of the center region, with the dotted curve corresponding to the equi-potential surface of the inner cusp. (b) presents a broader view of the vertical structure. The dashed curves represent the equi-potential surface going through the outer cusp in both plots.}
\label{fig:disk_a9_retro}
\end{figure}

The presented solution above is of Type 1a, with the equi-potential surface of the outer cusp enclosing the disk center. The main disk properties are similar to the prograde Type 1a disks. Nevertheless, for faster rotating KBHSB, the retrograde disk center moves outwards and the disks are less compact compared to other disk solutions. This is also reflected by the large scale structure (Fig. \ref{fig:disk_a9_retro} (b)), where the horizontal extension is greater compared to the prograde case for $a = 0.9$ due to the smaller swirling parameter $j$. A greater verticality would correlate with a higher value of the Swirling parameter; however, the retrograde disk parameter space regarding the swirling parameter is very limited in the case of faster rotating KBHSB, since for higher $j$ no stable circular geodesics and therefore no disk solutions exist.

\section{Comparison to Thick Disks around Schwarzschild Black Holes in a Swirling Background}
Ahead of the conclusion, we want to conduct a brief comparison to thick disks around a Schwarzschild black hole in a swirling background, which were studied in \cite{Chen2024}, and point out the effects of the black hole rotation and their consequences. We identify the spin-spin interaction of the rotating black hole with the swirling rotation of the spacetime background as a major differentiating factor for possible disk solutions. The spin-spin interaction influences the stability properties of circular geodesics and by that, it also affects the parameter space for thick disk solutions. The Kerr spin parameter determines the critical Swirling parameter, $j_c = j_c(a)$, up to which stable circular geodesics exist in the spacetime. For prograde motion, $j_c$ is an increasing function of $a$, while for retrograde motion, it is a decreasing function of $a$. If the value of the swirling parameter is fixed, the black hole rotation determines the spectrum of specific angular momenta for the disk particles, for which solutions exist (Fig. \ref{fig:W_j4}). In the case of prograde motion, a higher value of the Kerr spin parameter leads to a greater range of possible values of the specific angular momentum, while for retrograde motion the opposite holds. However, the vertical position of circular geodesics is mostly unaffected by the Kerr spin parameter, hence it is mainly determined by the Swirling parameter. Thus, the spatial distribution of circular geodesics stays similar to the Schwarzschild case for all Kerr spin parameters (Fig. \ref{fig:ell_K_2D}). Additionally, the positions of static orbits are mostly unaffected by the choice of the Kerr parameter (Fig. \ref{fig:static_orbits}). When it comes to the disk morphology and matter distribution, the Kerr parameter has a smaller influence, the disk shape is mainly determined by the Swirling parameter. However, disk solutions for Kerr spin parameter near the Kerr limit are more vertically bent and less horizontally outreaching. This applies to prograde as well as retrograde disks. Finally, we conclude that the major differences resulting from the spin-spin interaction appear in the range of the parameter space for which stable circular geodesic and by that also accretion structures can exist.

\section{Conclusion}
In this work, we investigated qualitatively geometrically thick disks around Kerr black holes in a swirling background. We classified different KBHSB solutions by their Kerr spin parameter $a$ and varied for fixed $a$ the swirling parameter $j$ to analyze the influence of the swirling background and the effects of the spin-spin interaction between the black hole and the background. For an exemplary analysis regarding the circular orbits and disk solutions, we chose the Kerr parameters $a = 0.2$, $a = 0.5$, and $a = 0.9$, for slow, medium, and fast rotating black holes, respectively. The swirling parameter was varied between $j = 10^{-5}$ and $5 \cdot 10^{-4}$, as for higher values of $j$ no stable circular orbits exist. Fruthermore, we have considered prograde and retrograde motion. In both cases, the properties of the orbits and disks are highly affected by the swirling parameter, as circular orbits deviate from the equatorial plane with increasing $j$. Prograde (retrograde) orbits get shifted into the upper (lower) hemisphere, where they are co-moving with the swirling background, and the verticality of the spatial distribution increases with $j$ for all possible values of $a$. The differences regarding the spatial distribution for the different Kerr parameters are negligible. The Keplerian specific angular momentum distribution develops in all cases an additional outer extremum, which marks an outer marginally stable orbit. The range of stable orbits is therefore bounded by the inner and outer marginally stable orbit, which also severely limits the parameter space of possible disk solutions.

In the case of prograde orbits, we found a stabilizing effect of the spin-spin interaction, since for higher values of $a$, stable circular orbits were possible for larger $j$. Faster rotating black holes therefore exhibit a broader parameter space for prograde disk solutions. For retrograde orbits, which are all located in the lower hemisphere, we found the opposite effect. In the lower hemisphere, the black hole is counter-rotating to the swirling background, which has a destabilizing effect on the circular orbits. Therefore, with increasing $a$, the range of stable circular orbits decreases for larger $j$ and the parameter space for disk solutions shrinks.

In all solutions, the rotation of the swirling background causes the emergence of a static orbit at which a test particle is at rest in the co-moving reference frame. Since the rest specific angular momentum has a higher absolute value for all orbits outside the static orbit, test particles on these outside orbits are counter-rotating in the co-moving reference frame of spacetime.

Due to the outer extremum of the Keplerian specific angular momentum, all possible accretion structures possess an outer cusp. In dependence of the KBHSB solutions and the chosen value of the disk momentum $\ell_0$, we found and classified four different possible disk types. Type 1 solutions are characterized by an inner cusp and an outer cusp, between which the disk center lies. Type 2 solutions only exist for prograde disks for the analyzed values of $a$ and are characterized by the absence of an inner cusp, they only possess an outer cusp. The Type 1 solutions can be further subclassified regarding their effective potential values at the cusps. For Type 1a solutions, the effective potential has a higher value at the inner cusp than at the outer cusp. This causes the equi-potential surfaces of the inner cusp to be open towards the center and towards the outer disk regions. The equi-potential surface of the outer cusp fully encloses the disk center region and is closed towards the KBHSB center but open towards the outer region. In the case of Type 1b solutions, the effective potential of the outer cusp has a higher value than the inner cusp. This causes the equi-potential surface of the outer cusp to be open towards the center and towards the outer region. The inner cusp potential surface now fully encloses the disk center, it is closed towards the outer region and open towards the KBHSB center. In the special case of Type 1c, the effective potential has the same value at the inner and outer cusp, the potential surface is fully closed between the inner and outer cusp, and it encloses the disk center but is open towards the KBHSB center and the outer disk region. All mentioned solutions have a high complexity of disk dynamics due to the different conceivable accretion mechanisms regarding the inner as well as the outer cusp. However, direct accretion onto the KBHSB, with the absence of inner-physical disk effects, is only possible for Type 1 solutions, as Type 2 do not possess an inner cusp and are closed towards the KBHSB center.

\newpage

\begin{acknowledgments}
We would like to thank Betti Hartmann and Jutta Kunz for many of the discussions that led to this paper.
R.C. is grateful to CAPES for financial support under Grant No: 88887.371717/2019-00, and would like to thank the University of Oldenburg for hospitality.
\end{acknowledgments}

\bibliography{literature}

\end{document}